\documentclass[useAMS,usenatbib,referee]{mn2e}
\usepackage{graphicx}
\usepackage{natbib}
\usepackage{times}
\usepackage{float}
\usepackage{amsmath}
\usepackage{epsfig}
\usepackage{xcolor}
\definecolor{red}{cmyk}{0,0,0,1}
\definecolor{blue}{cmyk}{0,0,0,1}
\newcommand{\bmth}[1]{\mbox{\boldmath${#1}$}}

\title[On the evolution of a binary system 
due to quasi-stationary tides]
{On the evolution of a binary system 
with  arbitrarily misaligned  orbital and stellar angular momenta 
due to quasi-stationary tides}
\author[ P. B. Ivanov and J. C. B. Papaloizou ]{P.B.Ivanov$^{1}$\thanks{E-mail:
pbi20@cam.ac.uk (PBI)}, J. C. B. Papaloizou $^{2}$\thanks{E-mail:
J.C.B.Papaloizou@damtp.cam.ac.uk (JCBP)}\\
$^{1}$Astro Space Centre, P.N. Lebedev Physical Institute, 84/32
Profsoyuznaya Street, Moscow, 117997, Russia  \\
$^{2}$ DAMTP, Centre for Mathematical Sciences, University of
Cambridge, Wilberforce Road, Cambridge CB3 0WA }

\begin{document}

\date{Accepted. Received; in original form}

\pagerange{\pageref{firstpage}--\pageref{lastpage}} \pubyear{2010}

\maketitle

\label{firstpage}

\begin{abstract}

We consider the evolution of a binary system  interacting due to tidal effects
without restriction on the orientation of the orbital,  and where significant, spin angular momenta,
and orbital eccentricity.
We work in the low tidal forcing frequency  regime in the equilibrium tide approximation.
 Internal degrees of freedom are fully taken into account for
one component, the primary. In the case of the companion the spin angular momentum is
assumed small enough to be  neglected but  internal energy dissipation is allowed for
as this can be significant for orbital circularisation in the case of planetary companions. 
We obtain a set of equations governing the evolution of the orbit resulting from tidal effects.
These depend on
  the masses and radii  of the binary components, 
 the form and
orientation of the orbit, and for each involved component,  the spin rate,   the Coriolis force,
the normalised rate of energy dissipation  associated with the equilibrium tide due to radiative processes and viscosity, 
and  the classical apsidal motion constant, $k_2.$ 
These depend on stellar parameters
with no need of additional assumptions or a  phenomenological approach as has been invoked in the past.
They can be used to determine the evolution of systems with initial significant misalignment of 
spin and orbital angular momenta  as  hypothesised for systems containing Hot Jupiters. 
The inclusion of the Coriolis force
may lead to evolution  of the inclination between orbital
and spin angular momenta and precession of the orbital plane 
which may have observational consequences.

\end{abstract}


\section{Introduction}
\textcolor{red} {Tidal interactions are important
in close binary systems where they determine the rates of  orbital  circularisation,
as well as synchronisation  and alignment of the component spins with the  angular momentum of the orbit
\citep[see][for a review]{O2014}.  Recent attention has focused on stars with planetary companions
such as  hot Jupiters where tides have been postulated to play an important role in shaping the system.
Tidal interactions have been considered in a low tidal forcing equilibrium tide regime
or alternatively in a regime where so called dynamical tides and the excitation of 
normal modes in one or both of the components is important.}  

\textcolor{red} {In this paper  we  consider  tidal interactions in the equilibrium tide regime in
a binary system consisting of a primary component with spin angular momentum that is arbitrarily misaligned 
with that of the binary orbit. The companion is assumed to be compact and initially with no internal degrees of freedom
though later this will be relaxed to allow for internal energy dissipation that can contribute to the circularisation of the orbit.
This is important if the model is to be applied to cases with planetary mass companions.}

\textcolor{red}{\citet{EKH},  hereafter EKH, derived  the force and couple on a binary
orbit that arises from dissipation of an assumed equilibrium tide under the imposed assumption that
the rate of dissipation of energy is  a positive definite function of the rate of change of the
primary quadrupole tensor  viewed in a frame rotating with the star.
Coriolis forces were neglected.
Their model  was found to lead to results obtained under the assumption of a constant time lag
between the tidal forcing and  response without explicit reference to dissipative processes as implemented
by \citet{Hut}.
They also determined a  way to connect the
hypothesised  relation between the  dissipation  rate and quadrupole tensor
to  a postulated  turbulent viscosity. However, this was done by determining the velocity field
from the continuity equation alone requiring an additional assumption connecting  the radial and non radial components
and so is  incomplete. In addition, EKH only considered tidal dissipation in one primary component.}

\textcolor{red}{We calculate the response of the primary to tidal forcing from first principles in the low tidal forcing frequency limit.
 In the leading approximation the stellar configuration is fully adjusted under the action of tidal forces. Then, we include dissipative processes and Coriolis forces as next order corrections, to first order in
the primary rotational frequency.
In the case of the former we include both radiative and viscous effects noting that while they may dominate 
in the low viscosity case radiative effects have not been considered previously in this context.
Notably our approach removes any need for ad hoc assumptions about these processes such as connecting them
with the behaviour of the quadrupole tensor and provides a complete form for the response displacement
without the need for assumptions about unknown functions.}

\textcolor{red}{In common with previous treatments, centrifugal distortion and 
the toroidal component of the  response displacement is neglected.
The latter is potentially important in the inertial regime where the tidal forcing frequency is less than
twice the rotation frequency  on account of the possible excitation of inertial modes \citep[e.g.][]{PP81, PS97,SP97, IP2007, IP2010}.
However, these may give only a minor contribution due to a weak overlap with the forcing potential.
We use the calculated response to determine the effect of the tidal interaction on the orbit
and go on to obtain equations governing the evolution of the orbit.  }

 A qualitative difference between our results and \textcolor{blue} {those} reported elsewhere 
{(e.g. \cite{EKH}, \cite{Barker2009}) }is the appearance of new terms
determined by \textcolor{blue} { rotational effects}.
Unlike the standard terms due to dissipative processes in
a star the new ones are readily calculated \textcolor{blue}{in detail}. In addition to the orbital parameters,  they are obtained from  properties
of the star, which are, in known in principal,
 being  the stellar density distribution  and angular velocity.

The new terms lead to precession of orbital plane \textcolor{blue}{, additional to that induced through
orbital torques and  stellar  centrifugal distortion, as well as  } 
 non-dissipative evolution of the angle
between  the orbital and spin angular momenta.
 Both effects are
most prominent for binaries with sufficiently large  \textcolor{blue} {companion}  masses and
eccentricities. In particular, a
typical change of the inclination angle over a timescale
determined by apsidal precession  \textcolor{blue}{ is potentially  significant} provided
the eccentricity is substantial, with optimal value $\sim 0.7$, the
orbital period is sufficiently small, while 
mass ratio and \textcolor{blue}{ primary angular velocity}  are sufficiently
large. This \textcolor{blue}{could}  lead to observational consequences. 
\textcolor{blue}{Furthermore}
  the effect  of \textcolor{blue}{ terms arising from the rotation} 
 should also be taken
into account when studying the evolution on the tidal friction time scale
  of the system when the apsidal precession is non-uniform, say, due 
to the presence of a third perturbing body.

\textcolor{red}{
Before giving a complete  plan of the paper we remark that those interested in
the equations we derived governing the orbital evolution but who wish to avoid the lengthy derivations
can, after reading the basic setup in Sections \ref{Basiceq} and \ref{Geometry} skip to Sections 
\ref{Detorbspinev} - \ref{inccomp} which contain the equations together with a summary  account of
the parameters involved.}

\textcolor{red}{The complete plan is as follows. In Section \ref{Basiceq} we give some  basic definitions and  equations.
In Section \ref{Geometry} we define the three coordinate systems
we use to represent the dynamics of the binary with misaligned orbital  primary component spin
angular momenta. One  is { a frame  with origin at the centre of mass of the primary}  with vertical axis aligned with the}
\textcolor{red}{conserved total angular momentum vector. A second, used to describe the orbit has vertical axis
aligned  with the orbital angular momentum vector, and the third, used to describe the primary
has vertical axis aligned with its spin angular momentum vector. In 
Section \ref{Jvecev}, based on angular momentum  conservation, 
we go on to derive equations  governing the evolution of  the angular momentum vectors
that are determined by torques exerted between the orbit and primary that we go on to calculate.}

\textcolor{red}{The calculation begins with  the specification of the perturbing  tidal potential and its representation in terms of spherical harmonics in
the orbit and {stellar spin } based coordinate systems
in Sections \ref{pertpoten} and \ref{Fourierpot}. Transformation of the representation between the systems is
facilitated with Wigner matrices.  We go on to calculate the tidal response of the primary
in the equilibrium tide limit in Section \ref{Respd}. The response  which includes
first order departures arising from Coriolis forces and dissipation is found in terms of the 
displacement associated with equilibrium tide in Sections 
\ref{Findingresp} -
\ref{Respoverlap}.}

\textcolor{red}{Having determined this,  the  induced torque acting on the star
is obtained in Section \ref{TorqueFind} and the induced 
 rate of change of orbital energy in Section \ref{Orbenrate}.
 These are time averaged and reduced to closed form in Sections
 \ref{Torquereduct}- \ref{Inteval}
with details given in appendix \ref{Aptima}.}

\textcolor{red}{Having obtained the time averaged torques and rate of change of orbital energy these are used
 in our set of equations for the determination of the orbital  and spin  evolution in Section \ref{Detorbspinev}
with expressions for the evolution of angular momentum vectors being given in Section \ref{Amvexp}
and. expressions for the evolution of the semi-major axis and eccentricity in Section \ref{andeev}.
The description is augmented by including energy dissipation in the companion under the assumption
of negligible spin angular momentum in Section \ref{inccomp}. This is an approximation appropriate for
low mass planetary companions.}

\textcolor{red}{A discussion of the evolution of orbital parameters  when only dissipative terms are included  (thus,  effects due to rotation} are neglected)
 is given in in \ref{eqsum}, where we relate our results to those of EKH in the appropriate limit.
 The potential contribution of terms  arising from { rotational effects including} the Coriolis force to the
orbital evolution is then  considered in Section \ref{NonrotT}.
Finally, in
Section \ref{Discussion} we review and discuss our results.

\section{Basic definitions and equations}\label{Basiceq}
\textcolor{red} {In this Section we  describe the basic model setup and outline the objective
of determining the tidal response  and its using that to  determine the exchange rates of angular
momentum and energy between the non compact stellar component and the orbit.
We go on to show how the evolution of the orbital  and stellar angular momenta may be determined. }
\subsection{Basic model}
\textcolor{red}{ For simplicity,} we consider a binary system for which one of the components \textcolor{red}{initially acts as } a point  mass, i.e. it has
no internal degrees of freedom and said to be \textcolor{red} {compact} and \textcolor{red}{ described as the companion.}
\textcolor{red} {Although later we shall relax this to allow for internal energy dissipation while still neglecting the
internal angular momentum content.} The other component, \textcolor{red} {labelled the primary,}  possesses a distributed mass and a spin angular momentum which 
is unrestricted in comparison to the orbital angular momentum. Thus, both these angular momenta
are allowed to evolve with the resultant total angular momentum being conserved.

We adopt a description of the system  of interest by making use  of quantities such as the tidal potential $U$ and  the Lagrangian 
displacement vector ${\mbox{{\boldmath$\xi$}}}$ associated with the non compact star etc. 
as in \cite{IP2011}. However, unlike that
 Paper we assume an elliptic 
orbit for the binary, and, accordingly,  represent such  quantities as  Fourier 
series in time  as in \cite{IP2004}. \textcolor{red}{  However, unlike these Papers, where it was assumed that 
only the  f-mode amongst  the spectrum of stellar normal modes is excited,  here we consider the response of the star influenced by tides 
in the so-called quasi-static or equilibrium tide approximation.
We note, however, that, formally, this approach can be shown
to coincide with  the one considered in those Papers after an appropriate redefinition of relevant quantities.}

In addition to determining the response  we aim  to show that 
the  evolution of the angular momentum vectors is  \textcolor{red}{fully } determined
by  \textcolor{red} { four simple governing} equations  
\textcolor{red}{following from the law  of conservation of angular momentum  provided 
the torques
acting on the star as a result of tides  have been determined.  Calculation of the energy exchange with the orbit 
and application of the law of conservation of energy  then enables a complete description of the
evolution of the system once a prescription to determine the evolution of the orbital apsidal line
is prescribed. }

\subsection{Coordinate system and notation}\label{Geometry}
We introduce three reference frames.
The first is a Cartesian coordinate system in {a frame with origin at the centre of mass of the primary,  and for which} the 
direction of the
conserved total angular momentum of the system,
${\bf J}$, defines the $Z''$ axis. The corresponding  $X''$ and $Y''$ axes are 
located in the orthogonal plane.

The second is a Cartesian  frame such that  the orbital  angular momentum, ${\bf L}$, 
defines the direction of the $Z'$ axis. This is inclined to the total angular momentum vector, ${\bf J}$,  with an 
inclination $i$ which need not be constant as the orbital angular momentum is not conserved.

\begin{figure}
\begin{center}
\vspace{1cm}
\includegraphics[width=14.0cm,height= 18.0cm,angle=0]{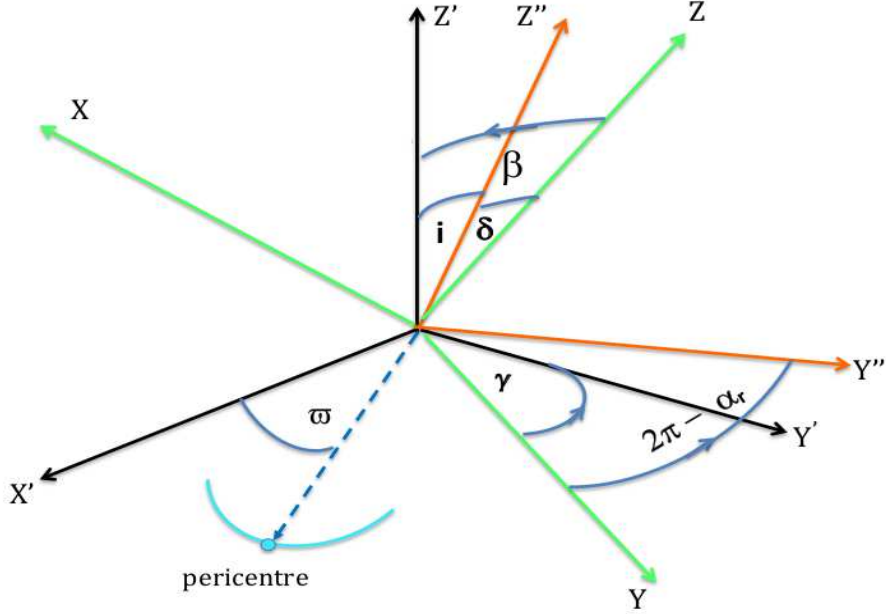}
\vspace{-7cm}
\caption{Illustration of the $ (X,Y,Z)$  and   $(X',Y',Z')$ coordinate systems
together with the direction of the total angular momentum, which coincides with the $Z''$ axis
of  a coordinate system that is fixed in { the primary centred}  frame .
Note that the $ X', Y', $ and $Y$ axes are coplanar as are the  $Z, Z' $ and $Z''$ axes .
 The angle between the angular momentum vectors ${\bf L},$ directed along the $Z'$ axis 
and ${\bf S}$ directed along the $Z$ axis  is $\beta.$ { The angle between  ${\bf L}$ and the $Z''$ axis directed along ${\bf J}$ is $i$
and $\delta=\beta - i.$.   The angle between the $Y''$ axis and the $Y$ axis $2\pi -\alpha_r.$
The apsidal line, the location of pericentre and  an orbital arc in its neighbourhood 
are shown.}
}
\label{coordinates}
\end{center}
\end{figure}

 The third
 $(X,Y,Z)$ coordinate system is defined as in \cite{IP2011} with $z$-axis being directed along the direction of the stellar 
angular momentum vector, ${\bf S}$.
The azimuthal angle associated with  both ${\bf J}$ and ${\bf S}$  measured in the
$(X',Y',Z')$ system is $\pi/2 -\gamma.$ 
The $Y$ axis  lies in the  orbital  plane and  defines the line of nodes as viewed in  the $(X,Y)$
plane as in  \cite{IP2011}.
Note that the $ X',Y',$ and $Y$ axes are coplanar as are the  $Z, Z' $ and $Z''$ axes.
\textcolor{red}{ For a Keplerian orbit with fixed orientation, the  line  of apsides can be chosen to coincide with the $X'$ axis.
In this case the angle between this line and the $X'$ axis, which we shall  more generally denote by $\varpi,$
will  simply be given by  
be $\varpi = 0 .$  Note that the angle between the apsidal line and the $Y$ axis, being the line of nodes is
quite generally given by  $ \varpi +\gamma - \pi/2$. }
The coordinate systems are illustrated in Fig. \ref{coordinates}.

The coordinates ${\bf r'} = (X',Y,'Z')$ in the orbit frame
are related to 
the coordinates ${\bf r} = (X,Y,Z)$ in the stellar frame
\textcolor{red}{by}
\begin{equation}
{\bf r}' = \boldsymbol{\mathsf{R}}_0 {\bf r} \label{jp1}
\end{equation}
where the rotation matrix is given by  $\boldsymbol{\mathsf{R}}_0 \equiv  \boldsymbol{\mathsf{R}} (0,\beta,\gamma) .$
This corresponds to a rotation through an angle, $\gamma,$ about the $Z$ axis 
followed by a rotation through an angle, $\beta,$ about the $Y$ axis.
Alternatively, it may be considered to be formed from a rotation  through an angle $\beta$ about the $Y$ axis
followed by a rotation through an angle $\gamma$ about the newly formed $Z'$ axis.
Similarly,
the coordinates ${\bf r''} = (X'',Y'',Z'')$ in the { primary centred }  frame frame defined by the total angular momentum
vector are related to
the coordinates ${\bf r} = (X,Y,Z)$ in the stellar  frame
through the relation
\begin{equation}
{\bf r''} = \boldsymbol{\mathsf{R}}_1 {\bf r}\label{jp2}
\end{equation}
where in this case  the rotation matrix $\boldsymbol{\mathsf{R}}_1$ is given by
$\boldsymbol{\mathsf{ R}}_1=\boldsymbol{\mathsf{R}} (0,0,\alpha) \boldsymbol{\mathsf{R}} (0,\beta - i,\gamma) .$
Here we note that we apply an additional rotation through an arbitrary angle $\alpha$ about the newly defined $Z''$ axis,
\textcolor{red}{which is such that $\alpha +\gamma$ can be used to define an angle of precession of ${\bf S}$ about ${\bf J}.$}
\textcolor{red}{We remark that  $\delta=\beta-i$ is the angle of inclination between the spin angular momentum vector ${\bf S}$ 
and the total angular momentum vector, the latter  being  fixed in the { primary centred}
frame.}
Thus,  $\boldsymbol{\mathsf{R}}_1$ is obtained from $  \boldsymbol{\mathsf{R}}_0$ 
by replacing $\beta$ by $\delta $ and $\gamma $ by $\gamma +\alpha.$


Recalling that the inverse of a rotation matrix is its transpose
we find from (\ref{jp1}) and (\ref{jp2}) that
\begin{equation}
{\bf r} =  \boldsymbol{\mathsf{R}}^T_1 {\bf r''}  \equiv  \boldsymbol{ \mathsf{R}}_2 {\bf r''} \label{jp3}
\end{equation}
 where 
 \begin{equation}
 \boldsymbol{\mathsf{R}}_2= 
  \boldsymbol{\mathsf{ R}} ^T(0,\delta, \gamma +\alpha)
  \label{j2}
\end{equation}
corresponding to taking the inverse of the transform related to a rotation through an angle  $\gamma +\alpha$ about the $Z$ axis followed
by a rotation through an angle $\delta$  about the $Y$ axis.
{ Definitions of parameters associated with the various coordinate systems and other symbols used later in the text
are summarised in table \ref{symbols1}.} 


        \begin{table}     
	\caption{Table of some parameters, variables and symbols  associated with the different coordinate systems
	         and the components of the binary } \label{symbols1}
        \vspace{-0.3cm}
 	\begin{center}
		\begin{tabular}
		{c p{12.00cm}}
		\hline
		Symbol & Definition \\
		\hline
		$(X'',Y'',Z'')$ &  Cartesian coordinate system  with the  $Z''$ axis  directed along the total angular momentum, \bmth{J}, ( $J={|\bf J}|)$ 
		and origin at the centre of mass of the primary with mass $M_*$\\
		$(X',Y',Z')$ & Cartesian coordinate system  with  $Z'$  axis directed along the orbital angular momentum, \bmth{L}, ($L=|{\bf L}|)$ - 
		 $(r,\theta',\phi')$ are the associated spherical polar coordinates\\
		$(X,Y,Z)$ & Cartesian coordinate system  with   $Z$  axis directed along the stellar spin angular momentum, \bmth{S}, ( $S=|{\bf S}|$)
		and $Y$ axis  in the $(X',Y')$ plane being the line of nodes - $(r, \theta,\phi)$  are the associated spherical polar coordinates\\
		$\gamma$&Angle between the Y' and Y axes\\
		$\varpi$&Angle between line of apsides and the $X'$ axis\\	
		$\varpi+ \gamma -\pi/2 $&Angle between line of apsides and the $Y$ axis\\
		$i$&Angle between $Z'$ and $Z''$ \\
		$\beta$&Angle between $Z$ and $Z'$ with $\delta=\beta -i$ \\
		$\alpha$&Angle of rotation used to specify the location of the $(X'',Y'')$ system in the plane orthogonal to the $Z''$ axis
		with ${\overline \alpha} = \alpha+\gamma $ and $\alpha_r = 2\pi -{\bar \alpha}$\\
		$(L^{i},L^{j},L^{k}) \equiv {\bf L}$&Components of the orbital angular momentum ${\bf L}$ in the $(I,J,K)$ coordinate system (eg. for $i=x'$ take $I\equiv X'$ etc.)\\
		$(S^{i},S^{j},S^{k}) \equiv {\bf S}$&Components of the stellar spin angular momentum ${\bf S}$ in the $(I,J,K)$ coordinate system (eg. for $i=x'$ take $I\equiv X'$ etc.)\\
		$(T^{i},T^{j},T^{k}) \equiv {\bf T}$&Components of the torque ${\bf T}$ in the $(I,J,K)$ coordinate system (eg. for $i=x'$ take $I\equiv X'$ etc.)\\
		$M_p, M_*, R_*, \Omega_r, q$& Respectively  the mass of the  perturbing companion component, mass. radius and angular velocity of the perturbed primary component, 
		and the mass ratio $M_p/M_*$ \\
				\hline
		\end{tabular}
 	         \end{center}
	          \end{table}

\section{Evolution equations for the angular momentum vectors}\label{Jvecev}
{ In this section we use the conservation of the total angular momentum to derive equations for the evolution
of $L,S,$ and the angles $\beta, \delta, $ and $\alpha_r =2\pi -{\bar \alpha}$ (see table \ref{symbols1}).}


\noindent \textcolor{red} {From the definition of $\beta$ given above } it  follows that 
\begin{align}
&\cos {\beta}={({\bf L}\cdot {\bf S})\over LS}, 
\end{align}
where $L$ and $S$ are the magnitudes  of ${\bf L}$ and ${\bf S}$, and
\begin{align}
 &\cos {i}={({\bf J}\cdot {\bf L})\over JL},
 \end{align} 
 where $J$ is the magnitude  of
${\bf J}$.
 We  introduce  the torque ${\bf T}$ exerted on the star due to the tidal interaction. 
From the constancy  of the total  angular momentum  ${\bf J}={\bf L}+{\bf S}$ it follows that in the primary centred frame
\begin{equation}
{\bf T}=\dot {\bf S}=-\dot {\bf L}, 
\label{e0}   
\end{equation}
where a dot over a quantity,  
  here and subsequently, indicates the time derivative of that quantity.  
  \textcolor{red}{In addition, we also have
\hspace{2mm}  $2{\bf J}\cdot {\bf L} =J^2+L^2-S^2 $ \hspace{2mm}  and \hspace{2mm}  $2{\bf L}\cdot {\bf S} =J^2- L^2-S^2.$ 
and, accordingly,
\begin{equation}
\cos \beta ={J^2-L^2-S^2\over 2LS}\hspace{3mm}{\rm and} \hspace{3mm}\cos i = {J^2+L^2-S^2\over 2JL}.
\label{e41}
\end{equation}
From these relations we can also  express the sines of $\beta$, $i$ and $\delta$ in terms of $J$, $L$ and $S$. We obtain
\begin{equation}
\sin \beta =\frac{\sqrt{(J^2-(L-S)^2)((L+S)^2-J^2)}}{2LS},\hspace{2mm} {\rm and}
\hspace{2mm} \sin i = \frac{\sqrt{(S^2-(J-L)^2)((J+L)^2-S^2)}}{2JL}           
\label{e39}
\end{equation}
together with
\begin{equation}
\sin \delta ={L\over S}\sin i = {L\over J} \sin \beta.
\label{e40}
\end{equation}
}

\textcolor{red} {{ We now make use of the rotation matrices 
 $\boldsymbol{\mathsf{R}} (0,\beta,\gamma)$
 and \newline  $\boldsymbol{\mathsf{ R}}_1~=~\boldsymbol{\mathsf{R}} (0,0,\alpha) \boldsymbol{\mathsf{R}} (0,\beta-i,\gamma)$ defined in Section \ref{Geometry}.
 These
  respectively
  enable the transformation of the components of any vector from representation in 
 the stellar frame to the orbit frame and 
 the stellar frame to the
 primary centred frame  as indicated by equations 
 (\ref{jp1})  and 
 (\ref{jp2}). From these  transformations and their inverses }
 it is straightforward to express the components of ${\bf L}$, ${\bf S}$
and the torque ${\bf T} = - \dot {\bf L}$  in the { primary centred}  frame in terms of  $L$, $S$ and the components of the torque
in the frame associated with the star. { We recall that by definition the components of ${\bf L}$ in the orbit frame are $(0,0,L)$
and the components of ${\bf S}$ in the stellar frame are $(0,0,S)$}.}
								 
\textcolor{red}{Thus the components of ${\bf L}$ in the { primary centred}  frame,  $( L^{x^{''}},   L^{y^{''}}, L^{z^{''}}), $ are found to be  given by
\begin{equation}
L^{x^{''}}= L\cos {\bar \alpha }\sin i , \quad L^{y^{''}}=-L\sin {\bar \alpha} \sin i  \quad {\rm and}  \quad L^{z^{''}}=L\cos i,
\label{e42}
\end{equation}
where ${\bar \alpha}= \alpha + \gamma.$ 
Similarly, the components of ${\bf S}$ in the { primary centred}  frame, $( S^{x^{''}},   S^{y^{''}}, S^{z^{''}}), $  are given by 
\begin{equation}
S^{x^{''}}= -S\cos{\bar  \alpha} \sin \delta , \hspace{2mm} S^{y^{''}}= S\sin {\bar \alpha }\sin \delta, \hspace{2mm} {\rm and}
\hspace{2mm}  S^{z^{''}}=S\cos \delta ,
\label{e43}
\end{equation}
and the components of ${\bf T}$ in this frame, $( T^{x^{''}},   T^{y^{''}}, T^{z^{''}}), $   are given by
\begin{equation}
T^{x^{''}}=\sin {\bar \alpha} T^y +\cos {\bar \alpha} T^{1}, \hspace{2mm} T^{y^{''}}=
\cos {\bar \alpha} T^y-\sin {\bar \alpha} T^{1}, \hspace{2mm} {\rm and} \hspace{2mm} T^{z^{''}}=\sin \delta T^x+ \cos\delta T^{z},
\label{e44}
\end{equation}
\begin{equation}
\hspace{-11.2cm} {\rm where}\hspace{2mm} T^{1}=\cos \delta T^x-\sin \delta T^z
\label{e44a}
\end{equation}
with the components of ${\bf T}$ in the stellar frame being given by
$(T^{x},T^{y},T^{z}).$}

\textcolor{red}{From equation (\ref{e0})  together with equations (\ref{e40}) - (\ref{e44a}) it is easy to obtain the following set of
equations 
\begin{equation}
{di\over dt}={1\over L}(-\cos \beta T^x +\sin \beta T^z), \quad {d L\over dt}=-\cos \beta T^z -\sin \beta T^x, \quad {d\over dt} {\bar \alpha }=
{T^y\over \sin i L}= {J\over LS}{T^y\over \sin \beta}, 
\label{e45}
\end{equation}
and
\begin{equation}
{d\over dt}\delta = -{T^x\over S}, \quad \dot S = T^z.
\label{e46}
\end{equation}
Note that we use (\ref{e40}) to obtain the last equality in (\ref{e45}). 
Also note that it is easy to check that $J^{i}=L^{i}+S^{i}$ are indeed  first integrals of
the set of equations  (\ref{e45}) and (\ref{e46}). 
 Using the first  two equations 
of (\ref{e45}) together with  (\ref{e46})  and the help of equations 
(\ref{e41}) and (\ref{e40}) it is straightforward to obtain the evolution equation for angle $\beta$:
\begin{equation}
{d\beta \over dt}=-\left({\cos \beta \over L}+{1\over S}\right)T^x+{\sin \beta \over L}T^z.
\label{en1}    
\end{equation}}
\textcolor{red} {Finally, we remark that the angle ${\bar \alpha}$ was defined through making a right handed rotation when transforming 
from  the $(X,Y,Z)$ system to the  $(X'', Y'', Z'')$ system. Accordingly, we define $\alpha_r = 2\pi -\bar\alpha$, so that
increasing $\alpha_r$ is associated with a right handed  rotation from  $(X'', Y'', Z'')$ to $(X,Y,Z).$
It may  thus  be used to describe precession of the stellar rotation axis in the conventional manner.
Clearly $d\alpha_r/dt = -d\bar\alpha/dt.$}   

Together with the energy conservation law equations (\ref{e45}) and (\ref{e46}) form a complete set for
our model. Note that an evolution equation for the angle $\gamma$ is absent. This is because, physically,
only the angle ${\bar \alpha}=2\pi -\alpha_r$  appears in the specification of the orientation of the angular momentum
vectors in the { primary centred}  frame. In addition,  only
 the angle between the apsidal line and the  projection of the stellar  spin angular momentum vector onto 
the orbital plane, $\varpi+\gamma  $
matters for the determination of the orbital evolution. The evolution of this angle should be found from other considerations which are
not based on the law of angular momentum conservation. This evolution is determined, in general, from 
tidal interactions, stellar flattening, General Relativity and/or a presence of other perturbing
bodies. We assume hereafter that this evolution is known. Also note that the components of the torque in
the stellar frame, which we are going to calculate below do not depend on the
angle $ \alpha_r$ used to specify
orbital precession with respect to the { primary centred} frame because it is ignorable in this context.
 Thus, the evolution equation for this angle
can be considered separately from the others. This equation 
contains only a term proportional to $T^y$  that is determined by tides. In addition to this term a standard contribution
due to stellar flattening must be added, as discussed below. The remainder of the equations governing the evolution
are determined by only two components of the torque in the stellar frame, $T^x$ and $T^z$.


\section{The perturbing tidal potential}\label{pertpoten}
{ In this section we develop the standard quadrupole form of the tidal potential appropriate for a near  Keplerian orbit
with arbitrary orientation  as a  series of spherical harmonics in the orbit frame each term of which is expressed as a Fourier series.
We transform the spherical harmonics expressed in the orbit frame to a representation in terms of spherical harmonics defined in the
stellar frame using Wigner matrices.
This is done to facilitate the calculation of the tidal response in the primary.
A list of some  of the parameters, variables and symbols associated with the developments in this Section is given in table \ref{symbols2}.}


The perturbing potential, $U$, can be readily found in the orbit frame $(X',Y',Z')$, 
where a spherical coordinate system $(r,\theta', \phi')$ is defined in the usual way
taking into account quadrupole terms only as
\begin{equation}
 U=-\frac{GM_pr^2}{R^3}P_2(\cos\psi),
\label{jpe0}
\end{equation}
where $R$ is distance between the binary components,  $P_2$ is the usual Legendre polynomial
and $\cos\psi =\sin\theta'\cos(\Phi-\phi')$.
Here the orbit is taken to be in the $\theta' = \pi/2$  plane with its line of apsides
located at $\phi' = \varpi ,$  with $\Phi$ being the  azimuthal angle of the line joining the binary components.
For convenience we shall measure both $\Phi$ and $\phi'$ from the $X'$ axis without loss of generality.
Equation (\ref{jpe0}) may also written as
\begin{equation}
 U=-\frac{GM_p r^2}{R^3}
 \left(\frac{4\pi}{5}\right) 
\sum^{'}_{m=0, |m|=2}Y_{2,m}(\theta', \phi') Y_{2,m}(\pi/2, 0)\exp(-{\rm i}m \Phi)
\label{jpe1}
\end{equation}  
where,  \textcolor{red}{ hereafter, the prime implies that only summations over $m=0, \pm 2$ is performed}, $ Y_{2,m}(\theta', \phi'),$  is the usual spherical harmonic.
\subsection{Fourier development  $U$ in terms of spherical harmonics defined in the stellar frame}\label{Fourierpot}
For an eccentric  Keplerian orbit
$R$ is a periodic function of time with period $2\pi/n_o$ and $\Phi = n_ot +v(t)+\varpi ,$
where $v$ s a periodic function of time with period $2\pi/n_o$ and zero time average with
$\varpi$ being the longitude of the apsidal line measured from the $X'$ axis. When  as for a strictly Keplerian orbit this remains fixed,
as remarked above we may choose the line to coincide with the $X'$ axis in which case $\varpi =0.$
 
For slightly non Keplerian precessing orbits,  the longitude of the  apsidal line, $\varpi, $ precesses at a rate 
given by  $d\varpi/dt ,$ the latter being determined by the deviation of the time averaged background
potential from proportionality to $1/r.$
Accordingly we have $ \varpi =\int^t_0 ( d\varpi/dt' )  dt'  + \varpi_0,$
where $\varpi_0$ is a constant reference angle, being the value of $\varpi$ at $t=0.$
When  the  rate  of precession is independent of the precession phase, we have  uniform precession
with $\varpi$ increasing linearly with time.
Then we have $ \varpi = (d\varpi/dt) t   + \varpi_0.$
In addition,  we have the Fourier expansion
\begin{equation}
\frac{\exp(-{\rm i}m (\Phi-\varpi)) a^3}{R^3}  = \sum_{k=-\infty}^{k=\infty} \phi_{k,m}\exp({\rm i}k
 n_o t),
\label{jpe2}
\end{equation}  
where $a$ is the orbital semi-major axis,  with the Fourier coefficients being given by
\begin{equation}
\phi_{k, m}   =\frac{n_o}{2\pi}\oint  \frac{\exp(-{\rm i}m (\Phi -\varpi)) a^3 }{R^3}  \exp(-{\rm i}
kn_o t) dt,
\label{jpe30}
\end{equation}  
where the integral is taken over an orbital  period  $2\pi/n_o.$
The coefficients, $\phi_{k,m},$ are related to the well known Hansen coefficients, $X^{q,m}_k,$
( see e.g. Branham 1990, Laskar 2005),
 
 \noindent through~$\phi_{k,m}~=~X^{-3,-m}_k$, see Appendix \ref{FourierU} for more details.
Practical prescriptions
for calculating Hansen coefficients have been provided by many authors ( e.g.  Branham 1990, Laskar 2005).
For small eccentricities a power law expansion in $e$ developed from (\ref{hansen} ) may be used. 
In this paper we  have followed the notation of Ivanov \& Papaloizou (2004),  who give a useful prescription
for calculating $\phi_{k,m}$ for eccentricities $e > 0.2$.
In  the stellar frame $(X,Y,Z)$  the associated spherical coordinate system is $(r,\theta, \phi).$
From the discussion  given in Appendix \ref{FourierU} we can write the potential in terms of these
in the form  
\begin{equation}
 U=-GM_p \left(\frac{4\pi r^2}{5 a^3}\right)\sum_{n=-2}^{n =2}Y_{2,n}(\theta, \phi)F_{n}(t) \equiv  r^2\sum_{n=-2}^{n =2}A_{n} Y_{2,n}(\theta, \phi),
\label{jpe7}
\end{equation}
with
$A_{n} =-4\pi GM_p F_{n}/(5a^3).$
Here
\begin{equation}
F_{n}(t)= \sum^{'}_{m=0, |2|}\sum_{k=-\infty}^{k=\infty} \phi_{k,m}D^{(2)}_{n,m} 
 Y_{2,m}(\pi/2, 0)\exp({\rm i}(k n_o t -m\varpi))),\label{jpe711}
\end{equation}
where the coefficients (Wigner matrix elements), $D^{(2)}_{n,m}$, are specified together
with some of their relevant properties in Appendices \ref{FourierU} and  \ref{Wigner}.
We remark that
\begin{equation}
D^{(2)}_{n,m} =  \exp(-{\rm i}m\gamma) d^{(2)}_{n,m}(\beta) 
\label{jpe51}
\end{equation}
where   $d^{(2)}_{n,m}$ is an element of
Wigner's (small) d-matrix  and is real, see e.g. Ivanov \& Papaloizou 2011. 

Note that $F_{n}(t)$
is in general {\it not} simply harmonically varying in time. Making use of  the  Fourier expansion expressed by  equation (\ref{jpe2}) it can be 
written in the alternative form
\begin{equation}
F_{n}(t)=\frac{a^3}{R^3} \sum^{'}_{m=0, |2|} D^{(2)}_{n,m}  Y_{2,m}(\pi/2, 0)\exp(-{\rm i}m\Phi ).\label{jpe711b0}
\end{equation}
We remark in addition that $\phi_{-k,-m}=  \phi^*_{k,m}$
and also that $Y_{2,-m}(\theta, \phi) = (-1)^{m}Y^*_{2,m}(\theta, \phi) $
 with  $ D^{(2)}_{-n,-m} = (-1)^{(n+m)}( D^{(2)}_{n,m} )^*$,
these relations together ensuring that the sum in (\ref{jpe7}) is real.
	         \begin{table}     
	          \caption{Table of some parameters, variables and symbols associated with the calculation of the response to tidal forcing} \label{symbols2}
                  \vspace{-0.3cm}
 	         \begin{center}
		\begin{tabular}
		{c p{12.00cm}}
		\hline
		Symbol & Definition \\
		\hline
		$a, n_o, e, R, U$& Respectively the semi-major axis, mean motion, eccentricity, distance between the components and perturbing tidal potential\\
		${\Phi}$& $\phi'$ coordinate of the perturber in its orbit in the $(r,\theta',\phi')$ system\\
		$r^2A_n= -4\pi G M_p F_nr^2/(5a^3)$&Coefficient in expansion of $U$ in spherical harmonics in the $(r,\theta,\phi)$ system, $n$ being 
		the azimuthal mode number, $G$ is the gravitational constant\\
		$r^2{\cal A}_{n,k},$ and $\omega_f=kn_o+n\Omega_r$&Fourier coefficient in the expansion  of the spherical harmonic  component of $U$ with azimuthal 
		mode  number $n$ ( or equivalently $r^2A_n$), $\omega_f$ is the  forcing frequency in the  rotating frame\\
		$\phi_{k,m} = X_k^{-3,-m} $& Fourier or Hansen coefficient occurring in the Fourier expansion of $\exp(-{\rm i}(\Phi-\varpi)(a/R)^3$\\
		$D^{2}_{n,m}, d^{2}_{n,m}$& Respectively the Wigner matrix element and small, $d,$ Wigner matrix element used to specify transformation of spherical harmonics
		between coordinate systems\\	
		$\bmth{\xi},  \bmth{\xi}_n,  \xi^r,  r\nabla_{\perp}\xi^S$&Respectively the Lagrangian displacement, the coefficient on expansion of this in spherical harmonics and the radial 
		and  angular components of that\\	 
		${\hat L},{\overline U},  {\hat L}_1$& The linear operator giving the adiabatic response to the $(n,k)$ component of the forcing potential, ${\overline U},$
		 for a spherical  body and the operator  correcting
		for rotation and dissipation\\
		$\bmth{\xi}_{n,k}, \bmth{\xi}_{eq,n,k},\bmth{\xi}_{eq1,n,k}$&Response displacement to  $(n,k)$ forcing, it's form for a spherical body in the equilibrium tide limit
		and its correction due to rotation and dissipation\\
		$ \omega^2_{eq}, N_0, Q_{eq}$& Square of eigenfrequency obtained with trial function $\bmth{\xi}_{eq,n,k},$ 
		 its norm (usually scaled to unity), and normalised overlap integral ( due to spherical symmetry, the spatial form of $\bmth{\xi}_{eq,n,k}$ is independent of $k$
		 which can be omitted)\\
		 ${\tilde\omega}^2, {\tilde Q}$ &Dimensionless forms of $\omega_{eq}^2,$ and $Q_{eq}.$ Thus ${\tilde\omega}^2 =(\omega_{eq}^2R_*^3)/(GM_*)$
		 and ${\tilde Q}= Q_{eq}/(\sqrt{M_*}R_*).$\\
		$\Gamma, {\hat \Gamma} $& Ratio of the rate of energy dissipation to four times the associated kinetic energy, ${\hat \Gamma}=\Gamma/\omega_{eq}$ \\
		$\beta_*, \lambda, {\hat \Omega}$& Parameter associated with the coriolis force,$-(\beta_*-1) n\Omega_r$
		 is the frequency shift in the rotating frame produced by a putative normal mode with displacement
		$\bmth{\xi}_{eq,n,k}$, $\lambda = \beta_*/(\beta_*+1/2)$  \\ 
		$\sigma,{\hat \Omega}$&   $\sigma=\Omega_r/(\lambda n_o),$ ${\hat \Omega} = \Omega_r/\omega_{eq}$\\		
		${\cal D}_{n,k}=-Q_{eq}{\cal R}_{n,k} {\cal A}_{n,k}$&The
		  contribution to $\bmth{\xi}_n$ from the  Fourier component with  frequency,  $kn_o,$ is ${\cal D}_{n,k}\bmth{\xi}_{eq,n}$ \\
		${\cal R}_{n,k}$& ${\cal R}_{n,k} =\omega_{eq}^{-2}(1-\delta_{n,k}+k^2\tilde\Omega^2) ,$ 
		where $\delta_{n,k}  ={\rm i}   (\delta_1k+\delta_2 n)
		+\delta_3(kn+n^2\sigma)$ \\
		$ \delta_1,\delta_2,\delta_3$&$\delta_1 = 2{\hat \Gamma}{\hat \Omega},\delta_2=\lambda\sigma\delta_1,
		\delta_3 = -2\beta_*\lambda \sigma{\hat \Omega}^2$
		($n$ is azimuthal mode number, $kn_0,$ is the frequency of the Fourier component). 
		 $\delta_i \rightarrow \delta_{i,p} $ signifies application to the perturber \\
		$b_n, E_{orb}$&$ b_n\bmth{\xi}_{eq,n}= \bmth{\xi}_n ,$ thus $b_n$ expresses the time dependence of $\bmth{\xi}_n,$ $E_{orb}$ is the orbital energy\\
        		\hline
		\end{tabular}
                  \end{center}
	          \end{table}

\section{Calculation of the Response displacement
}\label{Respd}
{ In this section we calculate the tidal response of the primary to  the forcing  tidal potential
by  solving for the Lagrangian displacement in the linear approximation.
The aim is to find the density perturbation that will subsequently be used to find tidal torques 
and the rate of energy transfer from the orbit.  A low frequency approximation is adopted
for which the response consists of an equilibrium tide with corrections arising from coriolis forces, inertia and
dissipative effects treated as small perturbations. }


We shall assume that the  effective stellar response is determined by
a displacement of spheroidal form appropriate to a  spherically symmetric background state.
We make the usual assumption that  the rotational  period  of the star
is much longer than the characteristic dynamical time scale  which allows us
 in the first instance  to regard  stellar rotation 
as being small and  adopt an expression for the Lagrangian displacement
 which is of spheroidal form which retains the angular dependence
for each azimuthal wavenumber, $n.$  
Thus, we write
\begin{equation}
\mbox{{\boldmath$\xi$}} =\sum_{n=-2}^{n=2} \mbox{{\boldmath$\xi$}}_{n},
\label{e1X}
\end{equation}  
where 
\begin{equation}
 \mbox{{\boldmath$\xi$}}_{n} =\xi^r(r,\theta, t, n)\exp({\rm i}n\phi){\bf e}_r+ r\nabla_{\perp} (\xi^S(r,\theta, t, n)\exp({\rm i}n\phi)), 
\label{e2X}
\end{equation}
where ${\bf e}_r$ is unit vector in the radial direction.
Note that,  \textcolor{red} {in general}, $\xi^{r}$ and $\xi^S$ \textcolor{red}{can} depend on $n.$
If the response was that of a non rotating strictly  spherically symmetric star
the angular dependence is through a factor $Y_{2,n}.$ 
\textcolor{red}{This extends to} the case here to lowest order in the stellar angular velocity $\Omega_r.$
As higher order effects are treated by perturbation theory, modification of this
dependence does not have to be considered. This procedure
neglects any contribution from displacements of toroidal form and 
the excitation of related normal modes (Papaloizou \& Pringle 1978). 

\noindent Finally we remark that 
as $\mbox{{\boldmath$\xi$}}$  is real we have
 $\mbox{{\boldmath$\xi$}}^{*}_{-n}   =   \mbox{{\boldmath$\xi$}}_{n}.$

To evaluate the response to tidal forcing we consider
an equation of the generic form
\begin{equation}
\omega^2_f \mbox{{\boldmath$\eta$}}=\hat L(\mbox{{\boldmath$\eta$}})+ {\cal L}(\mbox{{\boldmath$\eta$}})
+\nabla { \overline U }\label{Bs1}
\end{equation}
where $\omega_f$ is the forcing frequency, the forcing potential 
\begin{equation}
{\overline U}=  r^2  {\cal A}_{n,k}Y_{2,n}(\theta, \phi)
,\label{potpert}
\end{equation}
where
\begin{equation}
{\cal A}_{n,k}  =- \sum^{'}_{m=0, |2|} \frac{4\pi GM_p}{5 a^3} \phi_{k,m}D^{(2)}_{n,m}  Y_{2,m}(\pi/2, 0)\exp(-{\rm i} m\varpi).\label{jpe711e}
\end{equation}
Here  
we consider a single term contributing to the sum in equation (\ref{jpe7}) (see Section (\ref{Fourierpot}) and equations (\ref{jpe7}) - (\ref{jpe711b0})~)
and we consider the apsidal angle $\varpi$ 
to be sufficiently slowly varying that its variation may be neglected.
We  note 
    that a term 
associated with a particular forcing frequency, $kn_o$, and azimuthal mode number, $n$, of the form,
$\exp({\rm i}k n_o t) \mbox{\boldmath$\eta$},$
where  
$\mbox{\boldmath$\eta$}$
depends only on position
{will be contributed}
 to the  full Lagrangian displacement $ \mbox{{\boldmath$\xi$}}$ (see equation (\ref{e1X})). 
 Such terms as well as those corresponding to different $n$  may be combined through linear superposition.

The linear operator, $\hat L,$ is that applicable to an undisturbed spherical star
for which a  normal mode of oscillation, denoted by subscript, $k,$  satisfies
\begin{equation}
\omega^2_k \mbox{{\boldmath$\eta$}}_k=\hat L(\mbox{{\boldmath$\eta$}}_k).
\end{equation}
For a potential perturbation of the form (\ref{potpert})  the  frequency   
\begin{equation}
\omega_f =kn_o+n\Omega_r
\label{omf}
\end{equation}
is the Doppler shifted frequency as seen in the frame corotating with the star. 
The operator ${\cal L}$  contains the effects of   stellar rotation and damping to lowest order
and is considered to contribute a small perturbation (for more details see below).


\subsection {Equilibrium response}
For a spherical star
\begin{equation}
\hat L(\mbox{{\boldmath$\eta$}})\equiv \frac{1}{\rho}\nabla P' -\frac{\rho'}{\rho^2}\nabla P +\nabla\psi',
\end{equation}
where $P'$, $\rho'$ and $\psi'$ are respectively the pressure , density and  the potential perturbations arising from self gravity.
We have $\rho' = -\nabla\cdot ( \rho \mbox {\boldmath$\eta$})$  and we set $P' = P'_a + P'_{na},$ where

\noindent $P'_a = -\Gamma_1 P \nabla\cdot  \mbox {\boldmath$\eta$} -   \mbox {\boldmath$\eta$}\cdot\nabla P $
is the adiabatic component of the pressure perturbation and $P'_{na}$ is the non adiabatic part.

\noindent The latter may be obtained from the energy equation, which may be written as
\begin{equation}
P'+\Gamma_1 P \nabla\cdot  \mbox {\boldmath$\eta$} +  \mbox {\boldmath$\eta$}\cdot\nabla P = -\frac{(\Gamma_3-1)\nabla{\bf F}'}{{\rm i}\omega_f}.\label{Nonad}
\end{equation}
Here $\Gamma_1$ and $\Gamma_3$ are the standard adiabatic exponents and ${\bf F}'$ is the perturbed
\textcolor{red} {energy flux, ${\bf F}$, which may contain contributions from both radiative and convective transport,}
and perturbation to the
energy generation rate is neglected. Being second order in perturbations there is no contribution from viscous dissipation.
Note that (\ref{Nonad}),  after use of the equation of state,  can be regarded as an equation for $P'$ in terms of a specified $\mbox {\boldmath$\eta$}$
with there also  being a dependence  on  forcing frequency. Hence there  will be a dependence of $P'_{na}$ on forcing frequency.

\noindent For ${\cal L}$ we set 
\begin{equation}
{\cal L}(\mbox{{\boldmath$\eta$}}) \equiv    2{\rm i} \omega_f\Omega_r {\bf \hat{k}}\times\mbox {\boldmath$\eta$}    +  
 \frac{1}{\rho}\nabla P'_{na} + {\rm i}\omega_f D_{\nu}\mbox {\boldmath($\eta$}).\label{EL1}
\end{equation}


\noindent Here the first term on the right hand side arises from  the Coriolis force, 
and the final term gives the effect of viscosity.  Thus, the first  term  gives  a non dissipative contribution
and the others dissipative ones.
 
We  develop a solution of (\ref{Bs1}) by considering firstly  a solution for which { both the term $\propto \omega^2_f$ and $\cal L$ are} neglected and
which applies in the low forcing frequency limit
and so gives rise to an equilibrium tide,
and we need not neglect self-gravity (Cowling approximation). 


\noindent Setting
$\mbox{{\boldmath$\eta$}}= \mbox{{\boldmath$\xi$}}_{eq,n,k}$ in this case we see that
it  satisfies
\begin{equation}
\hat L(\mbox{{\boldmath$\xi$}}_{eq,n,k})=
-\nabla { \overline U },\label{Bs2}
\end{equation}
where we have attached the subscripts $n$  and $k$ to denote that a response corresponds  to  azimuthal mode number $n$ and forcing frequency $kn_o$.

\noindent We remark that the form of the equilibrium tide mentioned above 
has been discussed in detail when the Cowling approximation applies in 
\citet{Terquem}  and \citet{Bunting}.  In particular, the reader is referred
to the discussion in Appendix D of the latter paper. In order to relax the Cowling approximation the potential 
$ {\overline U}$ should be replaced by ${\overline U}+ {\overline U_{self}}$, where ${\overline U_{self}}$ is the contribution
from self gravity. This satisfies the Poisson equation
\begin{equation}
\nabla ^2{\overline U_{self}}  = \left(\frac {4\pi G}{g}\right) \frac{d\rho}{dr} ({\overline U}+ {\overline U_{self}}),\label{Poisson}
\end{equation}
where $g$ is the local acceleration due to gravity and $G$ is the gravitational constant.
We also note that in  radiatively  stratified regions the radial component of the equilibrium tide
displacement is given by
\begin{equation}
\mbox{{\boldmath$\xi$}}_{eq,n,k}\cdot{\hat {\bf r}} =  - \frac{({\overline U}+ {\overline U_{self}})}{g} .\label{radeq}
\end{equation}
However, this changes in regions that are convectively neutral and so barotropic (see Bunting et al. 2019).

\subsection{Finding the response  for a  given forcing frequency and value of $n.$}\label{Findingresp}
We now return to our original equation (\ref{Bs1}) { which we rewrite in the form}
\begin{equation}
\hat L(\mbox{{\boldmath$\eta$}})+\hat L_1(\mbox{{\boldmath$\eta$}})=
-\nabla {\overline U},\label{Cs1}
\end{equation}
where ${\hat L}_1({\mbox{\boldmath$\eta$}}) = {\cal L}({\mbox{\boldmath$\eta$}}) -\omega_f^2 {\mbox{\boldmath$\eta$}}.$


\noindent We write
$\mbox{{\boldmath$\eta$}}=\mbox{{\boldmath$\xi$}}_{eq,n,k} + \mbox{{\boldmath$\xi$}}_{eq1,n,k} $, so that (\ref{Cs1}) implies that
\begin{equation}
\hat L(\mbox{{\boldmath$\xi$}}_{eq1,n,k})=-\hat L_1(\mbox{{\boldmath$\xi$}}_{eq,n,k} +\mbox{{\boldmath$\xi$}}_{eq1,n,k} ),\label{Cs2}
\end{equation}
Noting that the tidal effects we are interested in are first order in $\hat L_1$ we remark
 that the term involving $\mbox{{\boldmath$\xi$}}_{eq1,n,k}$ on the right hand side of (\ref{Cs2})  can be regarded as 
second order. However, we shall retain it for now.
In that case $\mbox{{\boldmath$\xi$}}_{eq1,n,k}$ is the equilibrium tide corresponding to an external  force per unit mass,
${\bf f} = -\hat L_1(\mbox{{\boldmath$\xi$}}_{eq,n,k} +\mbox{{\boldmath$\xi$}}_{eq1,n,k} ).$
It is important to note that in requiring $\mbox{{\boldmath$\xi$}}_{eq1,n,k}$ to be of spheroidal form 
we are discarding the  toroidal component of ${\bf f}$  and hence neglecting toroidal and inertial modes
governed by rotation. Thus,  we make the implicit assumption that these do not play a significant role
enabling us to retain the spheroidal component alone.  
 
 \subsection{Calculation of the overlap integral}\label{overlapcalc}
 Our aim is to find the volume integral of the density perturbation  response with the forcing potential
 as this is directly related to  forces exerted  on the star as a result of the tidal perturbation.
 We begin by using the self-adjoint property of $\hat L$ to write
 \begin{equation}
\int\rho \mbox{{\boldmath$\xi$}}_{eq1,n,k}^*\cdot  \hat L(\mbox{{\boldmath$\xi$}}_{eq,n,k})dV=
-\int\rho\mbox{{\boldmath$\xi$}}_{eq1,n,k}^*\cdot\nabla {\overline U }dV = -\left(\int\rho\mbox{{\boldmath$\xi$}}_{eq,n,k}^*\cdot \hat L_1(\mbox{{\boldmath$\xi$}}_{eq,n,k} +\mbox{{\boldmath$\xi$}}_{eq1,n,k} )dV\right)^*.
\end{equation}
From  this it follows that
\begin{equation}
\int\rho_{eq1,n,k}^{'*} {\overline U} dV =\left( \int\rho\mbox{{\boldmath$\xi$}}_{eq,n,k}^*\cdot \hat L_1(\mbox{{\boldmath$\xi$}}_{eq,n,k} +
\mbox{{\boldmath$\xi$}}_{eq1,n,k} )dV\right)^*,
\end{equation}
where $\rho_{eq1,n,k}^{'*}$ is the density perturbation associated with $\mbox{{\boldmath$\xi$}}_{eq1,n,k}.$
Thus,
\begin{equation}
{\cal A}_{n,k} \int\rho_{eq1,n,k}^{'*}  r^2Y_{2,n}(\theta, \phi)dV =\left( \int\rho\mbox{{\boldmath$\xi$}}_{eq,n,k}^*\cdot \hat L_1(\mbox{{\boldmath$\xi$}}_{eq,n,k} +\mbox{{\boldmath$\xi$}}_{eq1.n,k} )dV\right)^*.
\end{equation}
Similarly, if $\rho_{eq, n, k}^{'*}$ is the density perturbation associated with $\mbox{{\boldmath$\xi$}}_{eq,n,k}$
we have noting  the right hand side of the following equation  is real that
\begin{equation}
{\cal A}_{n,k} \int\rho_{eq,n,k}^{'*}  r^2Y_{2,n}(\theta, \phi)dV = - \int\rho\mbox{{\boldmath$\xi$}}_{eq,n,k}^*\cdot \hat L(\mbox{{\boldmath$\xi$}}_{eq,n,k}  )dV.\label{Cs4}
\end{equation}
Setting $\rho_{n,k}^{'*}=  \rho_{eq,n,k}^{'*} +\rho_{eq1,n,k}^{'*},$ it follows from the above results that
\begin{align}
& {\cal A}_{n,k} \int\rho_{n,k}^{'*}  r^2Y_{2,n}(\theta, \phi)dV = \nonumber\\
&- \int\rho\mbox{{\boldmath$\xi$}}_{eq,n,k}^*\cdot \hat L(\mbox{{\boldmath$\xi$}}_{eq,n,k}  )dV
\left( 1-\frac{ \left( \int\rho\mbox{{\boldmath$\xi$}}_{eq,n,k}^*\cdot \hat L_1(\mbox{{\boldmath$\xi$}}_{eq,n,k} +\mbox{{\boldmath$\xi$}}_{eq1,n,k} )dV\right)^*}{N_0\omega_{eq}^2}\right).\label{Cs1s}
\end{align}
Here, the quantity multiplying ${\cal A}_{n,k}$ is what we define to be the overlap integral
and
 \begin{equation}
 N_0 = \int\rho | \mbox{{\boldmath$\xi$}}_{eq,n,k} |^2 dV\hspace{3mm}{\rm and}\label{DDs10}
 \end{equation}
 \begin{equation}
\omega_{eq}^2 = \frac{\int\rho\mbox{{\boldmath$\xi$}}_{eq,n,k}^*\cdot \hat L(\mbox{{\boldmath$\xi$}}_{eq,n,k}  )dV}{N_0}.\label{DDs1}
\end{equation}
The latter squared frequency can be regarded as being  obtained from the oscillation problem on having used $\mbox{{\boldmath$\xi$}}_{eq,n,k},$
which may be arbitrarily scaled,  as a trial function.
In fact, as the operator $\hat L$ is for a spherical star and it has no explicit frequency dependence,  when each  of the quantities   $\mbox{{\boldmath$\xi$}}_{eq,n,k},$ are scaled appropriately,  $N_0$ and $\omega_{eq}$
are independent of $n$ and $k.$ However, this is not the case  for $\hat L_1$ and  $\mbox{{\boldmath$\xi$}}_{eq1,n,k}.$

Making use of  equation (\ref{Cs4}) and with the help of equations  (\ref{DDs10} ) and  (\ref{DDs1}) equation (\ref{Cs1s})   may be rewritten as
\begin{align}
 &\int\rho_{n,k}^{'*}  r^2Y_{2,n}(\theta, \phi)dV =\nonumber\\
 &-{\cal A}_{n,k}^* \frac{ \left| \int\rho_{eq,n,k}^{'*}  r^2Y_{2,n}(\theta, \phi)dV\right|^2}{N_0\omega_{eq}^2}
\left( 1- \frac{  \left(\int\rho\mbox{{\boldmath$\xi$}}_{eq,n,k}^*\cdot \hat L_1(\mbox{{\boldmath$\xi$}}_{eq,n,k} +\mbox{{\boldmath$\xi$}}_{eq1.n,k} )dV\right)^*}{N_0\omega_{eq}^2}\right).\label{EE1}
\end{align}
We remark that the expression on the right hand is invariant to scaling the various displacements by an arbitrary complex constant.
This could be chosen to provide a normalisation such that $N_0=1.$ 
To proceed further we note that  equation (\ref{Bs1}) together with the self-adjoint property of $\hat L$ imply  that the quantity
 \begin{equation}
\frac{1}{2}\omega_f {\cal I}\left( \int\rho\mbox{{\boldmath$\eta$}}^*\cdot \hat L_1(\mbox{{\boldmath$\eta$}}  )dV\right) =
\frac{1}{2}\omega_f {\cal I}\left( \int\rho\mbox{{\boldmath$\eta$}}^*\cdot {\cal L}(\mbox{{\boldmath$\eta$}}  )dV\right) 
=\Gamma\omega_f^2
 \int\rho| \mbox{{\boldmath$\eta$}}|^2  dV,
\label{EDs1}
\end{equation}
{where ${\cal I}$ denotes that the imaginary part is to be taken,}
represents the rate of energy dissipation associated with the displacement $\mbox{{\boldmath$\eta$}}.$
Equation (\ref{EDs1}) also defines the quantity $\Gamma$, which \textcolor{red}
{is the ratio of this  dissipation rate to four times the kinetic energy associated with the disturbance. Introduced in this way it 
 would correspond to the decay rate~\footnote{Clearly,
$\Gamma$ is the decay rate, as assumed here, only for a stable mode, it is the growth rate 
for an unstable mode.}, were 
$\mbox{{\boldmath$\eta$}}$ a normal mode of the system with eigenfrequency $\omega_f.$}
In addition,  recalling that for $   \mbox{{\boldmath$\eta$}} =   \mbox{{\boldmath$\xi$}}_{eq,n,k} +\mbox{{\boldmath$\xi$}}_{eq1.n,k}$
and making use of the fact that $\hat L$ is self-adjoint, 
equation (\ref{Cs2})  implies that
\begin{equation}
{\cal I}\left( \int\rho\mbox{{\boldmath$\eta$}}^*\cdot \hat L_1(\mbox{{\boldmath$\eta$}}  )dV\right) =
{\cal I}\left( \int\rho\mbox{{\boldmath$\xi$}}_{eq,n,k}^*\cdot \hat L_1(\mbox{{\boldmath$\xi$}}_{eq,n,k} +\mbox{{\boldmath$\xi$}}_{eq1,n,k} )dV\right). 
\label{EDs2}
\end{equation}
Similarly, we write
 \begin{equation}
\frac{1}{2}\omega_f {\cal R}\left( \int\rho\mbox{{\boldmath$\xi$}}_{eq,n,k}^*\cdot \hat L_1(\mbox{{\boldmath$\xi$}}_{eq,n,k} +\mbox{{\boldmath$\xi$}}_{eq1,n,k}   )dV\right) = 
 {\cal X}\int\rho| \mbox{{\boldmath$\eta$}}|^2  dV
\label{Bs22}
\end{equation}
{ with ${\cal R}$ denoting that the real part is to be taken and  which may be regarded as defining ${\cal X}.$
To evaluate this we make use of (\ref{EL1}) and (\ref{Cs1}) which define ${\hat L}_1$ and,  noting that in evaluating the integrals in (\ref{Bs22}) we may consider only lowest order quantities,
we accordingly neglect $\mbox{{\boldmath$\xi$}}_{eq1.n,k}.$ In addition, we neglect contributions from the nonadiabatic dissipative terms, which
 we assume to be much less than those retained. Thus, we find that
\begin{align}
&\frac{1}{2}\omega_f {\cal R}\left(\frac{\int \rho \mbox{{\boldmath$\xi$}}_{eq,n,k}^*\cdot{\hat L}_1(\mbox{{\boldmath$\xi$}}_{eq,n,k})dV}
{\int\rho|\mbox{{\boldmath$\xi$}}_{eq,n,k}|^2 dV} \right)= \omega_f^2 \left(
\frac{\int {\rm i} \rho\Omega_r\mbox{{\boldmath$\xi$}}_{eq,n,k}^*\cdot( {\bf \hat{k}}\times\mbox{{\boldmath$\xi$}}_{eq,n,k}) dV}
{\int\rho|\mbox{{\boldmath$\xi$}}_{eq,n,k}|^2dV} -\frac{\omega_f}{2}\right)\equiv\nonumber \\
&={\cal X} = -n\omega_f^2(\beta_*-1)\Omega_r-{\omega^3_f}/{2}
\label{EDs3}
\end{align}
which can be regarded as defining the quantity $\beta_*.$
Making use of (\ref{EDs1}) - (\ref{EDs3}) we may write (\ref{EE1}) in the form
\begin{align}
 &\int\rho_{n,k}^{'*}  r^2Y_{2,n}(\theta, \phi)dV =\nonumber\\
 &-{\cal A}_{n,k}^* \frac{ \left|\int\rho_{eq,n,k}^{'*}  r^2Y_{2,n}(\theta, \phi)dV\right|^2}{N_0\omega_{eq}^2}
\left( 1- \frac{(2\omega_f (-n\beta_*\Omega_r- {\rm i} \Gamma) -k^2n_o^2-n^2\Omega_r^2)}{\omega_{eq}^2}\right)\label{EREV}
\end{align}
which gives an expression for the complex conjugate of the  overlap integral.
Here  we have again made the approximation that  
\begin{equation}
\int\rho| \mbox{{\boldmath$\eta$}}|^2  dV=N_0
\end{equation}
which is expected to be valid when the norm of $\mbox{{\boldmath$\xi$}}_{eq1.n,k} $ is much less than that of
 $\mbox{{\boldmath$\xi$}}_{eq,n,k} $ as  assumed.
 However,  in passing we remark that the retention of
 $\mbox{{\boldmath$\xi$}}_{eq1,n,k} $ when considering dissipative terms allows  its gradient to become significant
 when dissipative processes are concerned. 
 In addition, we remark that  
 \begin{align}
 -n(\beta_*-1)\Omega_r = \frac{\int {\rm i} \rho\Omega_r\mbox{{\boldmath$\xi$}}_{eq,n,k}^*\cdot( {\bf \hat{k}}\times\mbox{{\boldmath$\xi$}}_{eq,n,k}) dV}
{\int\rho|\mbox{{\boldmath$\xi$}}_{eq,n,k}|^2dV}\label{coribeta}
\end{align}
 gives the frequency shift produced by
 the Coriolis force as seen in the rotating frame for a putative normal mode with associated eigenfunction
 $\mbox{{\boldmath$\xi$}}_{eq,n,k}$ \citep{CD1998} and importantly $\beta_*$ does not depend on $n$
 (see next Section \ref{coriresponse}).}

 
 \subsection{Relating the response to the tidal forcing to the  overlap integral associated with the
 equilibrium tide}\label{Respoverlap}
 Taking the complex conjugate of (\ref {EREV}) we obtain
 \begin{align}
 &\int\rho_{n,k}  r^2Y_{2,n}^*(\theta, \phi)dV = -{\cal A}_{n,k} \times\nonumber\\
 &\frac{ \left(\int\rho_{eq,n,k}  r^2Y_{2,n}^*(\theta, \phi)dV\right)     
 \left(\int\rho_{eq,n,k}^*r^2Y_{2,n}(\theta, \phi)dV\right)}{N_0\omega_{eq}^2}
\left( 1- \frac{(2\omega_f (-n\beta_*\Omega_r+ {\rm i} \Gamma) -k^2n_o^2-n^2\Omega_r^2)}{\omega_{eq}^2}\right)
.\label{EREVCC}
\end{align}
Recalling that we can arbitrarily  scale    $\mbox{{\boldmath$\xi$}}_{eq,n,k} $  when evaluating the right hand side of  (\ref{EREVCC}) we now do this
so as to ensure $N_0=1$ and that  after specifying the angular dependence of the spheroidal  associated decomposition functions through (see equation (\ref{e2X}))
\begin{align}
&\xi^r(r,\theta, t ,n) \exp({\rm i} n \phi) = \xi^r(r)Y_{2,n}\exp({\rm i}kn_ot) ,   \nonumber\\
&\hspace{-4.3cm}{\rm and}\nonumber\\
& \xi^S(r,\theta, \phi , n) \exp({\rm i} n \phi) = \xi^S(r)Y_{2,n}\exp({\rm i}k n_ot)\label{sphdecomp}
\end{align}
  the associated  functions $\xi^r(r)$ and $\xi^S(r)$ are real. As these apply to a spherically symmetric
background they are in addition independent of $k$ and $n.$
Then we can write (\ref{EREVCC}) as 
  \begin{align}
 &\int\rho_{n,k}  r^2Y_{2,n}^*(\theta, \phi)dV =\nonumber\\
 &-{\cal A}_{n,k} \frac{ \left(\int\rho_{eq,n,k}  r^2Y_{2,n}^*(\theta, \phi)dV\right)     
 Q_{eq}}{\omega_{eq}^2}
\left( 1- \frac{(2\omega_f (-n\beta_*\Omega_r+ {\rm i} \Gamma) -k^2n_o^2-n^2\Omega_r^2)}{\omega_{eq}^2}\right),\label{EREVCC1}
\end{align}
where
{\begin{equation}
Q_{eq}=  \int\rho_{eq,n,k}^*r^2Y_{2,n}(\theta, \phi)dV= - \int\nabla\cdot (\rho {\mbox{\boldmath$\xi$}}_{eq,n,k}^*)r^2Y_{2,n}(\theta, \phi)dV  =2\int dr \rho r^3 (\xi^r+3\xi^S) 
\label{e7X}
\end{equation}  
is the overlap integral evaluated using the normalised displacement $(N_0=1)$ which as the subscript indicates  corresponds to $\mbox{{\boldmath$\xi$}}_{eq,n,k} .$
In evaluating this we have made use of the decomposition,  given by (\ref{e2X}) with (\ref{sphdecomp}), applied to $ {\mbox{\boldmath$\xi$}}_{eq,n,k},$
and note that there is in fact no dependence of this on $k,$ that the $n$ dependence is only through the spherical harmonics, and  that the factor
 $\exp({\rm i}kn_ot )$ may be dropped from perturbations as a result of cancelation.}
Note that if we restore the normalisation factor, $N_0,$ evaluated with the equilibrium tide, we see that $Q_{eq} \equiv Q_{eq}/\sqrt{N_0}$ has dimensions 
$\sqrt{M_*}R_*,$ with $M_*$ and $R_*$, respectively, being the mass and radius of the { primary.
} 


\noindent By  inspection of the way  the density perturbations appear in (\ref{EREVCC1})  and relating them back
to associated displacements we may make the identification
 \begin{align}
& \mbox{{\boldmath$\xi$}}_{n,k}  =  {\cal D}_{n,k}\mbox{{\boldmath$\xi$}}_{eq,n,k} \hspace{2mm}{\rm where}\hspace{2mm}
 {\cal D}_{n,k}=-{\cal A}_{n,k} \frac{ Q_{eq}}{\omega_{eq}^2}
\left( 1- \frac{(2\omega_f (-n\beta_*\Omega_r+ {\rm i} \Gamma) -k^2n_o^2-n^2\Omega_r^2)}{\omega_{eq}^2}\right).\label{EREVCC2}
\end{align}
Making the time dependence explicit and summing over $k$ 
we may write
\begin{align}
& \mbox{{\boldmath$\xi$}}_{n}  = \mbox{{\boldmath$\xi$}}_{eq,n} \sum^{\infty}_{k=-\infty} {\cal D}_{n,k} \exp({\rm i}kn_0 t)
 \label{EREVCC2X}
\end{align}
where we have removed the subscript $k$ from $ \mbox{{\boldmath$\xi$}}_{eq,n,k}$
as there is no such  dependence for this quantity.
We write equation (\ref{EREVCC2}) in a more compact form
\begin{eqnarray}
&{\cal D}_{n,k}=-Q_{eq} {\cal{ R}}_{n,k}{\cal A}_{n,k}  \hspace{2mm} {\rm where} \hspace{2mm}  \label{e15}\\
&{\cal R}_{n,k}=\omega_{eq}^{-2}(1-\delta_{n,k}+k^2\tilde\Omega^2), \quad {\rm with} \quad \delta_{n,k}=i2\tilde \Gamma\tilde \Omega(k+n\lambda\sigma)-
2n\beta_*\lambda\sigma \tilde \Omega^2 (k+n\sigma).\nonumber
\end{eqnarray}
\noindent Recalling that $\omega_f = kn_o+n\Omega_r,$ we have   
$\tilde \Gamma =\Gamma /\omega_{eq}$, $\tilde \Omega = n_o/\omega_{eq}$\quad  and \quad$\sigma = \Omega_r/(\lambda n_o) $
with $\lambda = 2\beta_*/(2\beta_*+1).$
{
\subsubsection{Evaluating the response}
Equation (\ref{e15}) relates the Fourier expansion coefficient ${\cal D}_{n,k}$ of the displacement to the corresponding Fourier expansion
coefficient in the forcing potential ${\cal A}_{n,k}$ through the factor $-Q_{eq}{\cal R}_{n,k}.$
This consists of a sum of terms $\propto $  $\delta_{n,k} $ and  $\propto k^2\Omega_r^2.$ 
The latter does not depend on $n$, and it  can be verified after summing over $k$ that it produces  a  density response  with the same angular 
dependence as $\partial^2 U/\partial t^2,$ as it can be written as a product of this and a function of $r.$ 
Given  that it is periodic in time this  response can readily be shown to produce no time-averaged  torques or energy dissipation after time averaging over an orbit. 
Accordingly, as it has no secular consequences we shall neglect this term from now on.{\footnote {The same conclusion follows if this term is dealt with
in the same manner as the others by making use of Parsevals theorem (see appendix \ref{Aptima}).}}
\subsubsection{Reduction of the integral determining $\beta_*$ and discussion of the related response}\label{coriresponse}
Here we discuss the contribution to the response from terms $\propto \beta_*$ that are also indeppendent of
$k$. Evaluating the integral equation (\ref{coribeta}) 
by making use of the decomposition,  given by (\ref{e2X}) with (\ref{sphdecomp}), applied to $ {\mbox{\boldmath$\xi$}}_{eq,n,k}$
as above, we obtain
 \begin{align}
 -n(\beta_*-1)\Omega_r = \frac{\int {\rm i} \rho\Omega_r\mbox{{\boldmath$\xi$}}_{eq,n,k}^*\cdot( {\bf \hat{k}}\times\mbox{{\boldmath$\xi$}}_{eq,n,k}) dV}
{\int\rho|\mbox{{\boldmath$\xi$}}_{eq,n,k}|^2dV}
=n\Omega_r\int r^2\rho(2\xi^r\xi^S+(\xi^S)^2)dr,
\label{coribetared}
\end{align}
where we have scaled the displacement to be normalised such that
\begin{align}
N_0 =\int r^2\rho((\xi^r)^2+6(\xi^S)^2)dr=1.
\label{coribetaredn}
\end{align}
Given that $\xi^r,$ and $\xi^S$ do not depend on $n,$  we confirm from (\ref{coribetared}) that neither does $\beta_*$ \citep[see eg.][]{CD1998}.

We now discuss the physical form of the response  induced by the contribution of  terms $\propto \beta_*$ that depend on $n$ but are independent of $k$
to the  Fourier expansion coefficients ${\cal D}_{n,k}$ of the response displacement  corresponding to the  Fourier expansion
coefficient in the forcing potential ${\cal A}_{n,k}.$ 
These terms arise from the  density response resulting from  the Coriolis force produced from the equilibrium tide. 
In particular from (\ref{e15})  we see that for a given Fourier component such a term contributes $-2n^2\beta_*\lambda \sigma^2{\hat \Omega}^2 = n^2\sigma\delta_3
$, so defining $\delta_3,$  to  $\delta_{n,k} .$
As there is no dependence on $k,$ on account of the factor $n^2,$ one can see that the angular response to each Fourier component
of the tidal potential will be misaligned with the angular dependence of that component in a way that depends on $\beta.$
In fact, when  $U$  is restricted to a particular Fourier component in time, the angular dependence of the density response
 associated with these terms
 is $\propto \partial^2U/\partial \phi^2.$
Thus because $U$ is not restricted to a single spherical harmonic, but rather a linear combination of them  in a misaligned system, non zero torques that are directed in the plane perpendicular to the rotation axis
may occur after time averaging. Such torques tend to cause a precession of the angular momentum vectors and because the magnitude of the orbital angular momentum is
affected but not the orbital energy (as can be inferred from the above discussion of the dependence on $\phi$) so is the orbital eccentricity. 
Important in this regard is that the direction of such torques depends on the orientation of the orbit in its plane and hence, $\varpi.$ This is unlike the situation for
the standard precessional torque resulting from rotational distortion considered in appendix \ref{newPrecession}. This is because in that case the time averaged
tidal potential is axisymmetric in the orbital plane which is a special feature of its quadrupole form that would not occur in more general cases.  }


 We remark that it is implied in the analysis presented in this  Paper 
that both $\tilde \Gamma \ll 1$ and $\tilde \Omega \ll 1$. 
On the other hand, $\sigma $ can be order of unity. It is convenient to represent $\delta_{n,k}$ as
\begin{equation}
  \delta_{n,k}={\rm i} ( \delta_1k + \delta_2 n) +(kn+n^2\sigma)\delta_3, \hspace{2mm}{\rm where} \hspace{2mm} \delta_1=2\tilde \Gamma \tilde \Omega,\quad \delta_2 = \lambda\sigma \delta_1 
  ,\quad {\rm and }\quad \delta_3=-2\beta_*\lambda \sigma \tilde \Omega^2.
\label{e17}
\end{equation}  
We remark that the $\delta_i$ have been defined  such that they are independent of $k$ and $n.$

 \section{Finding the  induced  torque acting on the star}\label{TorqueFind}
 { In this Section we use the tidal response calculated in the previous Sections to evaluate
 the components of the torque acting on the primary by performing  the appropriate volume integrals.
On account of the decomposition of the forcing tidal potential,  this naturally leads to results
expressed as double sums over terms arising from each Fourier component of the contribution for each azimuthal mode number,
which are then time averaged.  For the particular form of the response we calculate,
this summation can be performed by  use of Parseval's theorem  leading to expressions in closed form.
Some of the parameters and symbols that occur in the specification and calculation of these quantities
are listed in table \ref{symbols3}. }  


   \begin{table}     
	        \caption{Table of some  parameters, variables and symbols occurring  in the calculation of the components of the
	        torque and rate of change of orbital energy} \label{symbols3}
                 \vspace{-0.3cm}
 	         \begin{center}
		\begin{tabular}
		{c p{12.00cm}}
		\hline
		Symbol & Definition \\
		\hline
           	$\hat {\bf \Phi} $ & This is the operator   ${\bf r}\times \nabla.$  Thus $\hat {\bf \Phi} \equiv{\bf r}\times \nabla$ \\	
	        $\hat {\bf \Phi}^{+}, \hat {\bf \Phi}^{-}$ &These give the components of the operator   $\hat {\bf \Phi} $
	        in the $(X,Y,Z)$ Cartesian coordinate system through $\hat {\bf \Phi}~= ~( (\hat {\bf \Phi}^{-}- \hat {\bf \Phi}^{+})/\sqrt{2} ,  
	        -{\rm i}(\hat {\bf \Phi}^{+}+ \hat {\bf \Phi}^{-})/\sqrt{2},  \hat {\bf \Phi}^{0} )$ (see equation (\ref{jp182}))\\
	         $ {\cal W}_{n_1,n_2,m}^{(j)}(\beta)$& Quantity appearing in the expressions for  tidal torque components. It is constructed from Wigner small $d$ matrix elements 
	         and  is given by equation (\ref{WCRP})\\
	         $f_1,f_2$ &Quantities appearing on reduction of the torque components, given by $f_1=$ \newline$15\sin\beta\cos\beta(1+\cos 2{\hat \varpi}\cos 2(\Phi -\varpi)) /(16\pi),$ 
                  $f_2 =-15\sin\beta\sin 2{\hat \varpi} \cos 2(\Phi -\varpi)/(16\pi) $\\
                  $f_3,f_4$&Quantities appearing on reduction of the torque components, given by $f_3 =15\sin\beta/(8\pi),$  $f_4 =0$ \\
                  $f_5,f_6$&Quantities appearing on reduction of the rate of change of orbital energy, given by equation (\ref{f5f6}) \\
                   $\phi_1$& Factor given by equation (\ref{phi1}) containing the $e$ dependence of  the orbit integral $\oint{(1-e^2)^{-7}R^{-8}}dt$\\ 
		  $\phi_2$& Factor given by equation (\ref{phi2})  containing the $e$ dependence of  the orbit integral 
                   $\oint{(1-e^2)^{-5}R^{-6}}dt$ \\	
                    $\phi_3$& Factor given by equation (\ref{phi3}) containing the $e$ dependence of the orbit integral 
                    \newline $\oint{\cos2(\Phi-\varpi)}{(1-e^2)^{-5}R^{-6}}dt$ \\	
		 $\phi_4$& The $e$ dependence of the term $\propto \delta_1$ in $dE_{orb}/dt$ is given by $\phi_4/(1-e^2)^{3/2}$
		 (see equations (\ref{EDOTF}) and (\ref{jp18e1}). It is given by  (\ref{phi}) as $\Phi_4=1+{31}e^2/2+ {255}e^4/8+{185}e^6/16  +{25}e^8/64$\\ 
  		$\phi_5,\phi_6$& Respectively $(\phi_4-(1-e^2)\phi_1)/(9e^2)$ and $2(\phi_1-(1-e^2)\phi_2)/(11e^2)$ used in equation (\ref{ev5p}) for $de/dt$\\
		$\phi_7=(\phi_1\phi_4-\phi_2^2)/e^2$& The $e$  dependence of the contribution to $da/dt$ from the perturber is $e^2\phi_7/((1-e^2)^{3/2}\phi_2)$(equation(\ref{jp18e1p}))\\
		$\phi_8/(1-e^2)^5$& Gives  $e$ dependence of term $\propto \sigma$ in equation (\ref{Apse2}) for $d\varpi/dt$ with $\phi_8 =(12+46e^2+5e^4)/20$\\
		\hline
		\end{tabular}
 	\end{center}
	\end{table}

Substituting  the displacement given by  (\ref{e1X}) the torque acting on the 
star,  ${\bf T}$, 
is readily
calculated.
Working in the stars frame we have
\begin{equation}
{\bf T} = -\int_V \rho'{\bf r}\times \nabla U dV\label{jp18}
\end{equation}
where $\rho' = -\nabla\cdot (\rho \mbox {\boldmath$\xi$})$ is the density response and the integral is taken over the volume
of the star.
We denote the operator ${\bf r}\times \nabla \equiv \hat {\bf \Phi}.$ The components in the $(X,Y,Z)$ coordinate system are
\begin{equation}
\hat {\bf \Phi} =  \left(-\cot\theta \cos\phi
\frac{\partial}{\partial \phi}
-\sin \phi\frac{\partial}{\partial \theta} \ \  ,  \ \  -\cot\theta \sin\phi
\frac{\partial}{\partial \phi}
+\cos \phi\frac{\partial}{\partial \theta}  \ \  ,  \ \  \frac{\partial}{\partial \phi}   \right)
        \label{jp181}
\end{equation}
\begin{equation}
\hspace{-6.6cm}
{\rm  and \hspace{1mm} the  \hspace{1mm} torque  \hspace{1mm} is  \hspace{1mm}  given  \hspace{1mm} by}\hspace{4mm} {\bf T}
=   \int_V \nabla\cdot (\rho \mbox {\boldmath$\xi$}^{*} ) \hat {\bf \Phi}U dV ,\label{jp21X}
\end{equation}
where we remark that because $\mbox {\boldmath$\xi$}$ is real  we may take its  complex conjugate in (\ref{jp21X}).
Using equation (\ref{e1X})  for $\mbox {\boldmath$\xi$},$ after some algebra (see Appendix \ref{Torquecalcapp}) we obtain the components of the torque
expressed in terms 
of  the equilibrium tide components for a given $n$  specified by  equation (\ref{EREVCC2})
 in the form
\begin{equation}
{T}^{z} = -\sum_{n=0}^{n=2}2{\cal R}\left[\left(\frac{4\pi GM_p }{5a^3}\right)\int_V \nabla \cdot (\rho \mbox {\boldmath$\xi$}_{n}^* )  \left( {\rm i}n r^2
Y_{2,n}(\theta, \phi)F_{n}(t) \right)dV\right],\label{jp22iX}
\end{equation}
\begin{align}
& \hspace{-0.56cm}\frac{5a^3  T^x  }{2\pi GM_p}= \sum_{n=0}^{n=2}(\delta^{n}_{0}-2){\cal R}\left[ \int_V \nabla \cdot (\rho\mbox {\boldmath$\xi$}_{n+1}^*) 
   \left( {\rm i}r^2
   (  Y_{j,n +1}\sqrt{(j- n)(j +n+1)}   F_{n}\right) dV \right. \nonumber
   \\
&\hspace{-0.6cm}\left.  +\int_V \nabla (\cdot \rho   \left(\mbox {\boldmath$\xi$}_{n-1}^*\right )) 
  \left( {\rm i}r^2
     Y_{j,n -1}\sqrt{(j+n)(j -n+1)}   F_{n}\right) dV \right]
\label{jp23iX}
\end{align}
and
\begin{eqnarray}
&& \hspace{-0.56cm}\frac{ 5a^3{T}^{y}}{2\pi GM_p} = \sum_{n=0}^{n=2}(2-\delta^{n}_{0}){\cal R}
\left[ \int_V \nabla\cdot \left(\rho \mbox {\boldmath$\xi$}_{n-1}^*\right ) 
  \left( r^2Y_{j,n -1}\sqrt{(j+ n)(j -n+1)}   F_{n}\right) dV \right. \nonumber\\
&&\hspace{-0.6cm}-\left. \int_V \nabla \cdot \left( \rho \mbox {\boldmath$\xi$}_{n+1}^*\right )  
 \left( r^2  Y_{j,n +1}\sqrt{(j-n)(j +n+1)}   F_{n}\right) dV \right],
\label{jp24aiX}
\end{eqnarray}
where $\delta^{n}_{0}$ is the Kronecker $\delta.$

 We comment that when calculating the $X$ and $Y$ components
of the torque, the above equations imply that the azimuthal mode number of  a significant  response has to differ
from that of the  original forcing potential by $\pm 1.$
Apart from this,  the expressions  consist of contributions that are similar in form
to  that given by equation (\ref{jp22iX}) for the component of the torque in the $Z$  direction.

\subsection{The rate of change of orbital energy}\label{Orbenrate}
Working in the stellar frame the rate of
change of orbital energy is given by
\begin{equation}
\frac{dE_{orb}}{dt}=-\int_V  \rho' \left(\frac{\partial U }{\partial t} \right)dV.
\label{jp18ef}
\end{equation}
Note that, by taking the time derivative of  (\ref{jpe7}) we obtain
\begin{equation}
\frac{\partial U}{\partial t}=-GM_p r^2\left(\frac{4\pi}{5a^3}\right)\sum_{n=-2}^{n =2}Y_{2,n}(\theta, \phi)\frac{\partial F_{n}(t)}{\partial t}.
\label{jpe7e}
\end{equation}
In addition, from  (\ref{jpe711}), making the assumption that the orbital elements apart from
$\varpi$ are fixed,   we infer that
\begin{equation}
\frac{\partial F_{n}(t)}{\partial t} = \sum^{'}_{m=0, |2|} \sum_{k=-\infty}^{k=\infty} \left( {\rm i}k n_{o} -md\varpi/dt \right)\phi_{k,m}D^{(2)}_{n,m} 
 Y_{2,m}(\pi/2, 0)\exp({\rm i}(k n_o t -m\varpi))\label{jp18eg}
\end{equation}
Supposing initially that only  the terms in the sum (\ref{jpe7e})  corresponding to   a particular pair of values    $\pm n$ are retained, 
from (\ref{jp18ef}) -(\ref{jp18eg}) we obtain
\begin{equation}
\frac{dE_{orb}}{dt}=
-GM_p\left(\frac{4\pi}{5a^3}\right)\int_V r^2  (\nabla\cdot(\mbox {\boldmath$\xi$}_{n}^* ))
 \left(  Y_{2,n}(\theta, \phi)\frac{\partial F_{n}(t)}{\partial t} \right)dV + cc,
\label{jp18e}
\end{equation} 
where $cc$  denotes the complex conjugate.
 The  first term on the right hand side of the expression  (\ref{jp18e})
is such that when $n \rightarrow -n$ 
the complex conjugate is obtained.
Thus, when a sum of these terms over $n$ is made,   the result  is real.
As above we may consider only $n \ge 0$ and write the total 
rate of change of orbital energy as
\begin{equation}
\frac{dE_{orb}}{dt}=
-GM_p\left(\frac{4\pi}{5a^3}\right)\sum_{n=0}^{n=2} {\cal R}(2-\delta^{n}_{0})\left[\int_V r^2  (\nabla\cdot(\mbox {\boldmath$\xi$}_{n}^* ))
 \left(  Y_{2,n}(\theta, \phi)\frac{\partial F_{n}(t)}{\partial t} \right)dV\right].
\label{jp18e0}
\end{equation} 
The integral in the above  expression (\ref{jp18e0}) 
is of a similar form to that found in the expression 
for ${T}^{z}$ given by equation (\ref{jp22iX}). The latter involves
$F_{n}(t)$, while the former involves its time derivative.

\subsection{Reduction of the torque and rate of energy exchange  integrals}\label{Torquereduct}
Now we substitute our expression for the response displacement associated with
a  particular   $n$ given by  equation (\ref{EREVCC2X}) 
into the expressions for the components of the torques 
given by (\ref{jp22iX}) -  (\ref{jp24aiX}) and the rate of change of orbital energy given by 
(\ref{jp18e0}).   
Using the expression for the displacement given by (\ref{e2X}))
and the decomposition for the equilibrium tide response given by
(\ref{sphdecomp})  we have
\begin{equation}
\nabla \cdot (\rho  \mbox{{\boldmath$\xi$}}_{eq,n})= {\overline R}Y_{2,n}, \quad {\rm where} \quad
{\overline R}={1\over r^2}{d\over dr}(r^2\rho \xi^r)-{6\rho \over r}\xi^S.
\label{e5X}
\end{equation}  
Here  we imply the known properties 
of spherical harmonics   to obtain the second equality.

We then   make use of  (\ref{jpe711}) to allow us to eliminate the  $F_{n}$ in favour of
expressing the result in terms of the
coefficients $A_{n},$  perform the
integration over the volume of the star by parts  and the summation over $n$ to obtain

\begin{equation}
{T}^{z} = 2Q_{eq}  {\cal I}(A_1b_1^{*}+ 2b_2^{*}A_2),\label{jp30}
\end{equation}
where ${\cal I}$ denotes that the imaginary part is to be taken and (see equation (\ref{EREVCC2X}) )
\begin{equation}
b_{n} =  \sum^{\infty}_{k=-\infty} {\cal D}_{n,k} \exp({\rm i}kn_o t),\label{BEXP}
\end{equation}
where we have implied that $Q_{eq}$ is real.
 Similarly we find
\begin{equation}
{T}^{x} = 2 Q_{eq}{\cal I}\left(A_1b_2^{*} + b_1^{*} A_2 
    + \sqrt{3/2}\left(\frac{1}{2}(b_1^{*}A_0-b_{1}A_{0})  + b_0^{*}A_1 \right)\right),\label{jp31}
\end{equation}
and
\begin{equation}
{T}^{y} = -2Q_{eq} {\cal R}\left(A_1b_2^{*} - b_1^{*}A_2     + 
\sqrt{3/2}\left( \frac{1}{2}( b_1^{*}A_0 +b_{1}^{}A_0   ) - b_0^{*}A_1 \right) \right),\label{jp32}
\end{equation}
From the above two equations it follows that
\begin{equation}
{T}^{x} -{\rm i}{T}^{y} =
2 Q_{eq}{\rm i}(A_1^*b_2 - b_1^{*} A_2 
    + \sqrt{3/2}( b_{1}A_{0}  - b_0^{*}A_1 ),\label{jp3132}
\end{equation}
We remark that we have made use of the fact that $A_{-n}^* = (-1)^n A_n$ and $b_{-n}^{*} =(-1)^{n} b_{n}$


\noindent with the former implying that $A_0$ is real. Finally, the above process can be applied to  (\ref{jp18e0}) to determine the rate of
change of orbital energy as
\begin{equation}
\frac{dE_{orb}}{dt} = -2 Q_{eq} {\cal R}\left( b_1^{*}\frac{d A_1}{dt}+ b_2^{*}\frac{dA_2}{dt}   
  + \frac{1}{2} b_0^{*}\frac{d A_0}{dt} \right).\label{jp33}
\end{equation}

\subsection{Fourier decomposition of  the $A_{n}(t)$}\label{FdA}
Recalling that $A_{n} =-4\pi GM_p F_{n}/(5a^3) $
\textcolor{blue}{and using equations  (\ref{jpe7}),  (\ref{jpe711}) and  (\ref{jpe711e}) } we may write 
\begin{equation}
A_{n}(t)=\sum_{k=-\infty}^{k=\infty} {\cal A}_{n,k}\exp({\rm i}k n_o t )).\label{FdA1}
\end{equation}
We shall suppose that the orbit has an extremely slowly varying or fixed  apsidal line so that it can be regarded as being constant
when averaging over the fast orbital time scale. Even so the apsidal rotation rate may be rapid 
compared to the rate of tidal evolution.

\subsection{Time averages of relevant quantities}\label{Timeave}
\subsubsection{The case with non zero apsidal precession}\label{nonzeroapse}
From the above analysis we can evaluate quantities such as the time average of  the products $b_{n_1}^{*}(t) A_{n_2}(t).$
With the help of the expansions (\ref{BEXP}) and (\ref{FdA1})  as well as   (\ref{e15}) 
we find that when the longitude of pericentre varies sufficiently slowly that can be taken to be constant
during the time averaging process, we find

\begin {equation}
\langle b_{n_1}^{*}(t) A_{n_2}(t)\rangle = 
\sum_{k=-\infty}^{k=\infty} {\cal A}_{n_2,k}
{{\cal D}_{n_1,k}^{*}} 
=-\sum_{k=-\infty}^{k=\infty} {Q_{eq}}{{\cal R}_{n_1,k}^*}
\left. 
{\cal A}_{n_2,k}
{{\cal A}_{n_1,k}^{*}}, 
   \right.\label{aveKep}
\end{equation} 
{where the angled brackets denote time averaging.} Corresponding to this we have
\begin {equation}
\left\langle b_{n_1}^{*}(t) \frac{d A_{n_2}(t)}{dt}\right\rangle =- \sum_{k=-\infty}^{k=\infty} 
{\rm i}kn_oQ_{eq}
{\cal A}_{n_2,k}{\cal A}_{n_1,k}^{*} {\cal R}_{n_1,k}^*,\label{aveGenen}
\end{equation}

\subsubsection{First order departures from the equilibrium tide}\label{equidep}
Assuming that corrections to an equilibrium tide are small,  
$\delta_{n,k}$ is also small in magnitude. Thus, 
 terms only up to first order in this quantity are retained
recalling that 
\begin{equation}
{\cal R}_{n,k} = {\omega_{eq}^{-2}}(1-\delta_{n,k}). 
\label{e16ap}
\end{equation}


\noindent Accordingly, an expression such as for example   (\ref{aveKep}), which is  needed in order to evaluate
the torque components  becomes
\begin {equation}
\left\langle b_{n_1}^{*}(t) A_{n_2}(t)\right\rangle = -
\sum_{k=-\infty}^{k=\infty} \frac{Q_{eq}}{ \omega_{eq}^2}\left({\cal A}_{n_2,k}{{\cal A}_{n_1,k}^{*}}
 (1+{\rm i} ( \delta_1k + \delta_2 n_1) -(kn_1+(n_1)^2\sigma)\delta_3)\right).
\label{aveKepap}
\end{equation}


\subsection{Expressions for the components of  the torque in closed form}\label{closedtorque}
  The torque components and rate of orbital energy change  given by 
equations (\ref{jp30}) -(\ref{jp33}), that are required to enable calculation of the evolution of orbital elements
and stellar spin are expressed in terms of time averages of coefficient products,
that can be expressed as infinite sums as exemplified in equations (\ref{aveGenen}) - (\ref {aveKepap}).
As the terms in these summations are quadratic in $k$ these summations can be performed by
making use of sum rules obtained with help of Parseval's theorem applied to the Fourier coefficients
occurring the expansion of the perturbing potential as specified in Appendix \ref{Aptima}.

We should emphasise that, the property whereby the terms in the summations
involve  powers of $k$ less than 2 comes about from the assumed constancy of the decay rate
$\Gamma$, which has to be independent of forcing  frequency. While this is correct 
for standard viscosity (see Ivanov \& Papaloizou 2004), it is not true for radiative diffusion.
In such a case one cannot readily take advantage of the sum rules and must employ infinite 
summations to determine the evolution of the system.
Alternatively, a constant average value of $\Gamma$ could be assumed.
It can be assumed that has been adopted in what follows below.

\subsubsection{Evaluation of  time averaged torques and rate of change of orbital energy}\label{Inteval}
{ These quantities  are evaluated in appendix \ref{Aptima} 
with help of the integrals with respect to time  obtained there, that are used to obtain sum rules, that can be used to evaluate the
sums of the form specified in  (\ref{aveGenen}) and (\ref{aveKepap} ) that are needed to evaluate time averaged torques and  rate of change of orbital energy.
The latter are determined in appendix \ref{Aptima}  where the reader interested in the details is referred. The results for the
torque components given there are}
\begin{align}
&\hspace{-0cm}{T}^{z} = T_*\left(
2\delta_1 \cos\beta
\phi_1
-\delta_2(1-e^2)^{3/2}
\left(\left (1+\cos^2\beta\right)
\phi_2- \sin^2\beta\cos2{\hat \varpi} \phi_3  \right)\right)\hspace{3mm}{\rm and}\hspace{3mm}
\label{t1} 
\end{align}
\begin{align}
&\hspace{0.5cm}T \equiv {T}^{x}-{\rm i}{T}^{y} = \nonumber\\
&\hspace{0.5cm} T_*\sin \beta ((2\delta_1-{\rm i}\delta_3)\phi_1
-( 1-e^2)^{3/2}(\delta_2-{\rm i}\sigma\delta_3)
\left((\phi_2+\phi_3\cos(2{\hat \varpi}))\cos\beta - {\rm i} \sin(2{\hat \varpi} )\phi_3\right)),
\label{t2} 
\end{align}
and the  change of orbital energy is given by
\begin{align}
&\frac{dE_{orb}}{dt} =\dot E_*
\left(\delta_2\phi_1 \cos\beta 
-\frac{\delta_1}{(1-e^2)^{3/2}}
\phi_4 
 \right).
\label{eng1} 
\end{align}
Here
\begin{equation}
T_*= \frac{6\pi}{5}\left(\frac{ GM_pQ_{eq}}{ a^3(1-e^2)^3\omega_{eq}}\right)^2 = \frac{3k_2q^2}{1+q}
\left(\frac{R_*^5}{a^5}\right) \frac{M_*n_o^2a^2}{(1-e^2)^6} \quad {\rm and} \quad
\dot E_*=2n_o T_*,
\label{eqadd1}
\end{equation}
respectively, represent typical values of the torque and rate of change of  energy. Note that in the second equality we have used
equation (\ref{apsidal}) in Appendix \ref{newPrecession} with $N_0=1$ to relate $T_*$ to the apsidal motion constant.
{The quantities $\phi_1,\phi_2,\phi_3,$ and $\phi_4$ are functions of the eccentricity. They are specified both 
in table \ref{symbols3} and appendix \ref{Aptima}.}

\newpage

\section{Determination of the orbital  and spin  evolution }\label{Detorbspinev}
{ In this Section we use the time averaged torque and rate of orbital energy change obtained above and with the help of results in appendix \ref{Aptima}
to obtain equations governing the orbital and spin evolution that depend only on parameters required to specify them and quantities intrinsic to the primary 
star. We separately incorporate the standard  precession of its spin axis induced by rotational flattening that is otherwise not included in our discussion.
In addition we incorporate effects arising from dissipation in the compact companion under the assumption that it can only contain negligible angular momentum
and  thus instantaneously  adjusts its spin so as to attain  a condition of alignment with the orbit and net zero torque. }

\subsection{Expressions for the evolution of angular momentum vectors}
\label{Amvexp}
The rate of change of the absolute values of orbital and spin angular momentum vectors and the angles 
determining their orientation with respect to the { primary centred}  coordinate system follow from 
equations (\ref{e45}) and (\ref{e46}) after substitution of the components of the torque obtained from equations (\ref{t1}) and (\ref{t2}). 
Proceeding in this away we obtain the rates of change of the orientation  specifying angles in the form
\begin{align}
&\hspace{-2.8cm}{d i\over dt} = -(1-e^2)^{3/2}\sin \beta {T_*\over L}(\sigma \delta_3 \phi_3 \cos \beta \sin 2\hat \varpi 
+\delta_2 (\phi_2 - \phi_3 \cos 2 \hat \varpi)),    
\label{ev1}    
\end{align}
\begin{align}
&\hspace{0cm}\frac{d \delta}{dt} = -{T_*\over S}\sin \beta\left( 2\delta_1 \phi_1
-(1-e^2)^{3/2}(\delta_2 \cos \beta (\phi_2 + \phi_3\cos 2\hat \varpi )+\sigma \delta_3 \phi_3 \sin 2\hat \varpi )\right),
\label{ev3}    
\end{align}
\begin{align}
&\hspace{-16mm}\frac{d  \alpha_r}{dt} = - {JT_*\over SL}\bigg( (\delta_3\phi_1-
(1-e^2)^{3/2}(\sigma \delta_3\cos \beta (\phi_2 + \phi_3 \cos 2\hat \varpi )+
\delta_2 \phi_3 \sin 2\hat \varpi ))\nonumber \\
&\hspace{-12mm} +{1\over 3}(1-e^2)^{9/2}{1+q\over q}\sigma^2 \cos \beta \bigg)
\label{evn2}
\end{align}
and we note
that the rate of change of the  angle of inclination  between the spin and orbital angular momenta
is 
\begin{align}
&\hspace{-62mm}\frac{d\beta}{dt}  =\frac{d i }{dt}+\frac{d \delta}{dt} .\hspace{60mm}. 
\end{align}
The rate of change of the magnitudes of the orbital and spin angular momenta are given by
\begin{align}
&\hspace{-15mm}\frac{d L}{dt} = -2 T_* (\delta_1 \phi_1-(1-e^2)^{3/2}\delta_2\phi_2 \cos \beta)-(1-e^2)^{3/2}T_* \sigma \delta_3 \phi_3 \sin^2 \beta
\sin 2\hat \varpi,
\label{ev2}    
\end{align}
and 
\begin{align}
&\hspace{-5.8cm} \frac{d S}{dt} = T^z,\hspace{7.2cm}
\label{ev4}
\end{align}
where $T^z$ is given by equation (\ref{t1}). 

Note that in addition to the 
torque component $T^y$  that has been incorporated  in equation (\ref{evn2}) (see equation(\ref{e45}))  we have also included
an additional  torque component $T^y_{SF}$ arising
from the effect of stellar flattening due to rotation. 
It is calculated in a form convenient for our purposes  in Appendix \ref{newPrecession} ( see equation
(\ref{D2})).  It is represented in (\ref{evn2}) as the last
term \textcolor{red}{and has}  the factor $(1+q)/ q$, where $q=M_p/M_*$ is the mass ratio. While our 'standard' torque component $T^y$ is proportional to stellar rotational frequency $\Omega_r$, $T^y_{SF}$ is proportional to the square of $\Omega_r.$

\noindent \textcolor{red}{In addition, we recall that $J$ is the 
conserved total angular momentum of the system, while $i$ and $\delta$ are,  respectively, the angles of inclination
between this and the orbital and spin angular momenta. 
 The quantities $\phi_i$ are given by
equations (\ref{phi1})-(\ref{phi}) with $T_*$ and ${\dot E^*}$ being given by equation (\ref{eqadd1}).}

\subsection{Evolution of the semi-major axis and eccentricity}\label{andeev}
\noindent For Keplerian orbits the relationship between the rate of change of the 
semi-major axis and the rate of 
change of orbital energy is given by \textcolor{red}{
\begin{equation}
\frac{d a}{dt} = \frac{2a^2}{GM_pM_*}\frac{dE_{orb}}{dt}=  
\frac{2a^2}{GM_pM_*}\dot E_*
\left(\delta_2\phi_1 \cos\beta  
-\frac{\delta_1\phi_4}{(1-e^2)^{3/2}}\right),  \label{jp18e1}
\end{equation}
where we have used the expression for } $dE_{orb}/dt$  given by equation (\ref{eng1}).
The  rate of change of the
orbital eccentricity is given in terms of the  rates of
change of orbital angular momentum and  energy  by
\begin{equation}
\frac{d e}{dt} =\frac{a(1-e^2)}{GM_pM_*e} \left(\frac{dE_{orb}}{dt}-
\frac{d L}{dt}\frac{\sqrt{G(M_p+M_*)}}{a^{3/2}\sqrt{1-e^2}}\right)\label{jp1121}
\end{equation}
Substituting (\ref{eng1}) and (\ref{ev2}) in (\ref{jp1121})
we obtain
\textcolor{red}{
\begin{align}
&\dot e =-{3a e(1-e^2)^{-1/2}\dot E_*\over GM_pM_*}(3\delta_1 \phi_5-{11\over 6}\delta_2\phi_6(1-e^2)^{3/2}\cos \beta)-\nonumber\\
&{3\over 4}{a e (1+e^2/ 6)(1-e^2)^2\dot E_*\over
GM_pM_*}\sigma \delta_3 \sin^2 \beta \sin 2\hat\varpi ,
\label{ev5}    
\end{align}}
where we \textcolor{red}{ make use of equation (\ref{phi3}) to obtain} $\phi_3$ and
\begin{equation}
\phi_5=(\phi_4-(1-e^2)\phi_1)/(9e^2)=
1+{15\over 4}e^2+{15\over 8}e^4 +{5\over 64}e^6,
\label{ev6}    
\end{equation}
and
\textcolor{red}{
\begin{equation}
\hspace{-6mm} \phi_6=\frac{2(\phi_1-(1-e^2)\phi_2)}{11e^2}=1+{3\over 2} e^2
+{1\over 8}e^4.
\label{ev7}    
\end {equation}}


\textcolor{red}{
\subsection{Incorporating tidal dissipation in the companion}\label{inccomp}
So far we have regarded the companion of mass, $M_p,$ as acting like a point mass.
However, when it represents a giant planet, tides are expected to be significant in 
leading to orbital circularisation \citep[e.g.][]{IP2007}. During this process it is still possible
to neglect its angular momentum content with the consequence that the evolution of the angular momentum vectors
described above is unaffected. We further assume that the evolution of the companion spin  angular momentum
rapidly adjusts so that a zero torque  applies. However, energy dissipation still occurs which can lead to eccentricity damping.
In order to apply the zero torque condition we assume the companions spin is aligned with the  orbital
angular momentum and apply forms of equations (\ref{ev2})~-~(\ref{ev4}) and (\ref{eng1}), respectively, 
governing the spin up torque and rate of energy dissipation adapted to apply to the companion.
To do this we interchange $M_p$ and $M_*$ and  signify that quantities apply to the companion by adding a subscript, $p,$
so that  e.g. $\delta_i \rightarrow \delta_{i,p}$.  Thus, with assumed spin-orbit alignment and zero spin up torque (\ref{ev4})
implies that
$\delta_{1,p} \phi_1=(1-e^2)^{3/2}\delta_{2,p}\phi_2.$
Using this together with the adapted form of  equation (\ref{eng1}) gives
\begin{align}
&\left(\frac{dE_{orb}}{dt}\right)_p =-\dot E_{*,p}\frac{\delta_{1,p}e^2 \phi_7}{(1-e^2)^{3/2}\phi_2  } ,
\label{eng1p} 
\end{align}
where 
\begin {align}
\phi_7= \frac{\phi_2\phi_4 -\phi_1^2}{e^2}
 =\frac{7}{2}+\frac{45}{4}e^2+28e^4+\frac{685}{64}e^6+\frac{255}{128}e^{8}+\frac{25}{512}e^{10}\label{phi7}
 \end{align}
with
\begin{equation}
\dot E_{*,p}=2n_o T_{*,p}  \quad {\rm and} \quad  T_{*,p}= \frac{6\pi}{5}\left(\frac{ GM_*Q_{eq,p}}{ a^3(1-e^2)^3\omega_{eq,p}}\right)^2 
\label{eqadd1p}
\end{equation}
and we note that
 $\delta_{1,p} =2 \Gamma_p n_o/\omega^2_{eq.p}.$ 
\newline
\noindent We can now add $(dE_{orb}/dt)_p$ found above to $dE_{orb}/dt$ in equations (\ref{jp18e1}) and (\ref{jp1121})
in order to find the effect of the companion on the orbital evolution. Only the semi-major axis and eccentricity are affected.
Equation   (\ref{jp18e1}) for the rate of change of the semi-major axis  becomes
\begin{equation}
\frac{d a}{dt} =
\frac{2a^2}{GM_pM_*}\left(\dot E_*
\left(\delta_2\phi_1 \cos\beta  
-\frac{\delta_1\phi_4}{(1-e^2)^{3/2}}\right)
-\dot E_{*,p}\frac{\delta_{1,p}e^2 \phi_7}{(1-e^2)^{3/2}\phi_2  }
\right)
 \label{jp18e1p}
\end{equation}
Similarly, equation  (\ref{ev5}) for the rate of change of the eccentricity becomes}
\textcolor{red}{
\begin{align}
&\dot e =-{3a e(1-e^2)^{-1/2}\over GM_pM_*}\left(\dot E_*
(3\delta_1 \phi_5-{11\over 6}\delta_2\phi_6(1-e^2)^{3/2}\cos \beta)
+ \dot E_{*,p}\frac{\delta_{1,p}\phi_7}{3\phi_2}\right)- \nonumber\\
&{3\over 4}{a e (1+e^2/ 6)(1-e^2)^2\dot E_*\over
GM_pM_*}\sigma \delta_3 \sin^2 \beta \sin 2\hat\varpi .
\label{ev5p}    
\end{align}}
 
The evolution equations (\ref{ev1})-(\ref{jp18e1}) or (\ref{ev5}) form
a complete set for the system in which tides in the companion play no role. \textcolor{red}{Tidal energy dissipation can be
included under the assumption that the spin angular momentum may be neglected provided
 equation (\ref{jp18e1}) is replaced by equation (\ref{jp18e1p}), and equation (\ref{ev5}) by equation (\ref{ev5p}).
  In both cases } an equation for the evolution
of the \textcolor{red}{ orientation of}  the apsidal line characterised by the angle $\hat \varpi $ should be specified. 
The contribution of tidal distortion to this is readily estimated by noting that this is produced
by a deformation that follows the \textcolor{blue}{perturbing star}  and so it does not depend on the inclination of the orbit
or spin axis if the unperturbed  star is approximated as being spherical.
Accordingly, \textcolor{blue}{ in the first instance taking into account the distortion of the primary by the companion, we}  apply the standard theory \citep[e.g.][]{St1939} which gives
\begin{align}
\frac{d{\hat\varpi}}{dt} = \frac{d{\varpi}}{dt} =15 k_2n_0\frac{M_pR_*^5}{M_*(a(1-e^2))^5}\phi_6.\label{Apse}
\end{align} 
In writing the first equality in
(\ref{Apse}) we recall that  the orbital and spin angular momenta precess together about the fixed total
angular momentum vector with the angle $\gamma$ defined in Section \ref{Geometry} remaining constant.
In addition, $\phi_6$ is given by eq. (\ref{ev7}) 
 and the apsidal motion constant $k_2$ can be related to the parameters in our formalism
through equation (\ref{apsidal}) of Appendix \ref{newPrecession} which gives
\begin{align}
&\hspace{2cm} 2k_2 = \frac{4\pi G Q_{eq}^2}{5N_0 {\omega^2_{eq}}R_*^5}.\label{apsidal1}\hspace{7cm}.
\end{align}
 As this does not depend on dissipation it  leads to an evolution time scale that will be much shorter than \textcolor{blue}{processes}  that do.
 \textcolor{blue}{ Assuming that the effect of tidal distortion of the companion by the primary may be important in some circumstances we note that it can be taken into account by modifying (\ref{Apse}) in the form
 \begin{align}
\frac{d{\hat\varpi}}{dt} = \frac{d{\varpi}}{dt} =15 k_2n_0\frac{M_pR_*^5}{M_*(a(1-e^2))^5}(1+{\cal F}_{cp})\phi_6,\label{Apse1a}
\end{align}
 where 
  \begin{align}
{\cal F}_{cp}= \frac{k_{2,p}M_*^2R_{*,p}^5}{k_2M_p^2R_*^5},\label{Apse1aa}
\end{align}
with $k_{2,p}$ and $R_{*,p}$, respectively, being the apsidal  motion constant and radius of the companion. 
We remark that
 non dissipative contributions  arising from $\delta_3$  \textcolor{blue}{generally}  lead to significantly smaller effects and this is discussed in Section \ref{NonrotT}.
It is possible that other effects such as perturbations due to other orbiting bodies are more important 
and these may be added in.}
\textcolor{red}{Note too that equation (\ref{ev3}) for $\alpha_r$ decouples from the others
in that it is ignorable when solving  them for the evolution of the system. It can be integrated once that has been determined.}

Focusing on the primary while noting that a parallel
discussion applies to quantities associated with the companion, 
we recall that coefficients $\delta_i$, $(i=1,3)$, entering the above equations are defined through equation (\ref{e17}) { and also set out in table \ref{symbols2}
as $\delta_1=2\tilde \Gamma \tilde \Omega$, $\delta_2=\lambda \sigma \delta_1$ and $\delta_3=-2\beta_*\lambda\sigma 
{\tilde \Omega}^2$, where $\tilde \Gamma =\Gamma /\omega_{eq}$, $\tilde \Omega = n_o/\omega_{eq}$\quad  and \quad$\sigma = \Omega_r/(\lambda n_o).$ 
with $\lambda = \beta_*/(\beta_*+1/2),$ is the product of $\lambda^{-1}$ and the ratio of the stellar rotation frequency to the orbital mean motion. } 


The coefficient $\beta_*$
is defined in equations (\ref{EDs3}), { (\ref{coribeta}) and (\ref{coribetared}) with the influence of terms  $\propto \delta_3\sigma,$  which is $\propto\beta_*,$ 
on the tidal response being discussed in Section \ref{coriresponse}.} We remark that $\beta_*$
is expected to be smaller than, but of the order of unity. The quantity $2\Gamma$ defined through equation (\ref{EDs1}) represents
the ratio of the  rate of energy dissipation to twice the kinetic  energy associated with the tidally excited disturbance, which takes
the form of an equilibrium tide. It would be the decay rate of this disturbance were it to be a normal mode.
The quantities $\omega_{eq},$ and $Q_{eq}$ defined by equations (\ref{DDs1}) and (\ref{e7X}), respectively. 
represent a putative normal mode frequency and overlap integral associated with this disturbance.

\noindent  \textcolor{red}{A distinctive feature of the evolution equations is that
they contain two types of terms - those proportional 
to $\delta_1$ and $\delta_2$, which are, in turn, proportional to the decay rate $\Gamma$, 
when tides in the companion are included, $\delta_{1,p}$ which is proportional to $\Gamma_p$  is also involved,
and those 
proportional to $\delta_3$, which are independent of
the dissipation rate and are determined by rotational effects. We  respectively describe these terms as dissipative and rotational,
 and consider them in turn below.  }
 
However, we  note in passing that  apart from in equation (\ref{evn2}) for $\alpha_r,$
 which decouples from the rest,
 $\delta_3$ appears together with a factor $\sin2{\hat \varpi}$ or $\cos2{\hat\varpi} .$
 If the \textcolor{blue}{apsidal line of} 
 orbit precesses at a uniform rate with a period short  enough compared to other time scales  these
 quantities can be dealt \textcolor{blue}{ with following an averaging approach to determine  the role of terms $\propto \delta_3$ 
  in character of the orbital evolution.
 If the precession is not uniform 
 \textcolor{blue} {and/or the inclination angle $\beta$ has significant variations}
 these quantities can have a non zero time average even when the precession is fast, thus the value of $\delta_3$ may affect the orbital evolution. 
 This issue is discussed further  \textcolor{blue}{in  Section \ref {NonrotT} } where we estimate conditions under which 
 tidally driven apsidal precession is 
  rapid enough for averaging to be valid. }

\textcolor{red}{
\subsection{The evolution of orbital parameters  when only dissipative terms are included }\label{eqsum}
In this Section we formally neglect the contribution of rotational terms by setting $\delta_3$ to zero in the evolution equations 
given above. We shall also neglect tidal effects arising from the companion.
The   set of  equations  so obtained can then
be compared with those of EKH. These sets
can be seen to be  formally equivalent provided we take into account 
that EKH use different  angles to characterise
their coordinate systems 
\footnote{EKH use the angles $\eta$ and $\chi$ to characterise inclination and rotation of the orbital plane
with respect to an inertial coordinate system, respectively, and the 
angle $\psi$ to characterise position of the apsidal line.
It turns out that they are related to the angles used
in this Paper as $\eta =i$, $\chi =-\alpha -\pi/2\equiv \alpha_r -\pi/2$ and
$\psi= \hat \varpi - \pi/2$.} 
and relate the quantities defining the corresponding evolutionary timescale in our model to
the 'tidal friction' timescale,  $t_{TF}$,  adopted in EKH through
\begin{equation}
t_{TF}={\mu  a^2\omega_{eq}^2\over 8(1-e^2)^6T_*}{\tilde \Gamma}^{-1},
\label{ev8}    
\end{equation}
where  $\mu =M_pM_*/( M_p +M_*)$ is reduced mass.
It is instructive to substitute  the explicit expression for the 
typical torque $T_*$ given by equation (\ref{eqadd1}) in (\ref{ev8}).
We then express $Q_{eq}$ and $\omega_{eq}$ 
in terms of natural units by introducing the dimensionless 
quantities  $\tilde Q = Q_{eq}/(M_*^{1/2}R_*)$ and $\tilde \omega 
=\tau_* \omega_{eq}$, where $\tau_*=\sqrt{R_*^3/ (GM_*)}$
with $R_*$ being  the radius of the primary star.
 We then have
\begin{equation}
 t_{TF}={5\over 48\pi} {\tilde \Gamma}^{-1} {\tilde Q}^{-2}{\tilde \omega}^{4}{1\over q(1+q)}\left({a\over R_*}\right)^{{8}}.
\label{ev9}    
\end{equation}
\textcolor{red}{
We now use equation (72) of EKH 
to relate this 
to  quantities characterising their model and intrinsic to it, $\sigma_{EKH}$ determining the magnitude of the rate of energy dissipation and dimensionless parameter $Q_{EKH}$ 
characterising density distribution in the primary.
Doing this we find how these quantities relate to
$\tilde \Gamma$, $\tilde Q$ and $\tilde \omega$, thus  obtaining
\begin{equation}
\left({Q_{EKH}\over 1-Q_{EKH}}\right)^{2}\sigma_{EKH}
={16\pi \over 15}{\tilde \Gamma {\tilde Q}^{2}\over M_*R_*^2{\tilde \omega}^{4}}.
\label{ev10}    
\end{equation}}
}

\subsection{Contribution of the rotational terms to the
orbital evolution}\label{NonrotT}
{ In this Section we consider the effect of terms in the tidal response  that are $\propto \delta_3.$
In doing this we neglect tides in the companion and dissipative effects in the primary. The tidal interaction
then conserves orbital energy and consists of an interaction between the spin and orbital angular momenta
characteristically leading to changes in their mutual inclination accompanied by their  precession around 
 the total angular momentum vector.}

Thus we consider the evolution of orbital parameters 
due to the presence of rotational terms proportional
to $\delta_3$ formally setting $\delta_1=\delta_{1,p} = \delta_2=0$
in the evolutionary equations above. It is important to
stress that this approximation may be adequate \textcolor{red} {for sufficiently short time intervals, since the
dimensionless parameters
$\tilde \Gamma^{-1} $ and  $\tilde\Gamma_{p}^{-1}$ which  determine  the timescale of  evolution 
due to the presence of non conservative effects  are  expected to be 
quite large.}

When $\delta_1,$ $\delta_{1,p}$ and $\delta_2$ are set to zero the orbital
energy is conserved and the $Z$ component of torque $T^z=0$. In this case from equation (\ref{ev4}) it follows 
that the absolute value of rotational angular momentum,
$S$, is an integral of motion. Also, from \textcolor{red}  {the second  equation in the set 
(\ref{e45})   and (\ref{en1}) }it follows that there is additional integral
of motion
\begin{equation}
I={L^2\over 2S}-L\cos \beta,   
\label{ev11}    
\end{equation}
which is valid for any form of $T^x$.

Since eccentricity $e$ depends only on $L$ when the orbital
energy, and, accordingly, semi-major axis $a$ are fixed, from equations (\ref{en1}) \textcolor{red} {and (\ref{t2})}   it follows
that the evolution equation for $\beta$ is a function 
of $\beta$, $e$ and $\hat \varpi$ when $a$ and $S$ are fixed. 
Therefore,  if  the form of $\hat \varpi$  \textcolor{red}{ is known as a function of time,}  the evolution of all 
\textcolor{red}{ orientation specifying}  angles is reduced to finding 
solution of only one first order ordinary differential 
equation. We assume below that the rate of change of ${\hat\varpi}$ under classical tidal distortion is given by equation
(\ref{Apse1a}).
We now consider the contribution of  perturbations arising from $\delta_3$ with other $\delta_i=0$
in the limits of large and small $S$ which as indicated above is constant under these conditions.
In the limit of large $S,$ the spin axis coincides with the total angular momentum vector and the angle $\delta=0,$ with
 $i \equiv \beta.$ In this case $L \cos i$ is a constant of the motion and the system is Hamiltonian with two degrees of freedom.
 Accordingly, we  may write\textcolor{blue}{
\begin{align}
&\hspace{-15mm}\frac{d L}{dt} = -\frac{\partial {\cal R}_H}{\partial \varpi}
\label{evx1}    
\end{align}}
where
\begin{align}
&{\cal R}_H = - (1-e^2)^{3/2}T_* \sigma \delta_3 \phi_3 \sin^2 \beta
\cos 2\hat \varpi.
\label{evy1}    
\end{align}


\noindent As the system is Hamiltonian, we have \citep[see e.g.][]{MurrayD} \textcolor{blue}{ 
\begin{align}
&\hspace{-15mm}\frac{d \hat \varpi}{dt} = \frac{d \varpi}{dt} =\frac{\sqrt{1-e^2}}{ \mu en_oa^2}\frac{\partial {\cal R}_H}{\partial e}
\label{evx2}    
\end{align}}
after using this to find the contribution to the apsidal line advance rate arising from $\delta_3$ and adding it to the 
\textcolor{blue}{tidal} contribution given by (\ref{Apse1a}) 
we obtain
\textcolor{blue}{
\begin{align}
\frac{d{\hat\varpi}}{dt} = 15 k_2n_0\frac{M_pR_*^5}{M_*(a(1-e^2))^5}\left((1+{\cal F}_{c,p})\phi_6
-\phi_8\sigma\delta_3\sin^2\beta\cos2\hat\varpi\right),
\label{Apse2}
\end{align} 
where $\phi_8 = (12+46e^2+5e^4)/20.$} { We note that corrections to apsidal advance rate arising from 
the term $\propto \delta_3$ ( see table \ref{symbols2} for its definition)  will be small for slow subcritical rotation.}

We may also consider the limit of small $S$, normally the one of physical interest in the same way.
In  \textcolor{blue}{ the extreme limit of} 
that case the orbital angular momentum coincides with the total angular momentum vector, $i=0$, 
while the angle $\delta\equiv \beta$ determines the orientation of the spin. Regarding this as a given function of time,
although the system, being non autonomous is not  strictly Hamiltonian, equations (\ref{evx1}) and (\ref{evx2}) still apply
with ${\cal R}_H$ playing the role of a disturbing function. Hence we again obtain (\ref{Apse2}).
\textcolor{blue}{But note that although the orbital plane approaches coplanarity with the plane with normal 
to the total angular momentum vector that defines the { primary centred} frame, the apsidal line is measured relative to
a line that asymptotically rotates (precesses) with angular velocity $d\alpha_r/dt.$
Thus, if $\hat\varpi$ is instead measured relative to a fixed line in the { primary centred} frame, we have 
\begin{align}
\frac{d{\hat\varpi}}{dt} =\frac{d\alpha_r}{dt}+  15 k_2n_0\frac{M_pR_*^5}{M_*(a(1-e^2))^5}\left(\phi_6(1+{\cal F}_{c,p})
-\phi_8\sigma\delta_3\sin^2\beta\cos2(\hat\varpi -\alpha_r)\right),
\label{Apse2a}
\end{align} 
Returning to the equivalent equation (\ref{Apse2})}  we see that the correction arising from $\delta_3$  is of order $(\Omega_r/\omega_{eq})^2$, 
which is assumed to be small and in fact comparable to neglected effects arising from rotational distortion.
It also follows that a time average of  $\sin2\hat\varpi$  over a precession cycle is zero and that of $\cos2\hat\varpi$
is of order $(\Omega_r/\omega_{eq})^2.$ Thus, terms involving these and $\delta_3$ will give negligible contributions
\textcolor{blue}{ to (\ref{Apse2a}) }
if the evolution of the system occurs on a time scale significantly longer
than an apsidal rotation period and no other processes affect the evolution of $\hat \varpi.$

\textcolor{red}{We now consider  
the evolution of the inclination angles in more detail in the limit $S/L \ll 1$
when the absolute value of orbital angular momentum is also
conserved to within a variation of order  $S/L $ . In this case equation (\ref{en1}) is reduced to
\begin{equation}
 \dot \beta =-{T^x\over S}, \quad T^x=T_*(1-e^2)^{3/2}\sigma \delta_3 \phi_3 \sin \beta  \sin 2\hat\varpi,  
\label{ev12}    
\end{equation}
which can be readily integrated to give
\begin{equation}
\ln \left({1-\cos \beta \over 1+ \cos \beta}\right) = {1\over T_{\beta}} \int dt \sin 2\hat \varpi, \quad T_{\beta} = -{S\over 2 (1-e^2)^{3/2}\sigma \delta_3  \phi_3 T_*},
\label{ev13}    
\end{equation}}
{and we recall that $\delta_3 < 0$ and, therefore, $T_{\beta}$ is a positive quantity.} 

Now let us assume that the apsidal precession is uniform and $\hat \varpi =\omega_{a}t$, where we set an initial
value of apsidal angle to zero without loss of generality. Then, we have from (\ref{ev13})
\textcolor{red}{\begin{equation}
\tan {\beta \over 2} =\tan {\beta_0 \over 2}  \exp\left({1-\cos 2\omega_a t\over 4T_{\beta} \omega_a }\right),
\label{ev14}
\end{equation}}
where $\beta_0=\beta (t=0)$. In this case $\beta$ changes periodically, with the period of change being \textcolor{red}{
one half } the period of apsidal precession. A typical amplitude of change is inversely proportional to 
the product $T_{\beta}\omega_a$. 

It is instructive to represent $T^{-1}_{\beta}$ in terms of natural units \textcolor{red}{ in the form
\begin{equation}
T^{-1}_{\beta}=-{6L k_2\over S}\frac{\phi_3\sigma\delta_3}{(1-e^2)^5}\frac{M_pR_*^5}{M_*a^5}n_o,
\label{ev16}    
\end{equation}
where we have made use of the expression for $k_2$ given by (\ref{apsidal}) in Appendix \ref{newPrecession}.
Equations (\ref{Apse}) and (\ref{ev16}) state that $T_{\beta}^{-1}$ is the product of the classical apsidal advance rate
\textcolor{blue}{ induced by the companion} and a factor \textcolor{blue}{of magnitude}
\footnote{Let us stress that  \textcolor{blue}{here} we consider only tidal precession induced by tidally deformed primary.
 \textcolor{blue}{However, in  some 
realistic situations the contribution of the companion}  could be potentially 
\textcolor{blue}{dominant}, see e.g. the corresponding analysis in \cite{RW} for 
Hot Jupiter systems.} 
\begin{align}
f_{\beta}={4\over 5}\left(\beta_*+\frac{1}{2}\right) {\phi_3\over \phi_6}\frac{\Omega_r^2 R_*^3}{{\tilde\omega}^2GM_*} \frac{L}{S},\label{fbetadef}
\end{align}}
{ where $\tilde \omega =\omega_{eq}\sqrt{{R_*^3/ GM_*}}$ is expected to be order of unity.}

\noindent \textcolor{red} {In order for averaging to be valid  and a resulting small change in $\beta$
we require that $f_{\beta} \ll 1$ which will be violated for sufficiently small $S$ or moment of inertia
of the primary. In that case all terms involving $\delta_3,$  apart from in the equation for $d\hat\varpi/dt,$ should be retained.} 

\begin{figure}
\begin{center}
\vspace{1cm}
\includegraphics[width=10.0cm,height= 10.0cm,angle=0]{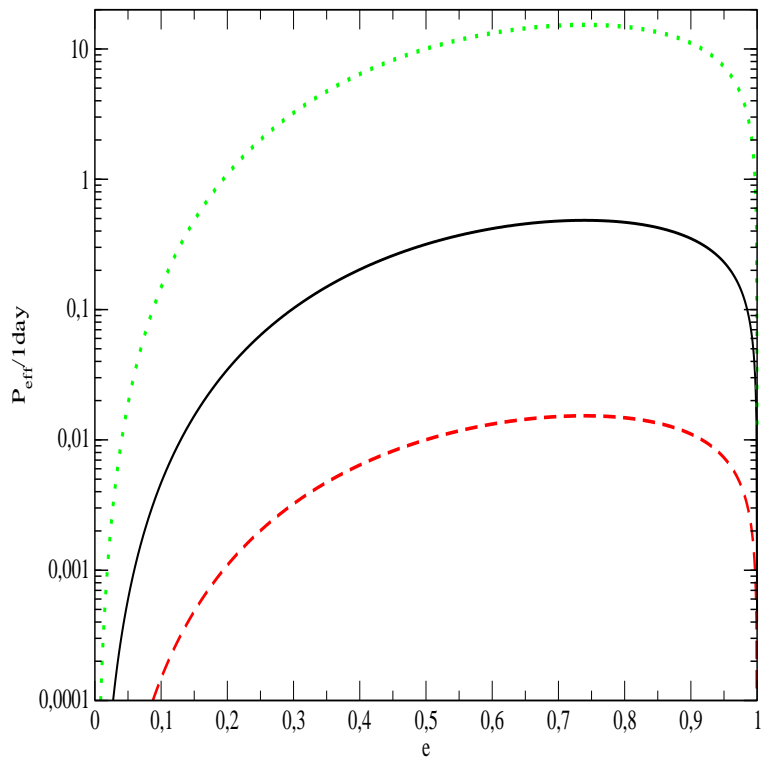}
\caption{The value of 'effective' period $P_{eff}$ defined in eq. (\ref{fb1})
as a function of eccentricity $e$, for a given value of $f_{\beta}$. Dotted, solid and  dashed curves correspond to 
$f_{\beta}= 0.1$, $1$ and $10$, respectively.}
\label{Period}
\end{center}
\end{figure}

\begin{figure}
\begin{center}
\vspace{1cm}
\includegraphics[width=10.0cm,height= 10.0cm,angle=0]{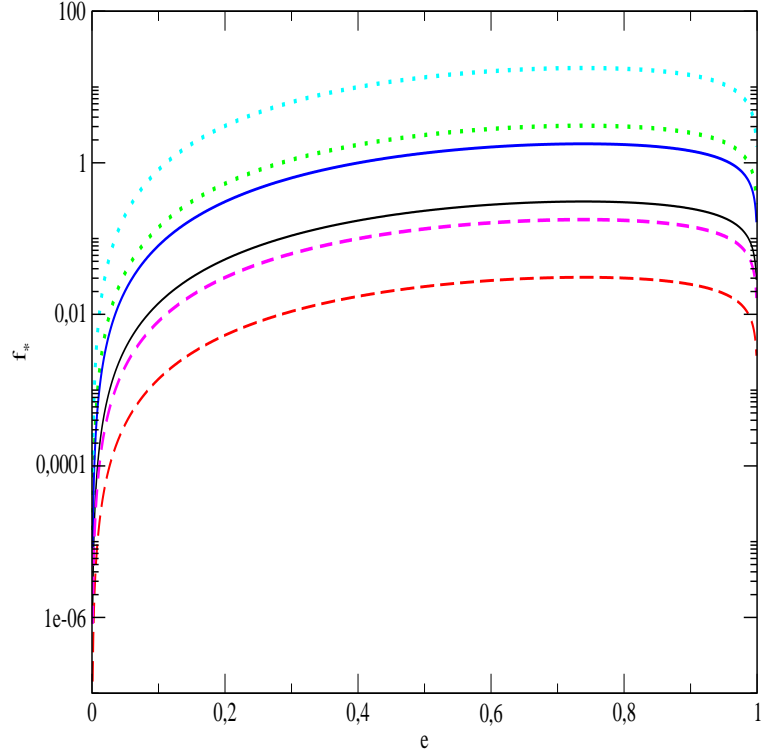}
\caption{$f_{*}$ defined in eq. (\ref{fbeta1}) as a function of eccentricity. Curves with smaller (larger) values of the argument correspond to 
the nominal solar model (the model of Kepler 91). Solid, dashed 
and dotted curves correspond to $P_r/P_{orb}=1$, $10$ and $0.1$.}
\label{fbeta}
\end{center}
\end{figure}

\begin{figure}
\begin{center}
\vspace{1cm}
\includegraphics[width=10.0cm,height= 10.0cm,angle=0]{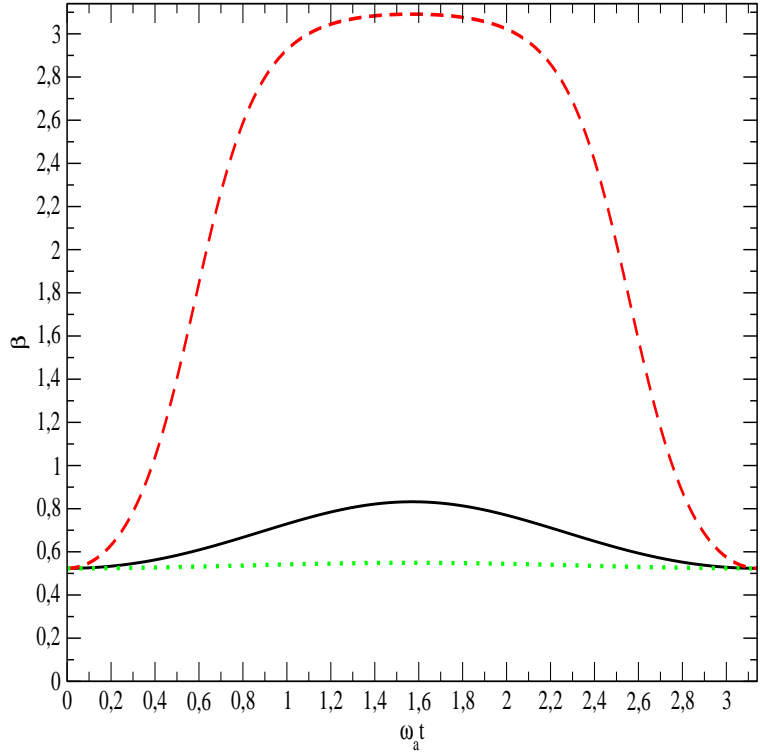}
\caption{The dependency of the angle $\beta$ (in radians) on time determined from equation (\ref{ev14}) for 
one half of the period of apsidal precession. Solid, dashed and dotted curve correspond to $f_{\beta}=1$, $10$ and $0.1$, respectively. For all curves the initial $\beta_0=\pi/6$.}
\label{beta}
\end{center}
\end{figure}

More generally the variation of $\beta$ over one  half of  an apsidal precession period is 
very roughly $ \sim f_{\beta}/2$ radians.

The situation where $f_{\beta}$ is significant  may  be  of  special importance, 
since it could lead to observational consequences and in this context note
that relatively small variations may be noticeable in eclipsing or transiting systems. To  further examine
 conditions leading to such situation we rewrite 
(\ref{fbetadef}) in the form
\begin{align}
 &f_{\beta} =1.9 {\phi_3\over \phi_6} \sqrt{1-e^2}  \left({1day \over P_{eff}}\right)^{2/3}, \quad {\rm where}\nonumber\\
&  P_{eff}= {(1+q)^{1/2}\over q^{3/2}}{\left ({I_{0.1}{\tilde \omega}^2\over   (\beta_*+1/2)}\right)}^{3/2} 
 {\left( {\bar \rho_* \over \bar \rho_{\odot}}\right)}^{1/2}\left(\frac{ P_r}{P_{orb}} \right)^{3/2}P_{orb}.
\label{fb1}    
\end{align}
 \textcolor{blue}{ Here}
       {$P_{orb}$ is orbital period, $I_{0.1}=I/(0.1M_*R_*^2)$
with $I$ being primary's moment of inertia, $\bar \rho_*$
and $\bar \rho_{\odot}$ \textcolor{blue}{are respectively  the mean } densities of 
the primary and of the Sun.

From (\ref{fb1}) it follows that 
\begin{equation}
\frac{P_{eff}}{1day} =  {\left( {1.9 \phi_3\sqrt{1-e^2}(1+q)^{2/3} \over f_{\beta} \phi_6}\right)}^{3/2}.
\label{fb2}
\end{equation}
{A plot of  $P_{eff}/(1day)$ as a function of $e$  for $q=1$ and  fixed $f_{\beta}=0.1$, $1$ and $10$ 
is shown in Fig. \ref{Period}. 
\noindent Accordingly, we infer } 
that rather large eccentricities $e\sim 0.7$ are needed to produce  \textcolor{blue}{ significant }  values 
of $f_{\beta}$. \textcolor{blue}{From (\ref{fb1}) it also follows  that} 
 in order  to have significant variations of $\beta$ \textcolor{blue}{ for a realistic orbital period}  a system 
should have \textcolor{blue}{in addition, 
 relatively large 
companion masses and small rotation 
periods.}  Also, stars with large central condensations
corresponding to smaller $I_{0.1}$ and small mean densities are favoured. 
}

{ To illustrate the dependency of $f_{\beta}$ on other parameters 
we represent it in the form
\begin{equation}
f_{\beta}=f_{*}\left ({1day\over P_{orb}}\right)^{2/3},
\label{fbeta1}    
\end{equation}
where an expression for $f_*$ directly follows from (\ref{fb1}). In Fig. \ref{fbeta} 
we plot $f_*$ as a function of $e$ for $q=1$, $\beta_*=0.5$, $\tilde \omega=1$ and three values of $P_r/P_{orb}$ for two stellar models. One 
is our 'nominal' solar model with $I_{0.1}=1$ and ${\bar \rho_* / \bar \rho_{\odot}}=1$, while the other represents a model
of an evolved star used in \cite{Chernov2017} to model the exoplanetary system Kepler 91. In this model $I_{0.1}\approx 1.4$ and
${\bar \rho_* / \bar \rho_{\odot}} \approx 5.2\cdot 10^{-3}$.
One can see from this Figure that the use of the star with smaller average density results in larger values of $f_{\beta}$ as expected. Also, smaller ratios $P_r/P_{orb}$
(and, accordingly, larger ratios of rotational to orbital
frequencies) are favoured.
Finally, we illustrate the result
of solution of equation (\ref{ev14}) for $f_{\beta}=1$, $10$
and $0.1$ in Fig. \ref{beta}. One can see from this Fig. 
that a typical change of $\beta$ over half  an apsidal period gets larger with increase of $f_{\beta}$ as expected.}


Now let us consider the evolution of the precessional angle $\alpha_r$. The corresponding equation directly follows 
from equation (\ref{evn2}), where we set $\delta_1=\delta_2=0$ and use the explicit expression for $\delta_3$
to obtain
\textcolor{red}{\begin{align}
&\hspace{-16mm}\frac{d  \alpha_r}{dt} = - {JT_*\over SL}\sigma\bigg((2\beta_*+1) {\tilde \Omega}^2 (\phi_1-
(1-e^2)^{3/2}\sigma \cos \beta (\phi_2 + \phi_3 \cos 2\hat \varpi ))\nonumber \\
&\hspace{-12mm} +{1\over 3}(1-e^2)^{9/2}{1+q\over q}\sigma \cos \beta \bigg)
\label{ev17}
\end{align}}
As discussed above, when the apsidal precession is uniform and  on a timescale much \textcolor{red}{smaller
 than $T_{\beta},$ }  the term proportional to $\phi_3$ can be neglected.
  Then the angle $\beta$ and the eccentricity $e$ may be considered as constants so that
the evolution of $\alpha_r$ has the character of uniform precession. 
It is of interest to estimate the condition 
under which  the new term proportional to $2\beta_*+1$ is larger than
the standard term responsible for precession due to rotational flattening of the star.
 The corresponding condition readily follows from (\ref{ev17}), in which, \textcolor{red}{in the assumed limit of
  small ${\Omega_r},$}
  we neglect the contribution of the term proportional
to $\phi_2$ and \textcolor{red}{insert the expression for}  $\phi_1.$ 
\textcolor{red}{It is found to be
\begin{equation}
 \sigma < {(6\beta_*+3) q\over (1+q) (1-e^2)^{9/2}\cos\beta } \left({n_0\over \omega_{eq}}\right)^2
 \left(1+\frac{15}{2}e^2+ \frac{45}{8}e^4+\frac{5}{16}e^6\right).
\label{ev18}    
\end{equation}
This can also be expressed as
\begin{equation}
 \frac{\Omega_r}{n_o}  < \left(\frac {M_pR_*^3}{M_* R_{peri}^3}\right)
 \frac{(6\beta_*+3) \left(1+\frac{15}{2}e^2+ \frac{45}{8}e^4+\frac{5}{16}e^6\right)}{(1-e^2)^{3/2}(1+e)^3\cos\beta } \left({GM_*\over R_*^2 \omega^2_{eq}}\right).
\label{ev19}    
\end{equation}
where $R_{peri}$ is the pericentre distance.}

From (\ref{ev19}) it follows that the new term prevails when
either rotation and, accordingly, $\Omega_r$, is quite small
and/or when either eccentricity or mass ratio is large.
While all factors in brackets apart from the first are arguably of order unity unless the eccentricity is  close to unity  or $\beta$ is close to $\pi/2$,
 the first factor is usually small,
especially for planetary mass ratios. Thus {unless $\beta$ is very close to $\pi/2$},  the new term could only be significant for eccentricities that are  extremely  close to unity
in that case.
However,  a qualitatively different analysis based on a sequence of parabolic encounters  and the possibility of tidal capture may be more appropriate in that limit,
which is beyond the scope of this paper.

\section{Discussion}\label{Discussion}
\textcolor{red} {In this paper  we  have considered the tidal interactions of
a binary system consisting of a primary component and a compact companion.
The primary component has  spin angular momentum which is misaligned with the orbital angular
momentum. There are no restrictions on the mutual inclination of these vectors or on the orbital eccentricity.
The companion is initially assumed to have no internal degrees of freedom but at a later stage
we allow for internal energy dissipation as would be important for a planetary mass companion.
This enables us to consider the evolution of systems such as those containing
hot Jupiters  in orbits with angular moments  significantly misaligned with the primary spin axis.}

\textcolor{red}{  
We derived equations  governing the evolution of  the angular momentum vectors in Section \ref{Jvecev} and
  calculated the tidal response of the primary
in the equilibrium tide limit in Section \ref{Respd}. 
Effects arising from both  dissipation  and { rotation} were considered.
The torque acting between the primary and orbit was determined in
 Section \ref{TorqueFind} and the
 rate of change of orbital energy in Section \ref{Orbenrate}, being { reduced to a}  final closed form
 in Sections \ref{Torquereduct} - \ref{Inteval}}.

\textcolor{red}{Using the above results we obtained  
equations enabling the determination of the orbital  and spin  evolution in Sections \ref{Detorbspinev}
 and \ref{Amvexp}. 
Energy dissipation in the companion under the assumption
of negligible spin angular momentum is  incorporated into the equations governing the evolution of the system in Section \ref{inccomp}.
This can be important for orbital circularisation  for planetary mass companions. The parameters occurring in these equations
is also reviewed in these Sections. Apart from the masses and radii  of the binary components, the rotation rate of the primary,
 and parameters specifying the form and
orientation of the orbit, the system of equations depends,  for each component, on $\Gamma,$   being 
the ratio of the  rate of energy dissipation to four times the kinetic energy associated with the disturbance in the form of
 the equilibrium tide ( see  equation (\ref{EDs1})), and
the quantities $\omega_{eq},$ and $Q_{eq}$ defined by equations (\ref{DDs1}) and (\ref{e7X})  representing 
a putative normal mode frequency and overlap integral associated with the equilibrium tide, respectively. But note that as an alternative to the last two parameters use can be made of the classical apsidal motion constant, $k_2,$ see equation \ref{apsidal}. There is also a dependence on
 the dimensionless quantity  $\beta_*$, which is associated with { rotational effects and } the Coriolis force acting on the primary and is
 defined through equations (\ref{EDs3}) { and (\ref{coribeta}) ( see also the discussion in Section \ref{coriresponse}). }
We remark that  \citet{IP2004} assumed only the $l=2$ $f$ mode contributed
to the tidal response. This corresponds to  replacing the displacement associated  with the equilibrium tide by that
associated with the $f$ mode  in our analysis while $\Gamma$ is replaced
by the decay rate of this  mode. But note that the forcing frequency dependence of $\Gamma$  
in the case of radiative damping \textcolor{blue}{ requires}  that this has been assumed to be the frequency of the $f$ mode in that 
\textcolor{blue}{scheme}.
A simple estimate based on equation (\ref{EDs1}) indicates $\Gamma$ varies as the inverse of the square of the forcing frequency
and so may be significantly  underestimated by that substitution.}

\textcolor{red}{It is also important to realise that the closed form of equations we obtain is valid only if the quantity $\Gamma$ is independent of tidal
forcing frequency.  This would be the case for a standard Navier Stokes viscosity. However, as we indicated a contribution from radiative
damping should also be considered and this is frequency dependent. In order to obtain our equations in their closed form,
an average value of $\Gamma$ has to be adopted.  In the first instance we  would suggest the value corresponding to the 
angular velocity at periastron  $\sim (1+e)^{1/2}(1-e)^{-3/2} n_o$ is appropriate as this location is where most of the tidal interaction 
takes place.  However, dealing with the frequency dependence associated with radiative damping as well as relaxing the equilibrium tide assumption to incorporate  dynamical effects, which could be important for determining  where most energy dissipation takes place, is an important issue
when considering planetary mass objects such as Hot Jupiters,  which should 
be a subject for future investigation.}

\textcolor{red}{Our equations for the tidal evolution include  both dissipative terms  and those associated with the  Coriolis force and { the dependence of forcing frequency on angular 
velocity on account of the Doppler effect,}
which we describe as rotational terms. If the latter are dropped we  may relate our results to those of EKH in the appropriate limit 
as done in 
 Section  \ref{eqsum}. There we were able to link the parameters in our equations to theirs thus showing how their parameter
 which was assumed to relate the energy dissipation rate to the rate of change of the quadrupole tensor associated  with the primary can
 be connected to the energy dissipation rate occurring in the proper solution of the tidal response  problem.}
 
 \textcolor{red}{The contribution of terms  arising from { rotation} to the
orbital evolution not included in EKH  takes two forms that were considered in Section \ref{NonrotT}.
The first is through effects on the evolution of the orbital parameters and orientation specifying angles
through terms $\propto ( \beta_*+1/2)$  and sines and cosines of the longitude of periapse.
These can be \textcolor{blue}{dealt with by an averaging procedure}  if the advance of the apsidal  line is uniform and on a time scale shorter than
the time scale of the tidal evolution.}
\noindent In general, the influence of other perturbing bodies may have to be
included to determine \textcolor{blue}{ if this is appropriate.}  However, we estimated a condition for this  to hold when the apsidal line advance was determined
purely by tidal effects in the form $f_{\beta}\ll 1$, see equations (\ref{fbetadef}-\ref{fb2}). 

 The \textcolor{blue} {situation when this is not satisfied } is of special interest
since it \textcolor{blue}{ may lead to significant}  variations of the inclination angle
over a half of an apsidal line  precession period 
and, therefore,  in a system with appropriate parameters this \textcolor{blue} {may be directly observable}. Under the assumption of uniform apsidal precession induced by tides
in the  primary the corresponding condition \textcolor{blue}{ for this can be obtained from} equation
(\ref{fb2}) \textcolor{blue}{ ( see also Fig. \ref{Period})}. 
 From this it follows that in order
to have  \textcolor{blue} {a significant}  effect the mass ratio and  \textcolor{blue}{primary angular velocity}
should be sufficiently large, while orbital period should be 
sufficiently small. An optimal value of eccentricity is close 
to $0.7$. Also, centrally condensed stars with small \textcolor{blue}{ mean} densities are
favoured 
and, clearly, spin and orbital angular momenta should be misaligned.
It should be 
interesting to \textcolor{blue}{investigate}  whether there are observed
systems with such properties.

The other phenomenon that is affected is the precession of the spin
and orbital angular momentum vectors about the total conserved  angular momentum vector.  A new rotational term contributes
a precession rate that is linear in the primary's angular velocity. This was compared with the classical precession rate driven by centrifugal distortion,
which is proportional to the square of the angular velocity. Formally, the new term dominates for sufficiently small rotational frequency
estimated in equation (\ref{ev18}). This indicates that a small angular velocity is needed  unless the mass ratio is around  unity 
and/or the eccentricity is close to unity.  Note 
that that in order to have the new term  dominate orbital
precession it is necessary to have large eccentricities 
and/or  \textcolor{blue}{a small angular velocity.  The latter  is the opposite of  what is  needed to produce significant  
 variations in} the inclination angle
discussed \textcolor{blue}{in Section \ref{NonrotT} }. However, in both cases sufficiently large 
mass ratios are favoured.
Further work is necessary to investigate these features more fully.

{ There is a simple physical explanation for these new non-dissipative effects. Namely, in the absence of rotational 
terms (and, of course, neglecting dissipation) tidal bulge is
aligned with the direction to perturbing body. When the rotation axis is misaligned with respect to orbital angular momentum, the presence of
rotational terms breaks that alignment, thus producing torques,
which lead to the evolution of the corresponding orbital elements.}

{ Also, it is important to stress that we have neglected a contribution of toroidal displacements to perturbation of the star
due to tides. Although it may be small due to relative smallness 
of the appropriate overlap integrals, this contribution should be
separately analyzed. A convenient framework for such an analysis 
would be the self-adjoint approach to the problem of tidal excitation of normal modes of any kind in a rigidly rotating star put forward in \cite{Papaloizou2005} and \cite{IP2007}.}
 
It is interesting to note that a different approach has been used to describe tidal evolution of 'elastic' bodies (say, rocky planets) for inclined systems
(see \cite{K} and e.g. \cite{BE} and references therein for a recent development). It would be of interest to establish how this approach and the formalism developed in this Paper are related to each other.

 Finally, it is important to note that in order to find
\textcolor{blue}{a complete self-consistent set of} equations for the evolution of the orbital
parameters on the tidal timescale one must consider dynamical  tides in addition to quasi-stationary ones.
 This Paper provides a convenient \textcolor{blue}{foundation}   for such a study.
  In  future work we \textcolor{blue} {will undertake}  to generalise the \textcolor{blue}{treatment}
of dynamical  tides in the so-called regime of moderately large viscosity developed in
\cite{IPCh} for the aligned case to tidally
interacting systems with \textcolor{blue}{orbital and spin axes arbitrarily misaligned.}

\section*{Acknowledgements}

PBI was supported in part by the grant 075-15-2020-780 'Theoretical
and experimental studies of the formation and evolution
of extrasolar planetary systems and characteristics of exoplanets'
of the Ministry of Science and Higher Education of the Russian Federation.
 We are grateful to A. J. Barker and K. A. Postnov for useful comments. { We thank the  anonymous reviewer for a careful reading of the manuscript and helpful
comments that led to a significant improvement of this paper,}

\section*{Data availability}
There are no new data associated with this article.


\begin{appendix}
\section{Fourier development  $U$ in terms of spherical harmonics defined in the stellar frame}\label{FourierU}
We begin with  the Fourier expansion given in Section \ref{Fourierpot} in the form
\begin{equation}
\frac{\exp(-{\rm i}n (\Phi-\varpi)) a^3}{R^3}  = \sum_{k=-\infty}^{k=\infty} \phi_{k,n}\exp({\rm i}k
 n_o t),
\label{jpeFU1}
\end{equation}  
The Fourier coefficients, $\phi_{k,m},$ are related to the well known Hansen coefficients $X^{q,m}_k$
 
 \noindent through~$\phi_{k,m}~=~X^{-3,-m}_k.$ 
A formal expression for $X^{q,m}_k$ is given by
\begin{equation}
X^{q,m}_k = (1+\beta^2)^{-q-1}\sum^{\infty}_{p=-\infty}J_p(je)H^{q,m}_{k,p},\label{hansen}
\end{equation}  
where $\beta = (1-\sqrt{1-e^2)})/e,$  $J_p$ denotes the Bessel function and
\begin{equation}
H^{q,m}_{k,p} =    \frac {(-\beta^2)^{k-p-m}\Gamma(q+2-m)}{\Gamma(k-p-m+1)\Gamma(q+2 -k+p)!}  \ _2F_1( k-p-q-1,-m-q-1,j-p-m+1,\beta^2) ,
\end{equation}  
for $j> p+m,$ and
 \begin{equation}
H^{q,m}_{j,p} =    \frac {(-\beta^2)^{-j+p+m}\Gamma(q+2+m)}{\Gamma(-j+p+m+1)\Gamma(q+2 +j-p)}  \ _2F_1( -j+p-q-1,m-q-1,-j+p+m+1,\beta^2) ,
\end{equation} 
 for $j< p+m,$ and  $ \ _2F_1( \alpha, \beta, \gamma, z)$ denotes the hypergeometric function 
(see Abramowitz \& Stegun 1964 for the definition and properties).

Practical prescriptions
for calculating Hansen coefficients have been provided by many authors ( e.g.  Branham 1990, Laskar 2005).
For small eccentricities a power law expansion in $e$ developed from (\ref{hansen} ) may be used. 
In this paper we  have followed the notation of Ivanov \& Papaloizou (2004)  who give a useful prescription
for calculating $\phi_{k,m}$ for eccentricities $e > 0.2.$
Inserting the expansion (\ref{jpeFU1})  into the expression (\ref{jpe1}) for $U,$
given in Section \ref{Fourierpot}  we obtain in a form appropriate to the orbit frame
\begin{equation}
 U=-GM_p \left(\frac{4\pi r^2}{5 a^3}\right)\sum_{m=0, |2|}^{'}
\sum_{k=-\infty}^{k=\infty} \phi_{k,n}
Y_{2,m}(\theta', \phi') Y_{2,m}(\pi/2, 0)
\exp({\rm i}k n_o t -m\varpi)
\label{jpe4}
\end{equation}  
\subsection{The potential expressed in the stellar frame}
In  the stellar frame $(X,Y,Z)$  the associated spherical coordinate system is $(r,\theta, \phi).$
Spherical harmonics in the orbit frame are connected to those in the stellar frame through a relation of the form
\begin{equation}
 Y_{2,m}(\theta', \phi')  = \sum_{n=-2}^{n =2}D^{(2)}_{n,m} ( 0,\beta,\gamma)Y_{2,n}(\theta, \phi),
\label{jpe5}
\end{equation}
where the coefficients (Wigner matrix elements) $D^{(2)}_{n,m} ,$  being determined by the  rotation
defined by equation (\ref{jp1}),
 depend on the  orientation specifying  angles $\beta$ and $\gamma.$  
 It is important to note that the arguments of $D^{(2)}_{n,m} ,$  hereafter taken as read, have the form they do because of our particular choice
 that the coordinate system $(X',Y',Z')$  is connected to the $(X,Y,Z)$ system by rotation through an angle  $\gamma$
 followed by an angle  $\beta$ about fixed  $Z$ and $Y$ axes, respectively 
(see Fig. \ref{coordinates}). 
  Accordingly, we have 
\begin{equation}
D^{(2)}_{n,m} =  \exp(-{\rm i}m\gamma) d^{(2)}_{n,m}(\beta) 
\label{jpe51a}
\end{equation}
where   $d^{(2)}_{n,m}$ is an element of
Wigner's (small) d-matrix  and is real  (see Ivanov \& Papaloizou 2011). 

\noindent Thus we have the perturbing potential expressed in terms of coordinates in the stellar frame

\begin{equation}
 U=-GM_p \left(\frac{4\pi r^2}{5 a^3}\right)\sum_{n=-2}^{n=2}Y_{2,n}(\theta, \phi) 
\sum_{m=0, 2}^{'} \sum_{k=-\infty}^{k=\infty} \phi_{k,m}
D^{(2)}_{n,m} 
Y_{2,m}(\pi/2, 0)
\exp({\rm i}k n_o t -m\varpi)
\label{jpe6}
\end{equation}
For convenience we write  (\ref{jpe6}) in the more compact form
\begin{equation}
 U=-GM_p \left(\frac{4\pi r^2}{5 a^3}\right)\sum_{n=-2}^{n =2}Y_{2,n}(\theta, \phi)F_{n}(t) \equiv  r^2\sum_{n=-2}^{n =2}A_{n} Y_{2,n}(\theta, \phi),
\label{jpe7U}
\end{equation}
with
$A_{n} =-4\pi GM_p F_{n}/(5a^3)$
 and
\begin{equation}
F_{n}(t)= \sum_{m=0, |2|}^{'} \sum_{k=-\infty}^{k=\infty} \phi_{k,m}D^{(2)}_{n,m} 
 Y_{2,m}(\pi/2, 0)\exp({\rm i}(k n_o t -m\varpi)))\label{jpe711U}
\end{equation}
 Making use of  the  Fourier expansion given by  equation (\ref{jpeFU1}) we have the alternative
expression
\begin{equation}
F_{n}(t)=\frac{a^3}{R^3} \sum_{n=0, |0|}^{'} D^{(2)}_{n,m}  Y_{2,m}(\pi/2, 0)\exp(-{\rm i}m\Phi )\label{jpe711b0U}
\end{equation}
\section{ Calculation of the torque in terms of the  induced Lagrangian displacement}\label{Torquecalcapp}
The torque acting on the 
star,  ${\bf T},$ is 
given by
\begin{equation}
{\bf T} = -\int_V \rho'{\bf r}\times \nabla U dV\label{jp18A}
\end{equation}
where $\rho' = -\nabla\cdot (\rho \mbox {\boldmath$\xi$})$ is the density response and the integral is taken over the volume
of the star.
The components of the operator ${\bf r}\times \nabla \equiv \hat {\bf \Phi}.$ in the $(X,Y,Z)$ coordinate system are
\begin{equation}
\hat {\bf \Phi} =  \left(-\cot\theta \cos\phi
\frac{\partial}{\partial \phi}
-\sin \phi\frac{\partial}{\partial \theta} \ \  ,  \ \  -\cot\theta \sin\phi
\frac{\partial}{\partial \phi}
+\cos \phi\frac{\partial}{\partial \theta}  \ \  ,  \ \  \frac{\partial}{\partial \phi}   \right)
        \label{jp181X}
\end{equation}
We find it convenient to
introduce the so-called cyclic coordinates
$X^{\pm 1}=\mp {1\over \sqrt 2}(X\mp iY)$, $X^{0}~=~Z$
 and utilise the known expressions for the action of the above operator on spherical harmonics
expressed in them. 
The components of $\hat {\bf \Phi}$ in these coordinates are
 given by
\begin{equation}
\hat {\bf \Phi}^{+}=\frac{1}{\sqrt{2}} \left(\cot\theta {\rm e}^{{-\rm i \phi}}
\frac{\partial}{\partial \phi}
+{\rm i} {\rm e}^{-{\rm i \phi}}\frac{\partial}{\partial \theta}\right)
\label{jp19a}
\end{equation}
and
\begin{equation}
\hat {\bf \Phi}^{-} =\frac{1}{\sqrt{2}} \left(-\cot\theta {\rm e}^{{\rm i \phi}}
\frac{\partial}{\partial \phi}
+{\rm i} {\rm e}^{{\rm i \phi}}\frac{\partial}{\partial \theta}\right)
\label{jp19}
\end{equation}
with
\begin{equation}
\hat {\bf \Phi}^{0}=\frac{\partial}{\partial \phi}.\label{jp20}
\end{equation}
In terms of these quantities, the components in the $(X,Y,Z)$ coordinate system can be written as
\begin{equation}
\hat {\bf \Phi} =  \left( \frac{1}{\sqrt{2}}(\hat {\bf \Phi}^{-}- \hat {\bf \Phi}^{+}) \ \  ,  \frac{-{\rm i}}{\sqrt{2}}(\hat {\bf \Phi}^{+}+ \hat {\bf \Phi}^{-}) \ \  ,  \hat {\bf \Phi}^{0}\ \    \right)
        \label{jp182}
\end{equation}

Expressing $\rho'$ in terms of  $ \mbox {\boldmath$\xi$},$  (\ref{jp18A}) may be written
\begin{equation}
{\bf T} = \int_V \nabla\cdot (\rho \mbox {\boldmath$\xi$} )( {\bf r}\times \nabla U )dV
=   \int_V \nabla\cdot (\rho \mbox {\boldmath$\xi$}^{*} ) \hat {\bf \Phi}U dV ,\label{jp21}
\end{equation}
where we remark that because $\mbox {\boldmath$\xi$}$ is real  we may take its  complex conjugate in (\ref{jp21}).
Considering only the contribution of  the  terms   with the pair of  values   $\pm n $  
for a specified $n$
in the sum (\ref{jpe7}) for the forcing potential  and then summing over $ n \ge 0$ 
(it turns out that  there is no contribution from $n=0$  in the case that follows),
we obtain from equations (\ref{jp18A}), 
(\ref{jpe7}), (\ref{jp20}), (\ref{jp182})   and (\ref{jp21})
the $Z$ component of ${\bf T}$ as
\begin{equation}
{T}^{z} = -2\sum^{n=2}_{n=0} {\cal R}\left[\left(\frac{4\pi GM_p }{5a^3}\right)\int_V 
  \nabla(\cdot \rho  \mbox {\boldmath$\xi$}_{n}^* )  \left( {\rm i}n r^2
Y_{2,n}(\theta, \phi)F_{n}(t) \right)dV\right],\label{jp22}
\end{equation}
where.we recall that ${\cal R} $ here denotes that the real part is to be taken.
We remark that on account of the form of the  forcing potential  given by equation
(\ref{jpe7})
 and, accordingly, the response to it being real
 (see the discussion just below that  equation),
 the expressions in the summation associated with a particular $n$, whose real part is to be taken
  to give  (\ref{jp22}), along with  similar expressions occurring  in the context of other
 torque  components below,  are  such that for
 $n \rightarrow -n$ the complex conjugate is obtained.
 These contributions must be added in order to get
the final total torque component,  which is therefore, accordingly, real.
Thus, we take twice the real part of the expression for
$n  > 0.$ To obtain the total torque component,  we then sum over only  positive 
 values of $n.$

Turning to the $X$ and $Y$  components of the torque,  with the help of  (\ref{jp182} ) and (\ref{jp21})
 we obtain


\begin{eqnarray}
&& \hspace{-0.56cm}{T}^{x} = \nonumber\\
&&\hspace{-0.6cm}-\left(\frac{4\pi GM_p }{5a^3}\right)\sum_{n=-2}^{n=2} 
\int_V \nabla\cdot \left( \rho   \left( \sum^{n'=2}_{n'=-2}\mbox {\boldmath$\xi$}_{n'}^*\right )\right)   \left( \frac{1}{\sqrt{2}}(\hat {\bf \Phi}^{-}- \hat {\bf \Phi}^{+}) r^2
Y_{2,n}(\theta, \phi )F_{n}(t)\right)dV,\label{jp23}
\end{eqnarray}
and
\begin{eqnarray}
&&\hspace{-0.6cm}{T}^{y} = \nonumber\\
&&\hspace{-0.6cm}\left(\frac{4\pi GM_p }{5a^3}\right)
\sum_{n=-2}^{n=2} 
\int_V \nabla\cdot \left( \rho   \left( \sum^{n'=2}_{n'=-2}\mbox {\boldmath$\xi$}_{n'}^*\right )\right)
\left(  \frac{{\rm i}}{\sqrt{2}}(\hat {\bf \Phi}^{+}+\hat {\bf \Phi}^{-}) r^2
Y_{2,n}(\theta, \phi )F_{n}(t) \right)dV,\label{jp23a}
\end{eqnarray}
We now  take note of the known result that for general $j$ we have
\begin{equation}
\hat {\bf \Phi}^{\mp} Y_{j,n} \equiv {\bf \Phi}^{\mp}_{j,n}
 =\pm \frac{{\rm i}}{\sqrt{2}} Y_{j,n\pm1}\sqrt{(j\mp n)(j\pm n+1)}
\label{jp25}
\end{equation}
and of course in our case we need to specify $j=2$ and this is implied from now on  below.
Making use of (\ref{jp25}) to express the action of the components
of the operator $\hat {\bf \Phi} $  in equations (\ref{jp23}) and (\ref{jp23a}), these then  become
\begin{eqnarray}
&& \hspace{-0.56cm}{T}^{x} = -\left(\frac{2\pi GM_p }{5a^3}\right)\sum^{n=2}_{n=-2} \int_V 
\nabla\cdot \left( \rho   \left( \sum^{n'=2}_{n'=-2}\mbox {\boldmath$\xi$}_{n'}^*\right ) \right)\nonumber\\
&&\hspace{-0.6cm} \left( {\rm i}
  r^2 (  Y_{j,n +1}\sqrt{(j- n)(j +n+1)}
     +Y_{j,n-1}\sqrt{(j+ n)(j - n+1)})
F_{n}\right) dV,\label{jp23b}
\end{eqnarray}
and
\begin{eqnarray}
&&\hspace{-0.6cm}{T}^{y} = \left(\frac{2\pi GM_p }{5a^3}\right)\sum^{n=2}_{n=-2} \int_V\nabla\cdot\left( \rho  \left( \sum^{n'=2}_{n'=-2} \mbox {\boldmath$\xi$}_{n'}^*\right)\right) \nonumber\\
&&\hspace{-0.6cm}\nabla  
\left(  r^2( Y_{j,n-1}\sqrt{(j+ n)(j - n+1)}
-Y_{j,n +1}\sqrt{(j - n)(j+ n+1)}) 
F_{n}(t) \right)dV.\label{jp23c}
\end{eqnarray}

  Remarking that the azimuthal mode number associated with
  $\mbox {\boldmath$\xi$}_{n'}$ is $n^{'}$  and this has to be the same as that associated with the spherical harmonic 
with which it combines,  we see that (\ref{jp23b}) and (\ref{jp23c}) may be also be written in the form

\begin{eqnarray}
&& \hspace{-0.56cm}{T}^{x} = -\left(\frac{2\pi GM_p }{5a^3}\right)\sum^{n=2}_{n=-2}
\left[ \int_V \nabla\cdot \left( \rho \mbox {\boldmath$\xi$}_{n+1}^*\right ) 
  \left( {\rm i}r^2
   (  Y_{j,n +1}\sqrt{(j- n)(j +n+1)}   F_{n}\right) dV \right. \nonumber\\
&&\hspace{-0.6cm}+\left. \int_V \nabla\cdot\left( \rho\mbox {\boldmath$\xi$}_{n-1}^*\right ) 
  \left( {\rm i}r^2
     Y_{j,n -1}\sqrt{(j+n)(j -n+1)}   F_{n}\right) dV \right]
\label{jp24}
\end{eqnarray}
and
\begin{eqnarray}
&& \hspace{-0.56cm}{T}^{y} = \left(\frac{2\pi GM_p }{5a^3}\right)
\sum^{n=2}_{n=-2}\left[ \int_V 
\nabla\cdot\left(\rho \mbox {\boldmath$\xi$}_{n-1}^*\right ) 
 \left( r^2
     Y_{j,n -1}\sqrt{(j+ n)(j -n+1)}   F_{n}\right) dV \right. \nonumber\\
&&\hspace{-0.6cm}-\left. \int_V  \nabla\cdot\left(\rho \mbox {\boldmath$\xi$}_{n+1}^*\right ) 
\left( r^2
   Y_{j,n +1}\sqrt{(j-n)(j +n+1)}   F_{n}\right) dV \right]
\label{jp24a}
\end{eqnarray}
Note that  if $n \rightarrow -n$ in each of the terms in the above sums,
 as remarked above the complex conjugates of the expressions are obtained.
Thus, total torque components are obtained by taking twice the real parts and summing over $n > 0$ and then adding in the contribution for $n=0$
which  can be seen to be real.  We thus write the torque components in the form


\begin{equation}
{T}^{z} = -\sum_{n=0}^{n=2}2{\cal R}\left[\left(\frac{4\pi GM_p }{5a^3}\right)\int_V 
\nabla \cdot (\rho\mbox {\boldmath$\xi$}_{n}^* )  \left( {\rm i}n r^2
Y_{2,n}(\theta, \phi)F_{n}(t) \right)dV\right],\label{jp22i}
\end{equation}

\begin{eqnarray}
&& \hspace{-0.56cm}\frac{5a^3{T}^{x} }{2\pi GM_p}= \sum_{n=0}^{n=2}(\delta^{n}_{0}-2){\cal R}\left[ \int_V \nabla \cdot (\rho \mbox {\boldmath$\xi$}_{n+1}^*) 
   \left( {\rm i}r^2
   (  Y_{j,n +1}\sqrt{(j- n)(j +n+1)}   F_{n}\right) dV \right. \nonumber\\
&&\hspace{-0.6cm}+\left. \int_V \nabla (\cdot \rho   \left(\mbox {\boldmath$\xi$}_{n-1}^*\right )) 
  \left( {\rm i}r^2
     Y_{j,n -1}\sqrt{(j+n)(j -n+1)}   F_{n}\right) dV \right]
\label{jp24i}
\end{eqnarray}
and
\begin{eqnarray}
&& \hspace{-0.56cm}\frac{ 5a^3{T}^{y}}{2\pi GM_p} = \sum_{n=0}^{n=2}(2-\delta^{n}_{0}){\cal R}
\left[ \int_V \nabla\cdot \left(\rho \mbox {\boldmath$\xi$}_{n-1}^*\right ) 
  \left( r^2Y_{j,n -1}\sqrt{(j+ n)(j -n+1)}   F_{n}\right) dV \right. \nonumber\\
&&\hspace{-0.6cm}-\left. \int_V \nabla \cdot \left( \rho \mbox {\boldmath$\xi$}_{n+1}^*\right )  
 \left( r^2  Y_{j,n +1}\sqrt{(j-n)(j +n+1)}   F_{n}\right) dV \right],
\label{jp24ai}
\end{eqnarray}
where $\delta^{n}_{0}$ is the Kronecker $\delta.$

 We comment that when calculating the $X$ and $Y$ components
of the torque, the above equations imply that the azimuthal mode number of  a significant  response has to differ
from that of the  original forcing potential by $\pm 1.$
Apart from this,  the expressions  consist of contributions that are similar in form
to  that given by equation (\ref{jp22}) for the component of the torque in the $Z$  direction.

\section{Useful sum rules, time integrals taken around the orbit and their use to calculate  time averaged components of the torque
and the rate of change of orbital energy }\label{Aptima}


We find it convenient to make use of several sum rules derived from Parseval's theorem
when evaluating sums of the form (\ref{aveKepap}).
To obtain these we recall the Fourier expansion  (\ref{jpe2}) defining $\phi_{kn}$, which may also be  written in the form
\begin{equation}
{\cal F}_n(t)= \frac{\exp(-{\rm i}n (\Phi-\varpi)) a^3}{R^3}  = \sum_{k=-\infty}^{k=\infty} \phi_{k,n}\exp({\rm i}k
 n_o t),
\label{jpe2again0}
\end{equation}  
With the help of equation (\ref{jpe711e}) we then reconstruct (\ref{FdA1}) in the form
\begin{equation}
A_{n}(t) =    \sum_{k=-\infty}^{k=\infty} {\cal A}_{n,k}\exp({\rm i}k n_o t)=-\frac{4\pi G M_p }{5a^3} \sum_{m=0, |2|}^{'}{\cal F}_nD^{(2)}_{n,m}Y_{2,m}(\pi/2,0)\exp(-{\rm i}m\varpi)  ,
\label{jpe2again}
\end{equation}  
By considering the time derivative we also have
\begin{equation}
\frac{ dA_{n}(t)}{dt} ={\rm i}n_o \sum_{k=-\infty}^{k=\infty} k A_{n,k}\exp({\rm i}k n_o t),
\label{jpe3again}
\end{equation}  
From these two expressions we obtain the sum rules
\begin{equation}
\frac{n_o}{2\pi}\oint A^*_{n_1}(t) A_{n_2}(t)dt = 
 \sum_{k=-\infty}^{k=\infty} {\cal A}^*_{n_1,k}{\cal A}_{n_2,k}
\label{jpe2againa}
 \end{equation}
 \begin{equation}
\frac{n_o}{2\pi}\oint A^*_{n_1}(t)\frac{d A_{n_2} (t)}{dt} dt = 
 {\rm i}n_o \sum_{k=-\infty}^{k=\infty} k{\cal A}^*_{n_1,k}{\cal A}_{n_2,k},
\label{jpe3againa}
\end{equation}
\begin{equation}
\frac{n_o}{2\pi}\oint  \frac{d A_{n_1}^{*}(t)}{dt}        \frac{d A_{n_2}(t)}{dt}     dt = 
n_o^2 \sum_{k=-\infty}^{k=\infty} k^2{\cal A}^*_{n_1,k}{\cal A}_{n_2,k},
\label{jpe4againa}
\end{equation}

\noindent Then,  making use of  (\ref{jpe2againa}) -(\ref{jpe4againa})  together with (\ref{jpe2again}) we readily obtain the sum rules

\begin{equation}
 \hspace{-0.4cm}\sum_{k=-\infty}^{k=\infty}{\cal A}_{n_1,k }^*{\cal A}_{n_2,k}
  = \left(\frac{4\pi GM_p}{5}\right)^2 \sum_{m=-4}^{m=4}\exp(-{\rm i}m(\varpi+\gamma)) {\cal W}_{n_1,n_2,m}^{(0)}\frac{n_o}{2\pi}\oint \cos(m(\Phi-\varpi))R^{-6} dt ,
 \label{jpe711ca}
\end{equation}
where
\begin{equation}
 {\cal W}_{n_1,n_2,m}^{(j)} = \sum_{n=max(-2,-2-m)}^{n=min(2,2-m)} (n+m/2)^j  d^{(2)}_{n_2 ,n+m} d^{(2)}_{n_1 ,n}Y_{2,n+m}(\pi/2, 0) Y_{2,n}(\pi/2, 0)\label{WCRP}
 \end{equation}
has the property, as can be verified directly, while noting that only even values of $n$ and $m$ give  non zero contributions to the sum and
 all the factors in (\ref{WCRP}) are real,  that  ${\cal W}_{n_1,n_2,m}^{(j)} = {\cal W}_{n_2,n_1,-m}^{(j)} .$
\begin{align}
&\sum_{k=-\infty}^{k=\infty} k {\cal A}_{n_1,k }^*{\cal A}_{n_2,k}
  =- \left(\frac{4\pi GM_p}{5}\right)^2\frac{\sqrt{G(M_p+M_*)a(1-e^2)}}{2\pi }\times \nonumber\\
  &\sum_{m=-4}^{m=4}\exp(-{\rm i}m(\varpi+\gamma))
   {\cal W}_{n_1,n_2,m}^{(1)} 
    \oint\frac{dt\cos(m(\Phi-\varpi))}{ R^{8}},  
 \label{jpe711cb}
\end{align}
\begin{align}
&\hspace{-0cm}\sum_{k=-\infty}^{k=\infty} k ^2{\cal A}_{n_1,k }^*{\cal A}_{n_2,k}
  =\left(\frac{4\pi GM_p}{5}\right)^2\frac{1}{2\pi n_o}\times\nonumber\\
  &\sum_{m=-4}^{m=4}\exp(-{\rm i}m(\varpi+\gamma))\oint \left({\cal W}_{n_1,n_2,m}^{(0)}{\cal P}_0+
\left( {\cal W}_{n_1,n_2,m}^{(2)} -\frac{m^2}{4}{\cal W}_{n_1,n_2,m}^{(0)} \right){\cal P}_2\right)\cos(m(\Phi-\varpi))dt,
 \label{jpe711cc}
\end{align}
with
\begin{align}
&{\cal P}_0=\frac{d}{dt}\left(\frac{3dR/dt}{R^7}\right)+ \frac{9(dR/dt)^2}{R^8},\hspace{2mm} {\rm and} \hspace{2mm} \nonumber\\
&{\cal P}_2 = \frac{G(M_p+M_*)a(1-e^2)}{R^{10}}.\label{P02eqa}
\end{align}

\subsection{Useful time integrals taken around the orbit}
For standard Keplerian elliptical orbits we also make use of the  { time integrals taken around the orbit
for $n$ equal to  a positive integer $>1.$}
\begin{align}
\oint\frac{dt}{R^{2n}}= \frac{2\pi(2n-2)!}{(a(1-e^2))^{2n-2}\sqrt{G(M_p+M_*)a(1-e^2)}}
\sum^{n-1}_{k=0} \frac{e^{2k}}{(2n-2-2k)! k!^22^{2k}},\label{tave1}
\end{align}
\begin{align}
\oint\frac{\cos(2(\Phi-\varpi)) dt}{R^{2n}}= \frac{2\pi(2n-2)!}{(a(1-e^2))^{2n-2}\sqrt{G(M_p+M_*)a(1-e^2)}}
\sum^{n-1}_{k=1} \frac{e^{2k}}{(2n-2-2k)! (k+1)!(k-1)!2^{2k}},
\label{tave2}
\end{align}
and
\begin{align}
\oint\frac{{\dot R}^2dt}{R^{2n}}= \frac{2\pi e^2(2n-2)!\sqrt{G(M_p+M_*)a(1-e^2)}}{(a(1-e^2))^{2n}}
\sum^{n-1}_{k=0} \frac{e^{2k}}{(2n-2-2k)! k! (k+1)!2^{2k+1}}\label{tave3}
\end{align}
which can be used to complete the evaluation of integrals involving quantities such as ${\cal P}_0$ 
and ${\cal P}_2$.


 By making use of the sum rules given by equations (\ref{jpe711ca})  - (\ref{jpe711cc})
 together with the time { integrals taken around the orbit}  expressed by equations (\ref{tave1}) - (\ref{tave3})
we can derive expressions for the { time averages of the  components of the torque given by equations (\ref{jp30}) - (\ref{jp32}), 
expressed as combinations of infinite sums
by making use  of  equations (\ref{aveGenen}) - (\ref {aveKepap}), in closed form. In a similar way we may find
 the rate of change of orbital energy given by (\ref{jp33})  with the help of the time derivative of (\ref{FdA1}). }

Thus the $Z$ component of the torque is  given by
\begin{align}
&{T}^{z} = \left(4\pi Q_{eq} GM_p/(5\omega_{eq})\right)^2\times\nonumber\\
&\left( \frac{\delta_1
\sqrt{G(M_p+M_*)a(1-e^2)}}{\pi}
\sum_{m=0}^{m=4}({\cal W}_{1,1,m}^{(1)}+2{\cal W}_{2,2,m}^{(1)})(2-\delta_0^m) \cos(m{\hat \varpi})
\oint\frac{\cos(m(\Phi-\varpi)) }{R^{8}} dt  \right.\nonumber\\
&\left.
-\delta_2\left(n_o/\pi\right)\sum_{m=0}^{m=4}({\cal W}_{1,1,m}^{(0)}+4{\cal W}_{2,2,m}^{(0)})(2-\delta_0^m)\cos (m{\hat \varpi})
\oint\frac{\cos(m(\Phi-\varpi)) }{R^{6}} dt \right),\label{TZ87}
\end{align}
where $e$ is the orbital eccentricity and 
 ${\hat \varpi}=\varpi +\gamma $. Here we remark that  only even values of $m$ contribute to the sums.

\textcolor{red} {Similarly,  working with $T\equiv {T}^{x}-{\rm i}{T}^{y}$ as given by (\ref{jp3132}) we find
\begin{align}
&T = - n_o Q_{eq} ^2/(\pi \omega_{eq}^2) \left(4\pi GM_p/5\right)^2 \left((\delta_2-{\rm i}\sigma\delta_3)
\oint (f_1+{\rm i}f_2)R^{-6} dt  
-\right.\nonumber\\
& \left. \frac{ \sqrt{G(M_p+M_*)a(1-e^2)}}{2 n_o }(2\delta_1-{\rm i}\delta_3)
\oint (f_3 +{\rm i}f_4) R^{-8} dt \right),\label{T88}
\end{align}
where}
\begin{eqnarray}
&\hspace{-8mm} f_1= \sum_{m=0}^{m=4}\left((3/2)({\cal W}_{1,2,m}^{(0)}+ {\cal W}_{2,1,m}^{(0)})+\sqrt{3/8} ( {\cal W}_{1,0,m}^{(0)}
+  {\cal W}_{0,1,m}^{(0)})\right)(2-\delta_0^m)\cos (m {\hat \varpi})\cos (m(\Phi -  \varpi)),  \nonumber
  \end{eqnarray} 
  \begin{eqnarray}
&\hspace{-3mm} f_2= \sum_{m=0}^{m=4}\left(  3({\cal W}_{2,1,m}^{(0)}- {\cal W}_{1,2,m}^{(0)})-\sqrt{3/2} ( {\cal W}_{0,1,m}^{(0)}
- {\cal W}_{1,0,m}^{(0)})\right)\sin (m {\hat \varpi})\cos (m(\Phi -  \varpi)), \nonumber
\end{eqnarray}
\begin{eqnarray}
&\hspace{-3mm} f_3= \sum_{m=0}^{m=4}\left(({\cal W}_{1,2,m}^{(1)}+{\cal W}_{2,1,m}^{(1)})
+\sqrt{3/2}({\cal W}_{1,0,m}^{(1)}+{\cal W}_{0,1,m}^{(1)})\right)(2-\delta_0^m)\cos (m {\hat \varpi})\cos (m(\Phi -  \varpi)),\nonumber
\end{eqnarray}
and
\begin{eqnarray}
&\hspace{-3mm} f_4= 2\sum_{m=0}^{m=4}\left(({\cal W}_{2,1,m}^{(1)}- {\cal W}_{1,2,m}^{(1)})-\sqrt{3/2} ( {\cal W}_{0,1,m}^{(1)}
- {\cal W}_{1,0,m}^{(1)})\right)\sin (m {\hat \varpi})\cos (m(\Phi -  \varpi)).  \nonumber
\end{eqnarray}

\noindent The rate of change of orbital energy is in turn given by
\begin{equation}
\frac{dE_{orb}}{dt} =\frac{ n_o 
Q_{eq}^2}{\pi \omega_{eq}^2 }\left(\frac{4\pi GM_p}{5}\right)^2
\left(\delta_2 \sqrt{G(M_p+M_*)a(1-e^2)}
  \oint f_5 R^{-8} dt 
-\delta_1\left(1/n_o\right)f_{6}\right),\label{EDOT90}
\end{equation}
where
\begin{align}
&\hspace{-3mm} f_5= \sum_{m=0}^{m=4}\left(({\cal W}_{1,1,m}^{(1)}+2{\cal W}_{2,2,m}^{(1)})\right)(2-\delta_0^m)\cos (m {\hat \varpi})\cos (m(\Phi -  \varpi)),\hspace{3mm}{\rm and} \nonumber
\end{align}
\begin{align}
&f_{6}= \sum_{m=0}^{m=4} \oint\left( \left(({\cal W}_{2,2,m}^{(2)}+{\cal W}_{1,1,m}^{(2)}   + \frac{1}{2}{\cal W}_{0,0,m}^{(2)} )
-\frac{m^2}{4}({\cal W}_{2,2,m}^{(0)}+{\cal W}_{1,1,m}^{(0)}   +\frac{1}{2} {\cal W}_{0,0,m}^{(0)} )\right){\cal P}_2+\right.\nonumber\\
&\left. \left ( {\cal W}_{2,2,m}^{(0)} + {\cal W}_{1,1,m}^{(0)} + {\cal W}_{0,0,m}^{(0)}/2\right){\cal P}_0\bigg) \right.(2-\delta_0^m)\cos (m {\hat \varpi})\cos (m(\Phi -  \varpi))dt,
\label{f5f6}
\end{align}
where
\begin{align}
&{\cal P}_0=\frac{d}{dt}\left(\frac{3dR/dt}{R^7}\right)+ \frac{9(dR/dt)^2}{R^8},\hspace{2mm} {\rm and} \hspace{2mm} 
{\cal P}_2 =\frac{G(M_p+M_*)a(1-e^2)}
{R^{10}}.\label{P02eq}
\end{align}



\subsubsection{Torque and rate of change of orbital energy 
in terms of the inclination $\beta$}\label{torqueWigner}
The components of the Wigner $d$ matrix are given in Appendix \ref{Wigner}
and  \textcolor{red}{evaluation}  of relevant ${\cal W}^{(j)}_{n_1,n_2}$ in 
Appendix \ref{WignerW}.
Using results provided there 
the $Z$ component of the torque is found to be
\begin{align}
&{T}^{z} = 
\frac{6}{5}\left( Q _{eq}GM_p/\omega_{eq}\right)^2
\left(
\delta_1 \sqrt{G(M_p+M_*)a(1-e^2)}
\cos\beta
\oint R^{-8} dt
\right.\nonumber\\
&\left.
-\delta_2\left(
\frac{ n_o}{2}\right)\left(\left (1+\cos^2\beta\right)
\oint R^{-6} dt - \sin^2\beta\cos2{\hat \varpi}   \oint \cos (2(\Phi -\varpi))R^{-6} dt  \right)\right).\label{TZF}
\end{align}
Similarly, from  Appendix \ref{WignerW} we find 
$f_1 =15\sin\beta\cos\beta(1+\cos (2{\hat \varpi})\cos (2(\Phi -\varpi)) )/(16\pi),\hspace{3mm}\\$ 
 $f_2 =-15\sin\beta\sin (2{\hat \varpi} )\cos (2(\Phi -\varpi))/(16\pi),\hspace{3mm}$
$f_3 =15\sin\beta/(8\pi),\hspace{3mm}$ and $f_4 =0.$ \textcolor{red}{Accordingly,
\begin{align}
&T =-\frac{ 3 n_o (GM_pQ_{eq}) ^2\sin\beta}{5\omega_{eq}^2}\times\nonumber\\
& \left(  (\delta_2-{\rm i}\sigma\delta_3)
\oint \frac{((\cos\beta \cos(2{\hat \varpi})-{\rm i}\sin (2{\hat \varpi}) )\cos(2(\Phi -  \varpi))+\cos\beta)}{R^{6} }dt  -
\right.\nonumber\\
& \left. 
\frac{ \sqrt{G(M_p+M_*)a(1-e^2)}}{2 n_o }(2\delta_1-{\rm i}\delta_3)
\oint \frac {2}{ R^{8} }dt \right).\label{TperpF}
\end{align}
}

\noindent \textcolor{red}{ In addition,} we find   that $f_5 = 15\cos\beta/(8\pi) \hspace{3mm} $   and $\hspace{3mm} f_6 = (5{\cal P}_0   + 15{\cal P}_2)/(8\pi),\hspace{3mm}$ 
and thus the rate of change of orbital energy is given by
\begin{align}
&\frac{dE_{orb}}{dt} =\frac{ n_o
Q_{eq}^2}{\pi \omega_{eq}^2 }\left(\frac{4\pi GM_p}{5}\right)^2
\left(\delta_2\sqrt{G(M_p+M_*)a(1-e^2)}
 \frac{15}{8\pi} \cos\beta \oint R^{-8} dt 
\right.\nonumber\\
&\left. 
-\delta_1\left(1/n_o\right) \oint\left( \frac{5}{8\pi}{\cal P}_0+
 \frac{15}{8\pi}{\cal P}_2\right)dt \right).\label{EDOTF}
\end{align}
\subsubsection{Evaluation of the time integrals around the orbit}\label{IntevalC}
The evaluation of the above quantities  is completed
with help of the integrals calculated above.
Making use of  (\ref{tave1})  - (\ref{tave3})  after taking note of (\ref{P02eq}) 
to evaluate the integrals in  (\ref{TZF}) - (\ref{EDOTF}),  we obtain
\begin{align}
&\hspace{-2.3cm}{T}^{z} = T_*\left(
2\delta_1 \cos\beta
\phi_1
-\delta_2(1-e^2)^{3/2}
\left(\left (1+\cos^2\beta\right)
\phi_2- \sin^2\beta\cos2{\hat \varpi} \phi_3  \right)\right)\hspace{3mm}{\rm and}\hspace{3mm}
\label{t1} 
\end{align}
\begin{align}
\hspace{-3mm}T = T_*\sin \beta ((2\delta_1-{\rm i}\delta_3)\phi_1
-( 1-e^2)^{3/2}(\delta_2-{\rm i}\sigma\delta_3)
\left((\phi_2+\phi_3\cos(2{\hat \varpi}))\cos\beta - {\rm i} \sin(2{\hat \varpi} )\phi_3\right)),
\label{t2} 
\end{align}
The  change of orbital energy is given by
\begin{align}
&\frac{dE_{orb}}{dt} =\dot E_*
\left(\delta_2\phi_1 \cos\beta 
-\frac{\delta_1}{(1-e^2)^{3/2}}
\phi_4 
 \right).
\label{eng1} 
\end{align}
Here
\begin{equation}
T_*= \frac{6\pi}{5}\left(\frac{ GM_pQ_{eq}}{ a^3(1-e^2)^3\omega_{eq}}\right)^2 = \frac{3k_2q^2}{1+q}
\left(\frac{R_*^5}{a^5}\right) \frac{M_*n_o^2a^2}{(1-e^2)^6} \quad {\rm and} \quad
\dot E_*=2n_o T_*,
\label{eqadd1}
\end{equation}
respectively, represent typical values of the torque and rate of change of  energy. Note that in the second equality we have used
equation (\ref{apsidal}) in Appendix \ref{newPrecession} with $N_0=1$ to relate $T_*$ to the apsidal motion constant.
 In addition, 
\begin{align}
\phi_1=1+\frac{15}{2}e^2+ \frac{45}{8}e^4+\frac{5}{16}e^6,\label{phi1}
\end{align}
\begin{align}
\phi_2=1+3e^2+ \frac{3}{8}e^4,\label{phi2}
\end{align}
\begin{align}
\phi_3=\frac{3}{2}e^2+ \frac{1}{4}e^4 \hspace{3mm}{\rm and}\label{phi3}
\end{align}
\begin{align}
\phi_4=1+{31\over 2}e^2+ \frac{255}{8}e^4+\frac{185}{16}e^6  +\frac{25}{64}e^8. \hspace{3mm}
\label{phi}
\end{align}

\textcolor{red}{
\section{Precession due to second order rotational distortion: A comparison with the first order contribution }\label{newPrecession}
We here compare the component of the torque in the $Y$ direction that is  $\propto \beta_*+1/2$ and first order in the
stellar rotation frequency  to the conventional precessional torque
generated by the coupling of the tidal forcing to the density perturbation produced by centrifugal distortion.
The action of these two torques, which may be added together causes precession of the rotation axis but does not
play any role in the dissipative tidal interaction.}

\textcolor{red}{We consider without loss of generality the simplified circumstance of an eccentric  orbit inclined by an angle $\beta$ to the $(X,Y)$ plane  with both the line of nodes 
and the line of apsides coinciding with
the $Y$ axis in the stellar frame. But note that it is not difficult to show that  that the time averaged component of the potential arising
from the companion that we require is independent of the longitude of the apsidal line.
The effective perturbing potential  arising from the companion that we need is the time averaged quadrupole component 
with $m=1$  as it is this component that gives rise to a precessional torque to lowest order in the tidal perturbation.
After performing a multipole expansion and taking a time average, this is given by  
\begin{eqnarray}
U = \frac{3GM_p}{2a(1-e^2)^{3/2}}\left(\frac{r}{a}\right)^2\sin\beta\cos\beta \sin\theta\cos\theta\cos\phi \label{UapD}
\end{eqnarray}
The interaction with the axisymmetric density distribution resulting from
rotational distortion produces a torque with $Y$ component
\begin{eqnarray}
\hspace{-1mm}T_{SF}^y=  -\int \rho({\bf r}\times \nabla U)_y dV = -\frac {3\pi GM_p}{2a^3(1-e^2)^{3/2}}\sin\beta\cos\beta \int \rho(r,\theta)(3\cos^2\theta -1) \sin\theta r^4d\theta dr \hspace{2mm}
\end{eqnarray}
or equivalently noting that $\rho(r,\theta)$ can be replaced by  $\rho'(r,\theta) $ being the response to the
perturbing potential (\ref{UapD}). In the case discussed here this is also equal to the equilibrium tide response
determined by equation (\ref{Bs2}),   thus $\rho'(r,\theta)\equiv \rho'_{eq} $ 
\begin{eqnarray}
T_{SF}^y =  -\frac {3 GM_p}{2a^3(1-e^2)^{3/2}}\sqrt{\frac{4\pi}{5}}\sin\beta\cos\beta \int \rho'_{eq} r^2Y_{2,0}(\theta)
dV\label{torqueCY}
\end{eqnarray}
The centrifugal distorting potential is
\begin{eqnarray}
U =  \frac {\Omega_r^2}{3}\sqrt{\frac{4\pi}{5}} r^2Y_{2,0}(\theta)-\frac {\Omega_r^2r^2}{3}.
\end{eqnarray}
The second spherically symmetric term plays no role in determining the torques acting and so may be dropped.
Aligning with the  notation in Section \ref{Respd} (see e.g. equation (\ref{potpert}) ) we write
\begin{eqnarray}
U =  A_0r^2Y_{2,0}(\theta)\hspace{3mm}{\rm where}\hspace{3mm} A_0 = \frac {\Omega_r^2}{3}\sqrt{\frac{4\pi}{5}}. 
\end{eqnarray}
We then determine the equilibrium tide displacement response $\mbox{{\boldmath$\xi$}}_{eq}$  
using the formalism of Section~\ref{Respd}.  Thus, we use equation (\ref{Bs2}) with the above forcing potential.
From the appropriately adapted (\ref{EREV}), noting that in this case the operator  $\hat L_1 \equiv 0,$ one  finds that
\begin{eqnarray}
A_0\frac{  \int r^2Y_{2,0}(\theta) \rho'_{eq} dV}{N_0\omega^2_{eq}}  = - 1 \label{fuff}
\end{eqnarray}
We note in passing that the integral in the above equation is proportional to the gravitational potential
perturbation at the stellar surface  and the ratio of this to the forcing potential there
is defined to be twice the apsidal motion constant $k_2$.
Accordingly, we have
\begin{align}
&\hspace{-7cm} 2k_2 = \frac{4\pi G Q_{eq}^2}{5N_0 {\omega^2_{eq}}R_*^5}= 
 \frac{4\pi {\tilde Q}^2}{5}\left( \frac{GM_*}{ {R_*^3\omega^2_{eq}}} \right) ,\hspace{3cm}\label{apsidal}
\end{align}
where $Q_{eq}$ has been expressed in terms of its dimensionless form ${\tilde Q}$ with $N_0$ chosen to be unity
in the second expression on the right.}\newline
\textcolor{red}{\noindent Using (\ref{fuff}) together with (\ref{torqueCY}) we obtain 
\begin{eqnarray}
T_{SF}^y =  \frac {3A_0 GM_p}{2a^3(1-e^2)^{3/2}\omega_{eq}^2N_0}   \sqrt{\frac{4\pi}{5}}\sin\beta\cos\beta\left( \int \rho'_{eq} r^2Y_{2,0}(\theta)
dV\right)^2.\label{D1}
\end{eqnarray}
In this form we remark that $\mbox{{\boldmath$\xi$}}_{eq}$ can be rescaled (normalised) such that $N_0 =1.$
Noting that last term in brackets on the right hand side of (\ref{D1}) is $Q_{eq}^2$ evaluated with $\mbox{{\boldmath$\xi$}}_{eq}$
(see equation (\ref{e7X}) ) we obtain
\begin{eqnarray}
T_{SF}^y =  \frac { 2\pi GM_p\Omega_r^2Q_{eq}^2}{5a^3(1-e^2)^{3/2}\omega_{eq}^2} \sin\beta\cos\beta.
\label{D2}
\end{eqnarray}
This torque can be added to $T^y$ in the third  equation in the set (\ref{e45}) which reads
\newline
${d{\bar \alpha}}/dt=-d\alpha_r/dt=T^yJ/(LS\sin\beta)$
in order to include the effects of second order centrifugal distortion.}

\section{  Elements of the Wigner ${\bf {\lowercase { d}} }$ matrix}\label{Wigner}
These can all be obtained from (see e.g. \cite{KMV})
\begin{align}
&d_{2,2}= \frac{1}{4}(1+\cos\beta)^2,\nonumber\\
&d_{2,1}= -\frac{1}{2}\sin\beta(1+\cos\beta),\nonumber\\
&d_{2,0}= \sqrt{\frac{3}{8}}\sin^2\beta,\nonumber\\
&d_{2,-1}= -\frac{1}{2}\sin\beta(1-\cos\beta),\nonumber\\
&d_{2,-2}= \frac{1}{4}(1-\cos\beta)^2,\nonumber\\
&d_{1,1}= \frac{1}{2}(2\cos^2\beta+\cos\beta -1),\nonumber\\
&d_{1,0}= -2\sqrt{\frac{3}{8}}\sin\beta\cos\beta,\nonumber\\
&d_{1,-1}= \frac{1}{2}(-2\cos^2\beta+\cos\beta +1),\nonumber\\
&d_{0,0}= \frac{1}{2}(3\cos^2\beta -1).
\end{align}
Note that for ease of notation we have dropped the superscript $2$ and that components not listed can be obtained from those listed by making use of 
the relations $d_{n_1,n_2}(\beta) =(-1)^{n_1-n_2} d_{-n_1,-n_2}(\beta)$
and $d_{n_1,n_2}(\beta) = d_{n_2,n_1}(-\beta).$
\section{Quantities required for torque and rate of change of orbital energy evaluation}\label{WignerW}
We here give explicit expressions for some  combinations of Wigner
matrix elements that are required for evaluation of torques
and the rate of change of orbital energy.
\subsubsection{Terms associated with $m=4$}
For $m=4,$ we note that from (\ref{WCRP}) we have
\begin{align}
&{\cal W}_{n_1,n_2,4}^{(0)}= \frac{15}{32\pi}d_{n_2,2}d_{ n_1,-2},\label{wigm4}
\end{align}
\begin{align}
 &\hspace{-4.8cm} {\rm while}  \hspace{2mm} {\rm for}  \hspace{2mm} j > 0, \hspace{2mm}
 {\cal W}_{n_1,n_2,4}^{(j)}=0  \hspace{3mm}   {\rm identically.}\hspace{3mm} {\rm From}\hspace{2mm}  (\ref{wigm4})  \hspace{3mm} {\rm we}\hspace{2mm} {\rm find}\nonumber
  \end{align}
   \begin{align}
 &  {\cal W}_{1,1,4}^{(0)}=  -\frac{15}{128\pi}\sin^4\beta \hspace{3mm} , {\cal W}_{2,2,4}^{(0)}=-{\cal W}_{1,1,4}^{(0)}/4 \hspace{2mm}{\rm and } 
\hspace{2mm} {\cal W}_{0,0,4}^{(0)}=-3{\cal W}_{2,2,4}^{(0)}/2. \nonumber
 \end{align}
In addition we find
\begin{align}
&{\cal W}_{2,1,4}^{(0)}= -\frac{1}{4}\sqrt{8/3} {\cal W}_{1,0,4}^{(0)}= \frac{15}{256\pi}\sin^3\beta(1-\cos\beta) \hspace{3mm} {\rm and}\nonumber\\
&{\cal W}_{1,2,4}^{(0)}= -\frac{1}{4}\sqrt{8/3} {\cal W}_{0,1,4}^{(0)}= - \frac{15}{256\pi}\sin^3\beta(1+\cos\beta).\nonumber
\end{align}
Using the above results it is readily found that through a series of cancellations there are no contributions to $f_1 - f_6$
and hence the orbital evolution equations
proportional to $\cos(4{\hat \varpi })$ or  $\sin(4{\hat \varpi }).$
\subsubsection{Terms associated with $m=2$}
For $m=2,$ we note that from (\ref{WCRP}) we have
\begin{align}
&{\cal W}_{n_1,n_2,2}^{(j)}= -\frac{5}{16\pi}\sqrt{3/2}( (-1)^jd_{n_2,0}d_{n_1,-2} +  d_{n_2,2} d _{ n_1,0} ).\label{wmig2}
\end{align}
Thus, coefficients with $j=2$ are identical to those with $j=0$ and so need not be considered  separately.
From (\ref{wmig2})  we obtain  considering first $j=1$
  \begin{align}
 &  {\cal W}_{1,1,2}^{(1)} = -2{\cal W}_{2,2,2}^{(1)}=  \frac{15}{32\pi}\sin^2\beta\cos\beta \hspace{3mm},\nonumber\\
& {\cal W}_{1,0,2}^{(1)} =-  \sqrt{2/3} {\cal W}_{2,1,2}^{(1)}= \frac{5}{32\pi}\sqrt{3/8} \sin\beta(1-\cos\beta) (1+3\cos\beta) \hspace{3mm}{\rm and} \nonumber
 \end{align}
   \begin{align}
&\hspace{-9mm} {\cal W}_{0,1,2}^{(1)} = -\sqrt{2/3}  {\cal W}_{1,2,2}^{(1)} = \frac{5}{32\pi}\sqrt{3/8} \sin\beta(1+\cos\beta) (1-3\cos\beta) . \nonumber
 \end{align}
  Considering  the case $j=0,$ from (\ref{wmig2}) we find
   \begin{align}
 &  {\cal W}_{1,1,2}^{(0)} =  \frac{15}{32\pi}\sin^2\beta\cos^2\beta,  {\cal W}_{2,2,2}^{(0)} = - \frac{15}{128\pi}\sin^2\beta(1+\cos^2\beta) , 
 {\cal W}_{0,0,2}^{(0)} = - \frac{15}{64\pi}\sin^2\beta(3\cos^2\beta-1), \nonumber\\
& {\cal W}_{1,0,2}^{(0)} =  \frac{5}{32\pi}\sqrt{3/8} \sin\beta(1-\cos\beta) (6\cos^2\beta+3\cos\beta-1) , \nonumber\\
& {\cal W}_{0,1,2}^{(0)}  = \frac{5}{32\pi}\sqrt{3/8} \sin\beta(1+\cos\beta) (1+3\cos\beta - 6\cos\beta^2) ,\nonumber\\
 &  {\cal W}_{2,1,2}^{(0)} =  -\frac{15}{128\pi}\sin\beta(1+\cos^2\beta-2\cos^3\beta)  \hspace{3mm}{\rm and}\nonumber\\
 &  {\cal W}_{1,2,2}^{(0)} =  \frac{15}{128\pi}\sin\beta(1+\cos\beta)(1-\cos\beta+2\cos^2\beta)  . \nonumber
 \end{align}
Using the above results the contributions to $f_1 - f_6$
and hence the orbital evolution equations that are $\propto$ $\cos{2\hat \varpi}$ and $\sin2{\hat \varpi}$ are readily found.
 
\subsubsection{Terms associated with $m=0$}
For $m=0,$  from (\ref{WCRP}) we have
\begin{align}
&\hspace{-6.3cm}{\cal W}_{n_1,n_2,0}^{(0)}= \frac{15}{32\pi}(d_{n_2,2}d_{n_1,2}+d_{n_2,-2}d_{n_1,-2})
+\frac{5}{16\pi}d_{n_2,0}d_{n_1,0},\nonumber
\end{align} 
\begin{align}
&\hspace{-4mm}{\cal W}_{n_1,n_2,0}^{(1)}= \frac{15}{16\pi}(d_{n_2,2}d_{n_1,2}-d_{ n_2,-2}d_{n_1,-2})\hspace{2mm}{\rm and}
\hspace{2mm} {\cal W}_{n_1,n_2,0}^{(2)}= \frac{15}{8\pi}(d_{n_2,2}d_{n_1,2}+d_{n_2,-2}d_{n_1,-2}).\label{wmigm0}
\end{align}
The above expressions lead to
\begin{align}
&\hspace{-3cm} {\cal W}_{1,1,0}^{(0)}= \frac{15}{64\pi}\sin^2\beta(1+3\cos^2\beta),\hspace{3mm}
{\cal W}_{1,0,0}^{(0)}= \frac{5}{32\pi}
\sqrt{\frac{3}{8}}
\sin\beta\cos\beta(5-9\cos^2\beta),
\nonumber
\end{align}
\begin{align}
&\hspace{-2.5cm} {\cal W}_{1,2,0}^{(0)}= \frac{15}{128\pi}\sin\beta\cos\beta(1+3\cos^2\beta),
\hspace{1mm} {\cal W}_{2,2,0}^{(0)}= \frac{15}{256\pi}(3+2\cos^2\beta+3\cos^4\beta ),
\nonumber
\end{align}
\begin{align}
&\hspace{-2.5cm} {\cal W}_{0,0,0}^{(0)}= \frac{5}{128\pi}(11-3\cos^2\beta-27\sin^2\beta \cos^2\beta ),\hspace{1mm}
{\cal W}_{2,2,0}^{(1)}= \frac{15}{32\pi}\cos\beta(1+\cos^2\beta),
\nonumber
\end{align}
\begin{align}
&\hspace{-2cm}{\cal W}_{1,1,0}^{(1)}= \frac{15}{16\pi}\sin^2\beta \cos\beta, \hspace{1mm}
\nonumber
{\cal W}_{1,0,0}^{(1)}= \frac{15}{16\pi}\sqrt{\frac{3}{8}}\sin^3\beta ,\hspace{1mm}
\nonumber
{\cal W}_{2,1,0}^{(1)}= \frac{15}{64\pi}\sin\beta(1+3\cos^2\beta),
\nonumber
\end{align}
\begin{align}
&{\cal W}_{1,1,0}^{(2)}= \frac{15}{16\pi}\sin^2\beta(1+\cos^2\beta),\hspace{1mm}
\nonumber
{\cal W}_{0,0.0}^{(2)}= \frac{45}{32\pi}\sin^4\beta, \hspace{1mm}{\rm and}\hspace{1mm}
\nonumber
{\cal W}_{2,2,0}^{(2)}= \frac{15}{64\pi}(1+\cos^4\beta +6\cos^2\beta) .
\nonumber
\end{align}
From these  one can verify the identities
\begin{align}
&\hspace{-8.2cm} {\cal W}_{2,2,0}^{(0)}+{\cal W}_{1,1,0}^{(0)}   + {\cal W}_{0,0.0}^{(0)}/2 =\frac{5}{8\pi},
\nonumber
\end{align}
\begin{align}
&\hspace{-8.2cm} {\cal W}_{2,2,0}^{(2)}+{\cal W}_{1,1}^{(2)}   + {\cal W}_{0,0,0}^{(2)}/2 =\frac{15}{8\pi},
\nonumber
\end{align}
\begin{align}
\hspace{-8.2cm}{\rm and}\hspace{3mm} 2{\cal W}_{2,2,0}^{(1)}+{\cal W}_{1,1,0}^{(1)} =\frac{15}{8\pi}\cos\beta.
\nonumber
\end{align}
Recalling  the symmetry condition that  ${\cal W}_{n_1,n_2,0}^{(j)}= {\cal W}_{n_2,n_1,0}^{(j)},$
the above coefficients expressed in terms of $\beta$ can be used to evaluate the contribution of terms with $m=0$
to the  quantities $f_1 - f_6$ and  thus the components
of the torque and the rate of change of orbital energy. 

\end{appendix}




\label{lastpage}

\end{document}